\documentclass{jfm}
\usepackage{graphicx}
\usepackage{epstopdf, epsfig}
\usepackage{slashbox}
\usepackage{xfrac} 

\usepackage{caption}
\usepackage{subcaption}
\usepackage{color}
\usepackage{mathtools}
\usepackage{siunitx}
\usepackage{todonotes}
\usepackage{multirow}
\usepackage{amsmath}
\usepackage{lscape}
\usepackage{float}
\usepackage{placeins}
\usepackage{nicefrac}
\usepackage[utf8]{inputenc}
\usepackage{dirtytalk}
\usepackage{dblfloatfix}
\usepackage{diagbox}
\usepackage{hyperref}
\hypersetup{colorlinks=true,urlcolor=blue,linkcolor=blue, anchorcolor=blue, citecolor=blue}

\usepackage{adjustbox}

\newcommand\cites[1]{\citeauthor{#1}'s\ (\citeyear{#1})}

\shorttitle{Micro-droplet impact onto a scratch}
\shortauthor{K. H. A. Al-Ghaithi, O. G. Harlen, N. Kapur, and M. C. T. Wilson}

\title{Morphologies and dynamics of micro-droplet impact onto an idealised scratch}

\author{Khaled H. A. Al-Ghaithi\aff{1},
  Oliver G. Harlen,\aff{2} Nikil Kapur\aff{3}, \and Mark C. T. Wilson\aff{3}\corresp{\email{m.wilson@leeds.ac.uk}}}

\affiliation{\aff{1}EPSRC Centre for Doctoral Training in Fluid Dynamics, University of Leeds, Leeds, LS2 9JT, UK
\aff{2}School of Mathematics, University of Leeds, Leeds, LS2 9JT, UK
\aff{3}School of Mechanical Engineering, University of Leeds, Leeds, LS2 9JT, UK}

\begin{document}

\maketitle

---------------------------------------------------------------------------------------------------------------
This article has been published in a revised form in the Journal of Fluid mechanics \href{https://doi.org/10.1017/jfm.2021.638}{https://doi.org/10.1017/jfm.2021.638}. This version is published under a Creative Commons CC-BY-NC-ND. No commercial re-distribution or re-use allowed. Derivative works cannot be distributed. © The Authors.  2021.

---------------------------------------------------------------------------------------------------------------
\begin{abstract}
\textbf{Abstract:} As inkjet technology develops to produce smaller droplets, substrate features such as accidental scratches or manufacturing defects can potentially affect the outcome of printing, particularly for printed electronics where continuous tracks are required. Here, the deposition of micro-droplets onto a scratch of commensurate size is studied. The scratch is considered as a groove of rectangular cross-section, with rectangular side ridges representing material displaced from the substrate, and seven equilibrium morphologies are identified as a result of inertial spreading, contact--line pinning, imbibition into the scratch and capillary flow. A regime map is constructed in terms of scratch depth and width, and theoretical estimates of the regime boundaries are developed by adapting droplet spreading laws for flat surfaces to account for liquid entering the scratches. Good agreement is seen with numerical results obtained using a GPU-accelerated three--dimensional multiphase lattice Boltzmann model validated against published experiments, and the influences of Reynolds number, Weber number and advancing and receding contact angles are explored. Negative and positive implications of the results for printing applications are discussed and illustrated via multiple-droplet simulations of printing across and along scratches.
\end{abstract}

\begin{keywords}
\end{keywords}

\section{Introduction}
The impact and coalescence of inkjet droplets on a solid substrate are of paramount importance to several industries including printed electronics, ceramic and tile decoration and printing biological materials. Inkjet technology in the printed electronics industry has received increasing interest due to the method's potential to reduce the manufacturing costs of some devices \citep{Soltman2011}. Example devices include passive circuit elements, organic transistors, organic light-emitting diodes, sensors and radio frequency identification tags \citep{Kwon2018, Soltman2011}. Small imperfections in the substrate surface can arise through small variations during manufacturing, or as a result of unintended damage, such as scratching during transportation and/or handling of the substrate. These can pose a challenge for printing continuous tracks to form electrical circuits \citep{Chilton2012}, particularly as the droplet sizes are progressively decreasing in the quest for higher resolutions. Topographical features can also be added to substrates to control the flow of the droplets \citep{seemann2005wetting,kant2017controlling}.  This makes understanding the behaviour of the fluid and the morphologies formed in the presence of such topographical features important.
\begin{figure}
		\centering
		\includegraphics[width=\textwidth, trim={0cm 0cm 0cm 0cm},clip]{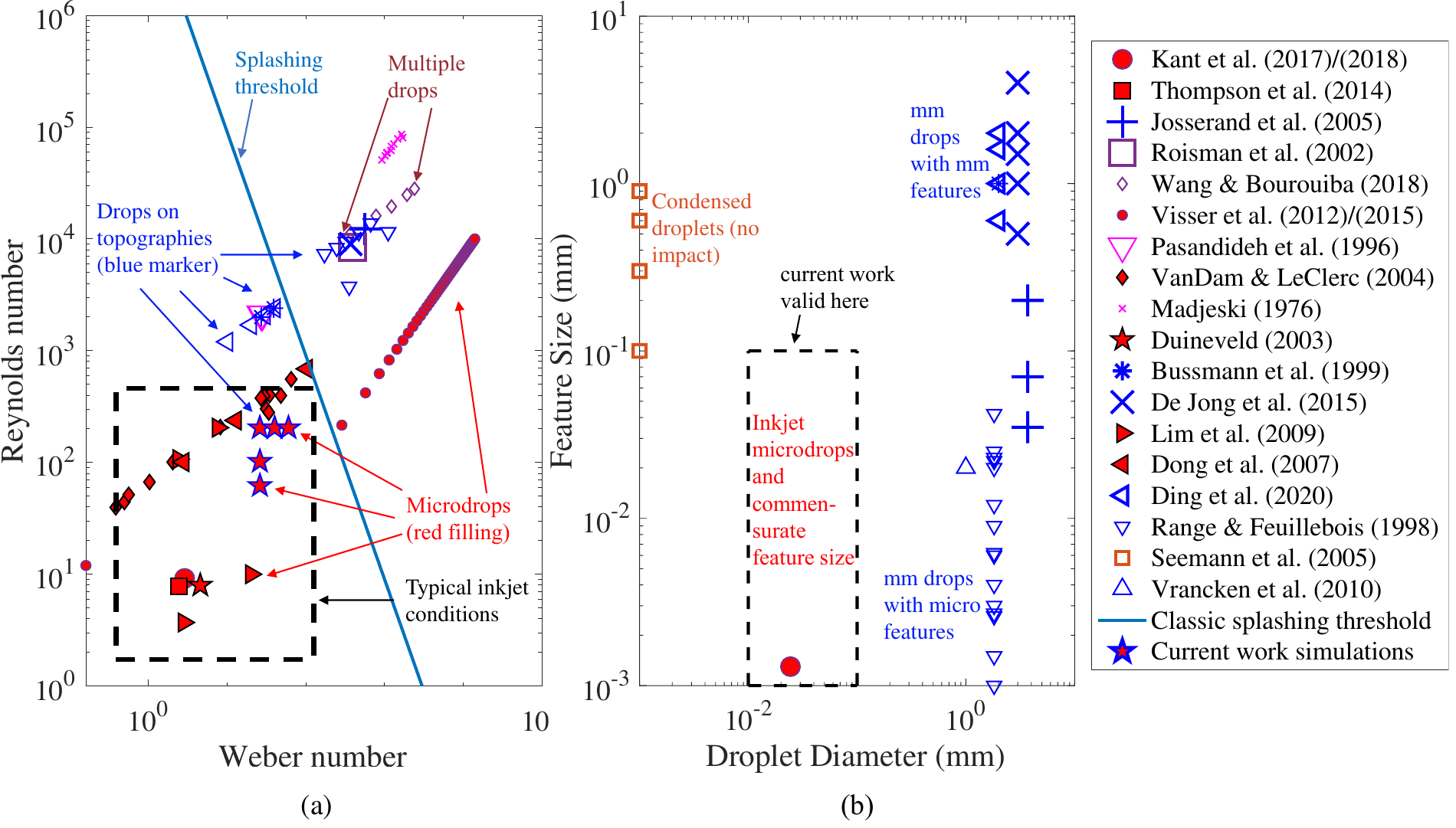}
	\caption{Maps of previous studies of single and multiple--droplet impact onto surfaces with and without topographical features in terms of: (a) Reynolds number and Weber number; and (b) droplet diameter and size of feature on the substrate.\phantom{ \citet{Josserand2005,Visser2015a, Dong2007,Ding2020,Range1998} } }
	\label{fig:Re-We-map}
\end{figure}

A significant amount of experimental, computational and theoretical work has focused on the impact of single and multiple droplets onto smooth or nominally flat rough substrates, making these parts of the problem relatively well studied; for reviews on the topic refer to \citet{Josserand2016}, \citet{Yarin2006} and \citet{Khojasteh20161}. The dynamics of a single droplet impacting a solid smooth surface can be classified into three stages. In the first stage, the droplet spreads due to inertia on an air layer which prevents direct contact with the substrate. This air layer is either expelled or trapped in the droplet as a bubble when the droplet eventually makes contact with the substrate. In the second stage, surface tension causes the droplet to oscillate or relax depending on the degree of viscous dissipation. In the final stage, the droplet spreads under capillary forces reaching an equilibrium that minimises free energy. The three stages were labelled by \citet{Rioboo2002} as kinematic and spreading, relaxation and wetting/equilibrium.  

Droplet impact conditions are typically described in terms of the Reynolds number and Weber number, defined respectively as
\begin{equation}
    Re = \frac{uD_0}{\nu} \textrm{~~~and~~~} We = \frac{\rho u^2 D_0}{\gamma},
    \label{e:ReWeDef}
\end{equation}
where $u$ is the impact velocity, $D_0$ is the in-flight droplet diameter, $\nu$ is the kinematic viscosity, $\rho$ the density and $\gamma$ the surface tension. To set the current work in context, figure \ref{fig:Re-We-map}(a) presents a $Re$-$We$ map of key previous studies of droplet impact onto solid surfaces.
By necessity, inkjet printing systems operate under non-splashing conditions, and the typical range of $Re$ and $We$ is indicated by the dashed rectangle. Micro-droplet studies are indicated by red-filled symbols in figure \ref{fig:Re-We-map}. Interestingly, micro-droplets (i.e.\ droplets with diameter below approximately 100 \si{\micro\meter}) have been shown not to splash, even at conditions well above the splashing threshold \citep{Visser2012,Visser2015a}. Hence micro-droplets do not necessarily behave in the same way as much larger droplets despite dynamic similitude. In fact, there is still uncertainty in how splashing is triggered at high speeds, though it is an active area of interest for aerospace applications \citep{Cimpeanu18}.  

There are many studies of the spreading of single droplets impacting on flat surfaces, resulting in a wide collection of models for important characteristics such as the maximum spreading diameter. These are discussed in \S\ref{s:overspill}, where they are used to develop estimates of critical conditions relevant to droplet deposition on a scratch.

The printing of multiple interacting droplets has received increasing attention in the literature. For example, the printing of elongated liquid beads has been investigated since the 1980s \citep{Davis1980, Sekimot1987, DUINEVELD2003, Stringer2010} --- for recent reviews on the topic refer to \citet{Thompson2014} and \citet{Kwon2018}. Key findings include: printed liquid beads/lines require contact angle hysteresis and contact--line pinning for stability \citep{Davis1980}, they are only stable for contact angles less than $90^\circ$ \citep{Davis1980} and that nano-inks used in inkjet printing have an approximately zero receding contact angle \citep{DUINEVELD2003}. Bulging instabilities can occur \citep{DUINEVELD2003} and can only be explained by considering viscous as well as capillary and inertial effects \citep{Thompson2014}. The present work therefore focuses on advancing contact angles below $90^\circ$ and a very low receding contact angle.

Other studies have also focused on the interactions of consecutively deposited droplets \citep{Yarin1995,Roisman2002b,AshokeRaman2017,Wang2020,Sykes2020a}. \citet{Yarin1995} experimentally investigated the impact of a train of droplets (without offset) on a surface, and studied the size distribution of secondary droplets generated from splashing of the primary droplets. \citet{Roisman2002b} considered the impact of two adjacent droplets with overlap and derived an expression for the thickness of the lamella and the maximum height. More recently, \citet{Wang2020} used a combination of experiments and modelling to show how a partially wetting surface leads to dramatically different regimes of coating and splashing from those in isolated impacts. They found four possible regimes: head-on collision, crescent-moon fragmentation, touch-and-flop collision and no collision. \citet{castrejon2013mixing} and \citet{Sykes2020a} focused on the internal mixing when a falling droplet impacts on a sessile one. All of these studies focused on millimetre-scaled droplets which are an order of magnitude larger than typical inkjet droplets that are the focus here.

Fewer studies have examined the effect of droplet impact on substrates with topographical features such as \citet{Ding2020,Josserand2005,Range1998,vrancken2010fully}. These are indicated in figure \ref{fig:Re-We-map} by blue symbols. With the exception of \citet{kant2017controlling,Kant2018}, these have considered millimetre-scale droplets interacting with millimetre- or micro-scale features, rather than micro-droplets impacting on micro-scale features. \citet{bussmann1999three} studied, experimentally and numerically, the impact of millimetre-sized droplets onto a substrate with a sharp step. They found that a droplet can split due to the presence of a corner. \citet{de2015exploring} studied experimentally the impact of similar-sized droplets near closed pits and open-ended pores and their effect on splashing. \citet{Rashidian2019} developed an analytical model and used lattice Boltzmann method (LBM) simulations to investigate how the presence of a small protrusion can cause the rupture of a droplet's spreading lamella and the effect of impact velocity, wettability and protrusion dimensions on this phenomenon. It was found that the presence of a small protrusion can rupture the lamella of the spreading droplet, possibly resulting in a non-continuous coating.

\citet{kant2017controlling} studied experimentally the spreading of a micro-droplet on a substrate with a recessed pixel and found that the presence of its sidewall can either enhance or hinder spreading depending on the gradient of the topography ahead. They found that topography can be used to restrict small volumes of liquids to a specific region; a droplet spreading with the topography ahead sloping downhill will be pinned. Using a model developed in \citet{Thompson2014}, \citet{Kant2018} then studied numerically the printing of sequential droplets into a recessed region and found that the presence of the walls enhances spreading but does not guarantee containment within the region. \citet{kant2017controlling,Kant2018} focused on the low $Re$ regime where the impact velocity is negligible. \citet{jackson2019droplet} used LBM simulations to explore the effects of misalignment between droplets and small cavities and the filling of the cavities. \citet{seemann2005wetting} experimentally studied the wetting of micro-structured surfaces using regular grooves separated with ridges, but using vapour condensation rather than droplet deposition.  Two main morphologies were observed, namely an overspilling droplet that extends onto the ridges and neighbouring grooves, and extended filaments that run parallel to the grooves.  

As droplets decrease in size, it is expected that substrate topographical features will have a greater effect on the printed morphology and product quality. There have been no studies of how such minor variations or defects on a substrate change the morphology of an impacting droplet. Therefore, we study the effect of a generic two--dimensional (2-D) feature (defined in \S\ref{s:geom}) that can be deliberately designed or accidentally caused by scratching. We begin (\S\ref{s:geom}) by considering possible single droplet impact outcomes based on capillary flows and contact-line pinning on topographical features, then adapt existing spreading models for droplets on uniform surfaces to account for the presence of a scratch, and use these to postulate a regime map. In \S\ref{methodology} we provide details and validation of a lattice Boltzmann numerical method that is used in \S\ref{s:mainresults} to test the theoretical estimates of critical scratch dimensions and to explore the associated dynamics of single micro-droplet impacts on the scratch. Section \ref{ss:EffectOfParameters} considers the influences of the flow and surface parameters on droplet morphologies. Finally, we discuss the implications of our findings for printing applications in \S\ref{s:Implications}, which includes multiple-droplet simulations, and we present conclusions in \S\ref{conclusionsSection}.

\section{Idealised scratch and anticipated single--droplet dynamics\label{s:geom}}
\begin{figure}
	\centering
	\captionsetup{width=1\linewidth}
	\includegraphics[width=0.8\textwidth, trim={0cm 0cm 0 0cm},clip]{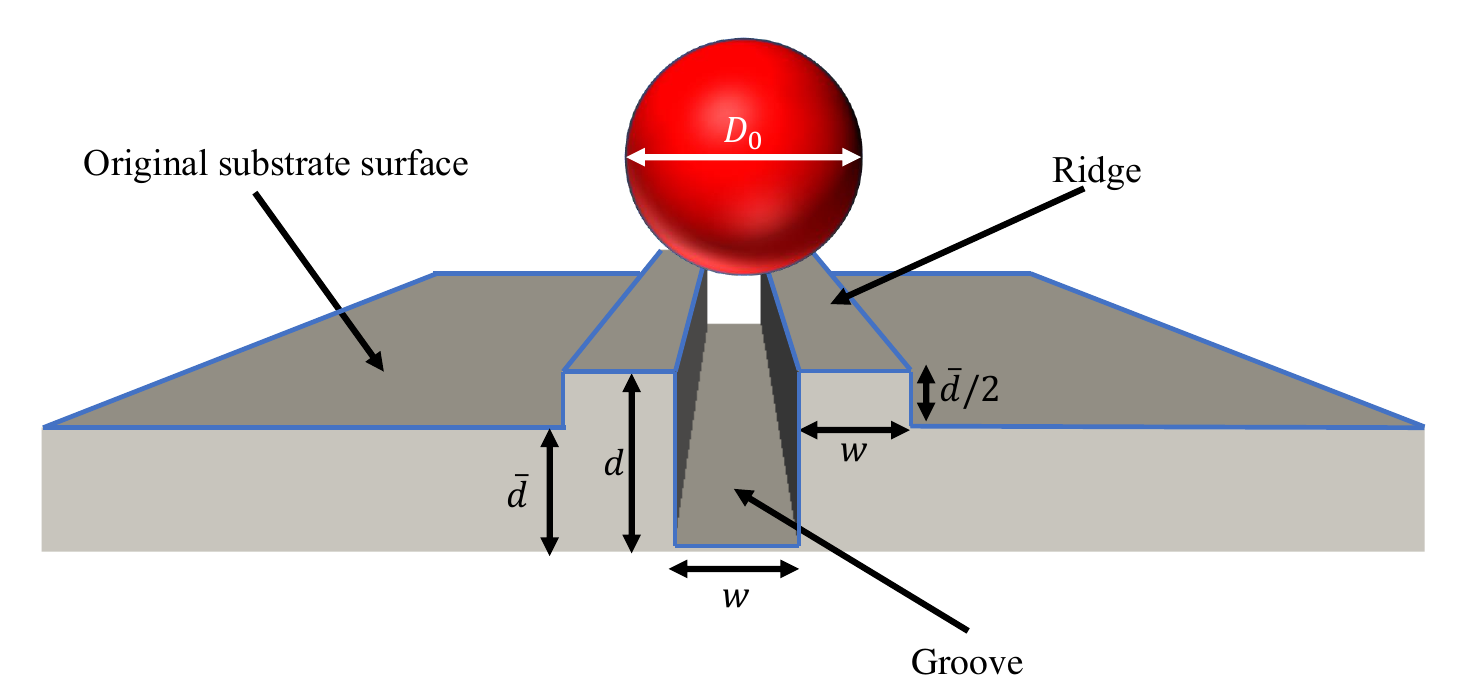}
	\caption{The geometry used to represent an idealised scratch. The groove has width $w$ and the side ridges are assumed to have the same width as the groove. The bottom of the groove is a depth $\bar{d}$ below the original substrate surface, and its total depth from the top of the side ridges is $d=\bar{d}+\tfrac{1}{2}\bar{d}$. These dimensions conserve the volume of displaced material.}
	\label{FigGeometry}
\end{figure}

  The specific surface geometry feature considered here is shown in figure \ref{FigGeometry}. A scratch on the substrate is idealised as a continuous uniform groove of rectangular cross-section, with a rectangular ridge on each side representing solid material displaced during formation of the scratch. Hence, the combined cross-sectional area of the side ridges matches that of the groove below the original substrate surface level. The relative dimensions of the groove and side ridges would in practice depend on the substrate material and scratching mechanism. Here, each side ridge is assumed to have the same width, $w$, as the groove, and a height of $\tfrac{1}{2}\bar{d}$, where $\bar{d}$ is the depth of the groove below the original surface. Hence the total depth of the groove from the top of the side ridges is $d=\tfrac{3}{2}\bar{d}$. As well as representing a scratch, the geometry also mimics micro-structured surfaces such as those studied by \citet{seemann2005wetting}, with the side ridges akin to the edges between two neighbouring grooves. Throughout this study, the groove width, $w$, and depth, $d$, are scaled by the in-flight diameter, $D_0$, of a droplet impacting on the solid surface. We will use the words `groove' and `scratch' interchangeably in what follows. Note that in practice scratch sizes can range from nanometres to tens of microns \citep{Brostow2004,Chen2008,Dasari2009}.

In considering possible outcomes of a single--droplet impact on this topography, there are some obvious limiting behaviours. If the groove is much wider than the droplet, i.e.\ $w\gg 1$, and the droplet lands away from the sidewalls, the impact is simply that of a droplet on a flat surface, which has been widely studied (see \S\ref{s:overspill}). If instead the droplet impacts the sidewall, the situation corresponds to impact on a step as previously considered by e.g.\ \citet{bussmann1999three} and more recently by \citet{jackson2019droplet} in the context of droplet deposition into square cavities. As the groove width becomes closer to the droplet diameter, i.e.\ $w\sim 1$, the dynamics are expected to become more complicated. If the droplet lands in the centre of the groove, full imbibition of the droplet into the groove would be expected, but only for sufficient groove depths and substrate wettability. At the other extreme, if $w\ll 1$ and the scratch is shallow, i.e.\ $d\ll 1$, it is expected to have a negligible effect on the spreading dynamics of the droplet, and we have again the well-studied scenario of droplet impact on a smooth flat surface. However, if $d$ is sufficiently large compared with the scratch width, one would expect capillary flow to occur along the narrow channel if the substrate is sufficiently wetting, i.e.\ $\theta < \pi/2$.

\begin{figure}
	\centering
	\includegraphics[width=0.9\textwidth, trim={0cm 0cm 0 0cm},clip]{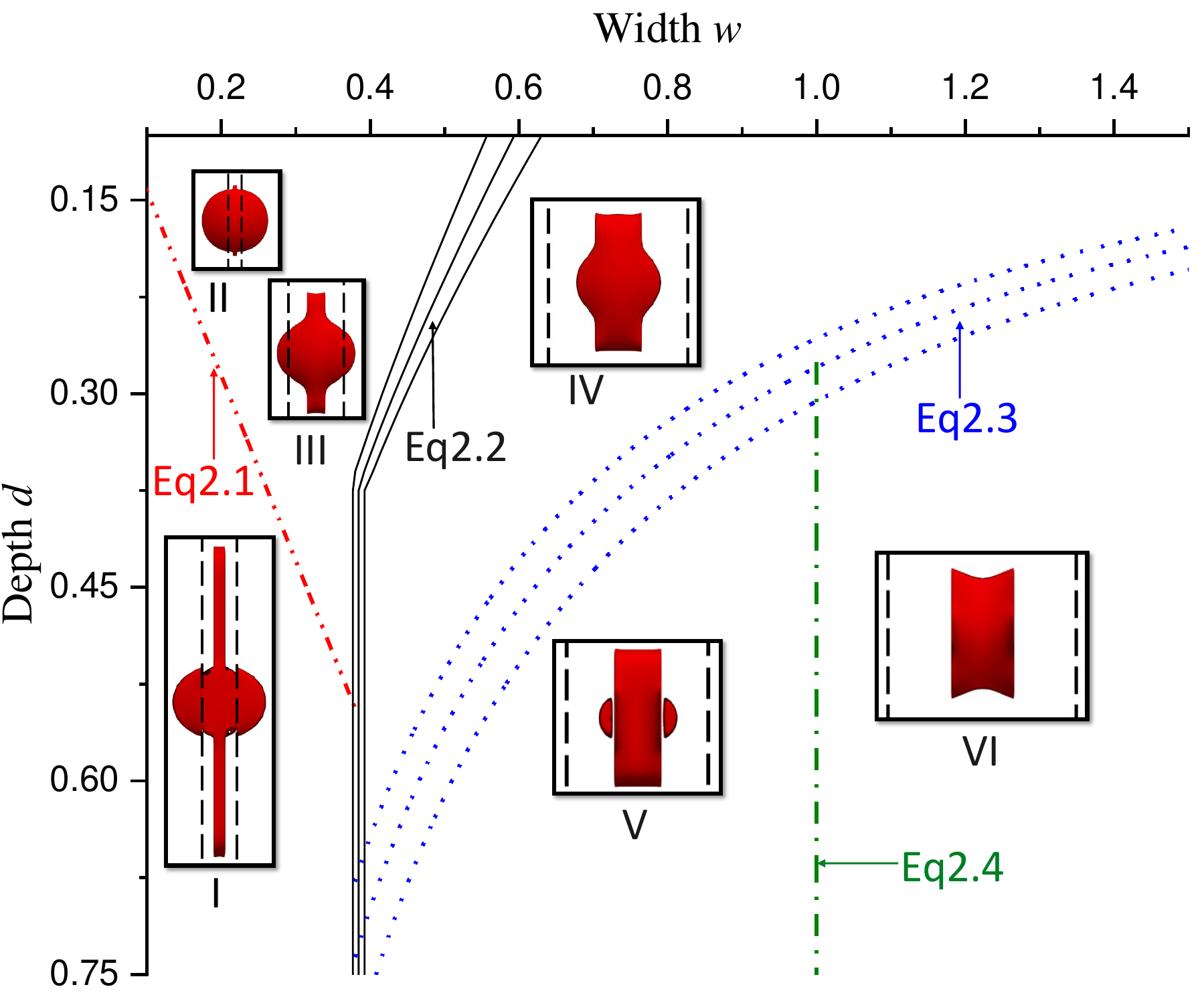}
	\caption{A postulated regime map for the outcome of a single--droplet impact centred on the idealised scratch shown in figure \ref{FigGeometry} in terms of the scratch width and depth scaled by the droplet's in-flight diameter. The lines and curves indicate theoretically estimated critical conditions for behaviours including capillary flow, contact-line pinning on edges, droplet splitting and full imbibition into the scratch. The inset top-view images within each region show examples of corresponding final printed droplet shapes predicted by numerical simulations, indicating possible different droplet morphologies. These are named: {\sf I} `capillary'; {\sf II} `quasi--spherical cap'; {\sf III} `inertial'; {\sf IV} `semi--imbibed'; {\sf V} `split semi--imbibed'; and {\sf VI} `fully imbibed'. Note that the inset images are shown at different scales. The vertical dashed lines in each image indicate the outer edges of the side ridges of the scratch. Equations \eqref{eq:CriticalOverspilling} and \eqref{eq:ImbibitionLine} employ predictions of maximum spreading diameters indicated by asterisks in table \ref{t:betamax}, and the three curves for each equation are obtained using the mean of those predictions and one standard deviation either side. }
	\label{fig:EightMorphos}
\end{figure}

In this work we therefore focus on the range $0<w\leqslant 1$, i.e.\ where the droplet diameter is of a similar size to or larger than the scratch. The droplet impact regime is taken to be non-splashing, non-bouncing deposition as required for a successful inkjet printing operation. It is also important to note that the advancing static contact angle is assumed to be less than 90$^\circ$ and the receding contact angle to be close to zero, again as is typical in printing systems. In all cases, the droplet initially spreads on impact due to inertia. Given the presence of ridges and edges in the geometry, it is expected that overspill, splitting and/or pinning of the droplet contact line on different edges will occur for certain conditions. A postulated regime map showing outcomes of single-droplet deposition on the centre of the scratch for different widths and depths is given in figure \ref{fig:EightMorphos}. The different droplet morphologies and estimates of the conditions under which they occur are explored and developed in the following subsections.

\subsection{Capillary flow along narrow scratches\label{s:capflow}}
A scratch much narrower than the impinging droplet diameter ($w\ll 1$) is expected to have little effect on the early stage inertial spreading of the droplet. However, once deposited, the drop becomes a source of liquid from which capillary flow can occur over a longer time scale along sufficiently narrow and deep scratches. Capillary flow has been extensively studied, including in cylindrical micro-channels, since the early twentieth century \citep{washburn1921dynamics, bell1906flow, lucas1918ueber}. 
With the advent of microfluidics, recent attention has been paid to other micro-channel geometries; most relevant to our geometry is the work on open rectangular micro-channels by  \citet{yang2011dynamics} who, based on experiments and a model balancing capillary and viscous forces, proposed a critical channel width below which capillary flow occurs
\begin{equation}
w_{cap} = 2d\cos\theta_A/(1-\cos\theta_A).
\label{eq:CriticalYang}
\end{equation}
The line given by \eqref{eq:CriticalYang} is shown in figure \ref{fig:EightMorphos}, with $\theta_A$ having a representative value of $75^\circ$. Below this line, i.e.\ where $w<w_{cap}$, the final printed droplet shape will consist of a localised droplet (since the small receding contact angle prevents full contraction of the droplet) with liquid filaments extending in both directions along the scratch. An example numerical simulation result (discussed later) is shown in figure \ref{fig:EightMorphos} as image {\sf I}. As a result of the mechanism producing it, we label this the `capillary' morphology. The long filament within the scratch has been observed by \citet{seemann2005wetting}, but not the liquid outside the scratch because they used vapour condensation rather than droplet deposition. Above the line \eqref{eq:CriticalYang}, i.e.\ for shallow, narrow scratches for which $w>w_{cap}$, there will be only a small deviation from perfect sphericity, as indicated by image {\sf II} in figure \ref{fig:EightMorphos}. This shape will be referred to as the `quasi-spherical cap'. 

\subsection{Predicting edge pinning and overspill\label{s:overspill}}
The impacting droplet will spread over the top of the side ridges while at the same time penetrating the scratch. Depending upon the width and depth of the scratch, overspill from the side ridges onto the original substrate may occur. To develop an estimate of conditions under which this will happen, we need to consider the expected maximum spreading diameter, $D_{max}$, of the droplet on a flat substrate with the same advancing and receding contact angles.

For a droplet impacting a planar surface, the maximum spreading diameter (normalised as $\beta_{max}=D_{max}/D_0$) has been extensively studied and found to depend on the impact velocity and fluid properties captured in the Reynolds and Weber numbers \eqref{e:ReWeDef}. \citet{Pasandideh-Fard1996} showed, using experiments and numerical simulations, that for low Weber and Reynolds numbers, surface wettability also becomes a significant variable. \citet{Clanet2004} showed experimentally that there are two regimes for droplet deposition: one is the capillary regime with low $We$ and high $Re$, where viscous effects are negligible, and the other is the viscous regime with high $We$ and low $Re$, where capillary forces can be ignored. Early studies reported two conflicting $\beta_{max}$ scalings with $We$ in the capillary regime: studies that used an energy balance predicted a $We^{1/2}$ dependence \citep{MADEJSKI19761009,Chandra1991,Bennett1993}, while those using a momentum balance predicted $We^{1/4}$ \citep{Clanet2004}.

In the viscous regime, most studies predict a $Re^{1/5}$ dependence \citep{MADEJSKI19761009,Chandra1991,Bennett1993}. \citet{Eggers2010} showed, using a dynamical model with a viscous boundary layer, that if the $We^{1/2}$ scaling holds, then $\beta_{max}= Re^{1/5}F(P)$, where $F$ is a function of $P=We Re^{-2/5}$. \citet{Laan2014a} used experiments with three different viscosity fluids and showed that $We^{1/2}$ holds rather than $We^{1/4}$ and used Padé interpolation to approximate $f(P)$ for $We>10$. \citet{Lee2015} extended this to $We>1$ by incorporating surface wettability. In fact, some models developed earlier than \cite{Eggers2010}, including \citet{Roisman2009} and \citet{Scheller1995}, can be written in the form $\beta_{max}= Re^{1/5}F(P)$. \citet{Wildeman2016a} recently used extensive simulations and modelling to bridge the energy balance and the momentum balance approaches and developed a model for $\beta_{max}$. All these models (summarised in table \ref{t:betamax}) have shown agreement with experimental and numerical data for certain (different) parameter ranges \citep{Josserand2016,Wildeman2016a,Lee2015}, see also the rightmost columns of table \ref{t:betamax}. We will compare the models with new numerical data for micro-droplets in \S\ref{ss:AnalyticalValid}.

Although originally developed for axisymmetric conditions, maximum spreading laws like the above can be modified by introducing a correction factor to account for the presence of the scratch, and hence used to estimate when overspill occurs. For ease of implementation we consider only models that can be written explicitly as $\beta_{max}=f(We,Re)$ or $\beta_{max}=f(We,Re,\theta_A)$. In particular, we use the models of \citet{Scheller1995}, \citet{Pasandideh-Fard1996}, \citet{Roisman2009}, \citet{Laan2014a} and \citet{Lee2015} listed in table \ref{t:betamax}.

\begin{table}
\begin{adjustbox}{angle=90}

\renewcommand{\arraystretch}{3.0} %
\begin{tabular}{lllll}
number & Study & $\beta_{max}$ ($=D_{max}/D_0$) & We  & Re \\
1 &\citet{Chandra1991}        &  $\dfrac{3We}{2Re}\beta^4_{max}+(1-\cos\theta)\beta^2_{max}-\left(\dfrac{We}{3}+4 \right)\approx 0$ &43  & 2300 \\
2* &\citet{Scheller1995}        & $0.61 Re^{1/5}(P)^{1/6}$ & 50 -- 2500  & 20 -- 16400\\
3* &\citet{Pasandideh-Fard1996} & $\sqrt{\dfrac{We + 12}{3(1-\cos\theta_A)+4(We/\sqrt{Re})}}$ & 20 -- 700  & 200 -- 36000\\
4 &\citet{Ukiwe2005} & $(We+12)\beta_{max}=8+\beta^3_{max}\left( 3\left[1+\cos(\theta_d)\right] + 4\left[We/\sqrt{Re}\right]\right)$  & 20-200  & 700-7000\\
5* &\citet{Roisman2009}         & $Re^{1/5}(0.87-0.4(WeRe^{-2/5})^{-1/2})$ \\
6* &\citet{Laan2014a}           & $Re^{1/5}\dfrac{P^{1/2}}{1.24+P^{1/2}}$ & 10 -- 1700  & 70 -- 17000\\
7* &\citet{Lee2015}             & $\sqrt{\dfrac{We^{1/2}Re^{1/5}}{7.6+We^{1/2}} + \beta_{eq}^2}$ & 1-1500  & 40-18000\\
8 &\citet{Wildeman2016a}             &$\dfrac{3(1-\cos\theta)}{We}\beta^2_{max}+\dfrac{0.7}{\sqrt{Re}}\beta^2_{max}\sqrt{\beta_{max}-1}=\dfrac{12}{We}+\dfrac{1}{2}$ & 30 -- 5000  & 10-5000\\
\end{tabular}

\end{adjustbox}
\caption{A selection of previously published models for predicting the maximum spreading diameter of a droplet impacting on a flat surface. Here $P=We Re^{-2/5}$ and $\beta_{eq}$ is the equilibrium spread diameter. $\theta_d$ in the model by \citet{Ukiwe2005} is the dynamic contact angle (here assumed to be equal to the advancing contact angle $\theta_A$). The ranges of $Re$ and $We$ are of experiments used to validate these studies. The asterisks indicate the explicit models used in constructing figure \ref{fig:EightMorphos}.}
\label{t:betamax}
\end{table}

Assuming that the droplet spreads on the tops of the side ridges and the inner scratch walls at the same rate, a correction factor can be derived by subtracting from the initial droplet volume the volume of liquid that will go into the scratch before the droplet reaches the outer edges. The new volume available for spreading can then be used to derive a new equivalent `initial' droplet diameter to be used in the maximum spreading laws. Assuming that $d\leq w$, the dimensional volume to be subtracted is approximately $dD_0\times wD_0\times D_0 = wdD_0^3$, so the new volume of liquid available for spreading over the side ridges is $V_{new}=\tfrac{1}{6}(\pi D_0^3-6dwD_0^3)=\tfrac{1}{6}D_0^3\left(\pi-6dw\right)$. This is equivalent to a free droplet of dimensional diameter $D_{0,new}= \pi^{-1/3}D_0\left(\pi-6dw\right)^{1/3}$.  The maximum spreading laws are of the form $D_{max}/D_0=f(Re,We,\theta_A)$, so using the new diameter $D_{0,new}$ to account for the presence of the scratch gives $D_{max}/D_{0,new}=D_{max}/\left(\pi^{-1/3}D_0\left(\pi-6dw\right)^{1/3}\right) = f(Re,We,\theta_A)$. Hence, the new dimensionless maximum spreading diameter allowing for liquid entering the scratch is given by 
\[
\beta_{max} = D_{max}/D_{0} =\pi^{-1/3}D_0\left(\pi-6dw\right)^{1/3}f(Re,We,\theta_A).
\]
This holds when $d\leq w$, otherwise the droplet reaches the outer edges of the side ridges before the entire depth is covered by the impacting droplet. If $d>w$, the dimensional volume to be subtracted will be $w^2D_0^3$ and hence $D_{0,new}$ will not depend on $d$, and the adjusted $\beta_{max}$ will be the same value as that when $d=w$ for the entire range $d>w$. The condition for the droplet contact line reaching the outer edges of the side ridges is $\beta_{max}=3w$, i.e.\ the dimensionless width of the scratch and the two side ridges. Hence, the outer edges of the side ridges will be reached only for scratches with widths $w_{pin}$ satisfying the condition

\begin{equation}
3w_{pin} =\pi^{-1/3} f(We,Re,\theta_A)\cdot
 \begin{cases}
                                   \left(\pi-6dw_{pin}\right)^{1/3} & \textrm{if} \hspace{0.1cm} d\leq w \\
                                   \left(\pi-6w_{pin}^2\right)^{1/3} & \textrm{if} \hspace{0.1cm} d > w 
 \end{cases}.
 \label{eq:CriticalOverspilling}
\end{equation}

The five different \emph{explicit} spreading models listed (with asterisks) in table \ref{t:betamax} give different predictions for $f(We,Re,\theta_A)$. Therefore, in plotting the values of $w_{pin}$ in figure \ref{fig:EightMorphos}, the three solid black lines correspond to the mean and standard deviation of the five predictions.
 
Note that equation \eqref{eq:CriticalOverspilling} simply provides an estimate of scratch widths for which the droplet can reach the outer edges of the side ridges. Since we assume that the receding contact angle is close to zero, the contact line will not recede from the edges if it is able to reach them. However, the droplet contact line may continue spreading beyond the outer edges of the ridges if it has sufficient momentum. If $\beta_{max}>3w$, the droplet will continue spreading horizontally while the contact line remains pinned on the edge. This will increase the dynamic contact angle $\theta_d$ (i.e.\ the angle of the free surface measured from the horizontal top of the ridges) until either the droplet stops spreading and recoils or $\theta_d>\theta_A+90^\circ$, i.e. $\theta_d$ exceeds the Gibbs criterion for pinning. In the latter case, the droplet will then advance down the vertical outer walls of the side ridges and spill onto the original substrate surface; the final droplet morphology will look like image {\sf III} in figure \ref{fig:EightMorphos}. This will be referred to as the `inertial' morphology, since the droplet has sufficient inertia to spill over the side ridges. Hence equation \eqref{eq:CriticalOverspilling} provides an estimate for conditions under which the droplet contact line will pin on the outer edge of the side ridges. This will be discussed again with the benefit of numerical simulations in \S\ref{ss:morphs}. 

\subsection{Droplet splitting and imbibition\label{s:splitting}}

For $w>w_{pin}$, the droplet does not cross the outer edges of the side ridges. For sufficiently shallow scratches, its final shape is expected to be one where the continuous liquid volume rests on top of side ridges while also filling the scratch. This morphology is labelled `semi-imbibed', and an example is shown in image {\sf IV} in figure \ref{fig:EightMorphos}. As the scratch depth increases, more of the liquid volume will occupy the scratch. Given that the low receding contact angle prevents significant contraction of the droplet contact line from the side ridges, for a deep enough scratch the droplet may split along the \emph{inner} edges of the side ridges such that the liquid on the top of the side ridges and that in the scratch become separated.  Splitting along the inner edges is expected when the cross-sectional area of the scratch multiplied by $D_{along}$, the length of spreading along the scratch, results in a greater volume than the droplet. Thus the critical width corresponds to when $wD_0\times dD_0 \times D_{Along}=\tfrac{4}{3}\pi(D_0/2 )^3$. Assuming that $D_{Along}=\beta_{max}D_0$, where $\beta_{max}$ is again the maximum spreading on a flat surface, this gives the following estimate of the critical width for which splitting will occur
\begin{equation}
  w_{split}= \frac{\pi}{6d\beta_{max}}.
  \label{eq:ImbibitionLine}
\end{equation}

Again the different explicit models asterisked in table \ref{t:betamax} have been used to give a spread of estimates of $\beta_{max}$ (for $\theta_A=75^\circ$) and these result in the blue dotted curves in figure \ref{fig:EightMorphos}. For depths below these curves and $w<1$, the final droplet shape will be a `split semi-imbibed' morphology where most of the liquid occupies the scratch, but separate small droplets remain on the upper surfaces of the side ridges. See the example image {\sf V} in figure \ref{fig:EightMorphos}.  

The droplet fully imbibes into the scratch without spreading on the side ridges when $w\geqslant 1$ and the scratch depth is sufficient to contain the entire liquid volume. The result is a `fully imbibed' morphology as indicated in image {\sf VI} in figure \ref{fig:EightMorphos}. By a similar argument to that above for droplet splitting, the condition for full imbibition into the scratch is
\begin{equation}
  w \geq 1, \hspace{0.25cm} \textrm{and} \hspace{0.25cm} d > \frac{\pi}{6\beta_{max}w_{split}.}
  \label{eq:FullyImbibed}
\end{equation}
This corresponds to the region to the right of the vertical dash-dot line in figure \ref{fig:EightMorphos}.

\section{Computational methodology}\label{methodology}

There are several possible methods that could be used to simulate deforming droplets on surfaces, see e.g.\ \citet{Wilson2016} for a review. A key requirement of any such method is the ability to represent and determine the shape of the liquid free surface as it deforms. One possibility is to use an interface tracking approach, where the computational mesh is fitted to and deforms with the free surface --- for example as in the finite element technique used by \citet{Feng15}. While this approach provides excellent sharp representations of free surfaces, it cannot track surfaces that break apart or intersect without remeshing, which can lead to instability \citep{Furlani2015}. This makes the method challenging when simulating systems with critical phenomena such as the break up or coalescence of small droplets. A more common approach is to use an interface capturing method, of which there are many types, such as the volume-of-fluid (VOF), level-set and phase-field methods --- see \citet{Mirjalili2017} for a classification and review. In such methods, the liquid--fluid interface moves through the computational mesh, allowing greater flexibility in terms of severe interface deformations and topological changes in the liquid volume, subject to sufficient mesh resolution.

VOF is used frequently for droplet deposition onto surfaces, with work by \citet{Wildeman2016a} being a recent example. However, a key limitation in VOF approaches --- and Navier-Stokes based approaches in general --- is that the \emph{dynamic} contact angle $\theta_d$ usually has to be prescribed as an input. For relatively well-behaved systems, empirical relations between $\theta_d$ and the contact-line velocity $U$ can be used \citep{Yokoi2009,Sykes2020}. However, for a complex 3-D geometry evolving in time, as considered here, this is not straightforward.

The LBM is increasingly used to simulate the fluid mechanics of multiphase systems. The mesoscopic nature of the method, which retains some molecular-scale physics via probability density distributions, makes it well suited to multiphase simulations, with interface motion, break-up and coalescence readily captured in three dimensions. Multiphase LBM does not require a relation between $\theta_d$ and $U$, but just the static contact angle $\theta_s$ in the case without contact angle hysteresis (CAH), and the advancing $\theta_A$ and receding $\theta_R$ contact angles in cases with CAH. The dynamic contact angle results naturally from the statistical mechanical nature of the method. LBM is localised and lends itself to parallel computing with GPUs \citep{Kruger2016}; we make use of this feature to run an extensive parametric study here. A disadvantage of LBM is that macroscopic variables such as density, viscosity, velocity and surface tension are derived quantities, see below. It is also a memory intensive method, putting limits on domain size when using GPUs, and local grid refinement is still a developing area. These disadvantages did not affect this work.

There are various approaches to including multiphase and/or multicomponent effects into the LBM framework, including methods based on `colour gradients' \citep{gunstensen1991lattice}, free energy functionals \citep{Swift1995}, mean-field approximations \citep{he1999lattice} and the `pseudo-potential' model \citep{Shan1993}. For a comprehensive review refer to \citet{Huang2015}.

In this paper, we use the single-component pseudo-potential model due to its relative simplicity and efficiency, but incorporate several modifications suggested in the literature to allow simulations under real inkjet printing conditions. These include the form of pseudo-potential required to include arbitrary equations of state \citep{Yuan2006}, specifically the Carnahan-Starling equation of state which enables high density ratio and manipulation of surface tension, the force correction by \citet{Li2013b} which was later extended to three dimensions in \citet{Li2018}, and the collision matrix in \citet{Li2018}. More detail of these, and the method in general, is provided in section \ref{mthodsubsec1}. To capture the wetting phenomena and CAH, a geometric boundary condition \citep{Ding2007,Connington2013} is employed and this is described in section \ref{mthodsubsec2}.

\subsection{The LBM solver}\label{mthodsubsec1}
LBMs solve the Boltzmann equation on a lattice that discretises the spatial and velocity domains to form the lattice Boltzmann equation so that the probability distribution density function $f$ at position $\boldsymbol{x}$ and time $t$ evolves according to
\begin{equation}
f_\alpha(\boldsymbol{x}+\boldsymbol{e}_\alpha \delta t,t+\delta t) =f_\alpha(\boldsymbol{x},t) -
\Omega_{\alpha\beta} (f_\beta(\boldsymbol{x},t) - f_\beta^{eq}(\boldsymbol{x},t)) + \delta t (S_\alpha(\boldsymbol{x},t) -0.5\Omega_{\alpha\beta} S_\alpha(\boldsymbol{x},t) ).
\label{mainequation}
\end{equation}
Here, $\delta t$ is the time step,
$\boldsymbol{e}_\alpha$ are velocities chosen according to the velocity discretisation used in the model such that $\boldsymbol{x}+\boldsymbol{e}_\alpha \delta t$ moves the particle distribution density function to a neighbouring lattice site and  $\alpha$ and $\beta$ are indices for the velocity vector space. In this paper, we use the D3Q19 velocity discretisation (that is 3 space dimensions and 19 velocity vectors) detailed in Appendix \ref{appA}. Also, $\Omega$ in equation \eqref{mainequation} is the collision operator, which mimics the effect of molecular collisions and is how viscosity is captured; for more detail on the role of the collision operator, refer to \citet{Kruger2016}. The multiple relaxation time (MRT) collision operator used here is written as
\begin{equation}
\Omega_{\alpha\beta} = (\mathsfbi{M^{-1}\Lambda M})_{\alpha \beta},
\end{equation}
where $\mathsfbi{M}$ is the collision matrix and $\mathsfbi{\Lambda}$ is a diagonal relaxation matrix used to relax the various moments of the distribution density function at various rates. The most commonly used collision matrix is that derived using the Gram--Schmidt procedure \citep{DHumieres2002}, however, an equivalent but more efficient and easier to implement matrix has been derived by \citet{Li2018}, which is used in this work. Instead of a single relaxation time for all moments, moments are relaxed at different rates in MRT simulations. The corresponding diagonal relaxation matrix is given by
\begin{equation}
\boldsymbol{\Lambda} =\textup{diag}(1,1,1,1,s_\zeta, s_\nu,s_\nu,s_\nu,s_\nu,s_\nu,s_q,s_q,s_q,s_q,s_q,s_q,s_\pi s_\pi, s_\pi,s_\pi),
\label{relaxationMatrix}
\end{equation}
where $s_\zeta$ and $s_\nu$ determine the bulk and kinematic shear viscosities respectively; $s_\pi$ and $s_q$ are relaxation rates for non-hydrodynamic moments that can be tuned to ensure the stability of the simulation. The term $S_\alpha(\boldsymbol{x},t)$ in equation \eqref{mainequation} is a source term used here to incorporate inter-molecular forces needed to produce coexistence of liquid and vapour phases. To simulate multiphase flow, \citet{Shan1993} introduced local attractive interaction forces written in discrete form as,
\begin{equation}
\boldsymbol{F}(\boldsymbol{x},t)=-G\psi(\boldsymbol{x},t) \sum_\alpha w_\alpha \psi(\boldsymbol{x}+\boldsymbol{e}_\alpha\delta t, t+\delta t)\boldsymbol{e}_\alpha,
\end{equation}
where $G$ is the strength of the interactive force, $\psi(\boldsymbol{x},t)$ is the  pseudo-potential used to prevent the force from diverging at high densities, and $w_\alpha$ are weights that depend on the velocity directions;  they are $w_0=1/3$, $w_1,\ldots,w_6=1/6$ and $w_7,\ldots,w_{18}=1/12$. This interaction force is incorporated differently into $S_\alpha(\boldsymbol{x},t)$ in the original \citet{Shan1993} model compared with \citet{Li2013b}, where the latter solves the thermodynamic inconsistency in the former. The source term used in this work is that proposed by \citet{Li2013b}, and is detailed in appendix \ref{appA}.  This model results in a pressure, $p=c_s^2\rho+\tfrac{1}{2}\psi^2 c_s^2G$, supporting two phases, where $c_s=1/\sqrt{3}$ is the speed of sound of the lattice \citep{Shan1993}. \citet{Yuan2006} showed that this equation of state, with the choice of $\psi(\boldsymbol{x},t)=\rho_0(1- \exp{(-\rho/\rho_0))}$ originally proposed by \citet{Shan1993}, is limited in terms of achievable density ratio between the liquid and vapour phases and instead proposed using a different expression for $\psi(\boldsymbol{x},t)$,
\begin{equation}
\psi(\boldsymbol{x},t)=\sqrt{\frac{2(p-c_s^2\rho(\boldsymbol{x},t))}{c_s^2G}}.
\label{pseudopotential for original SC}
\end{equation}

This enables using different equations of state for pressure such as the Carnahan-Starling equation of state used in this work,
\begin{equation}
p = \rho RT \frac{1+b\rho/4+(b\rho/4)^2-(b\rho/4)^3}{(1-b\rho/4)^3} - a\rho^2,
\label{CSEOS}
\end{equation}
where $T$ is temperature, $R$ is the universal gas constant (set to 1), $a=0.49963R^2T_c^2/p_c$ and $b=0.18727RT_c/p_c$, with $T_c$ and $p_c$ being the critical temperature and pressure, respectively. The parameter $a$ physically represents the strength of the molecular interaction in the equation of state and lowering it results in a thicker interface and a more stable simulation at higher density ratios \citep{Li2013b}. Reducing $T$ in \eqref{CSEOS} increases the density of the liquid and lowers that of the gas, hence increasing the density ratio. The parameter $b$ represents the volume occupied by the material's atoms and is chosen arbitrarily and kept constant. In this study, we use equation \eqref{CSEOS} with $a=0.05$, $b=4$ and $T=0.00472$ which permit a density ratio of \textit{O}(1000); see appendix \S\ref{appendix:constantEOS} for more details on $a$ and $b$.
The macroscopic variables of density, pressure and velocity are calculated from the moments of the probability density function using the following equations, respectively,
\begin{equation}
\rho (\boldsymbol{x},t) = \sum_\alpha f_\alpha(\boldsymbol{x},t),
\end{equation}
\begin{equation}
p(\boldsymbol{x},t) = c_s^2 \rho(\boldsymbol{x},t) + \frac{c_s^2G}{2}\psi^2(\boldsymbol{x},t),
\end{equation}
\begin{equation}
\rho(\boldsymbol{x},t)\boldsymbol{u} (\boldsymbol{x},t) = \sum_\alpha \boldsymbol{e}_\alpha f_\alpha(\boldsymbol{x},t) + 0.5\delta t \boldsymbol{F},
\end{equation}
while the kinematic viscosity is given by
\begin{equation}
\nu = c_s^2(s_\nu^{-1} - 0.5\delta t).
\end{equation}

\subsection{Wetting and CAH boundary condition}\label{mthodsubsec2}
Various methods can be used to prescribe a contact angle at a wall boundary, such as introducing an interaction force between the solid and fluid nodes (see for example \citet{Li2014}) or prescribing a constant density at the wall to achieve a predetermined contact angle (see for example \citet{castrejon2013mixing}). These methods work well at relatively low density ratios but become unstable at higher values. The prescription of a density at the wall also requires calibration to give a specific contact angle whenever the density ratio or the equation of state is changed \citep{Wilson2016}. An alternative is to use the geometric boundary condition developed by \citet{Ding2007}, which has been used to simulate inkjet printed droplets at high density ratio without loss of stability, and hence it is used in this study. This condition was originally developed for the volume of fluid method and adopted in the phase field multiphase LBMs by \cite{Connington2013} and has been used for the pseudo-potential multiphase models by \citet{Zhang2018}. A derivation detailed in Appendix \ref{AppendixGeomCondition} results in the equation
\begin{equation}
\tan\left(\frac{\upi}{2} - \theta\right) = \frac{-\boldsymbol{\nabla}\rho \boldsymbol{\cdot} \boldsymbol{n}}{|\boldsymbol{\nabla}\rho- (\boldsymbol{\nabla}\rho \boldsymbol{\cdot} \boldsymbol{n})\boldsymbol{n} | }
\label{ContGeomCond1},
\end{equation}
where $\boldsymbol{n}$ is the unit normal to the solid surface and $\rho$ is the density. This can be discretised and used to calculate a density to assign to the solid wall locally to satisfy a given contact angle. The discrete form for our geometry is detailed in Appendix \ref{AppendixGeomCondition}.

CAH was implemented by calculating the local contact angle using the inverse of equation \eqref{ContGeomCond1}. If the value of the local contact angle is lower than the receding contact angle ($\theta_R$) then $\theta$ is replaced with $\theta_R$ and similarly, if the local contact angle is higher than the advancing contact angle, it is replaced with $\theta_A$. Assigning the density on the wall controls the interaction pseudo-potential in equation \eqref{pseudopotential for original SC} to satisfy the contact angle $\theta$. In order to capture CAH, the wall lattice sites are initialised with $\theta$ in equation \eqref{ContGeomCond1} set to $\theta_A$. This allows the droplet to spread provided that the contact line forms a local contact angle of $\theta \geq \theta_A$. Once a lattice site has been wetted, the value of $\theta$ in equation \eqref{ContGeomCond1} is replaced with $\theta_R$ for this lattice site. This will stop the droplet from dewetting or the contact line receding unless the local contact angle at the contact line is $\theta \leq \theta_R$. This is implemented by rearranging equation \eqref{ContGeomCond1} for $\theta$ and calculating in every time step locally in every wall lattice site. Note that the values of $\theta$ assigned at the wall are used to control the interaction force, while the contact-line can have any contact angle.
The \citet{Shan1993} model with these additions can capture a range of contact angles, contact angle hysteresis, coalescence and breakup and contact-line dynamics at high density ratio up to $10^3$. The simulation is numerically stable for $45^\circ \leq \theta_A \leq 140^\circ$ in the smooth surface case and  $55^\circ \leq \theta_A \leq 130^\circ$ in the case with the scratch. 

To enable the extensive parametric study presented in section \ref{s:mainresults}, the code was accelerated using CUDA to run on GPUs. A simulation (of size 256$\times$256$\times$576 lattice cells) that would take up to 3 days  on a single CPU then ran in under two hours. On a 24-core CPU high-performance computing node (Intel Xeon Gold 6138 @ 2GHz) speeds of 5 million lattice updates per second (MLUPS) were achieved, compared with 200 MLUPS on an NVIDIA P100 GPU and 500 MLUPS on an NVIDIA V100 GPU.

\subsection{Validation of simulations against experimental data}\label{subSecExperimentalValidation}
\begin{figure}
	\centering
	\includegraphics[width=0.8\textwidth, trim={0cm 0cm 0 0cm},clip]{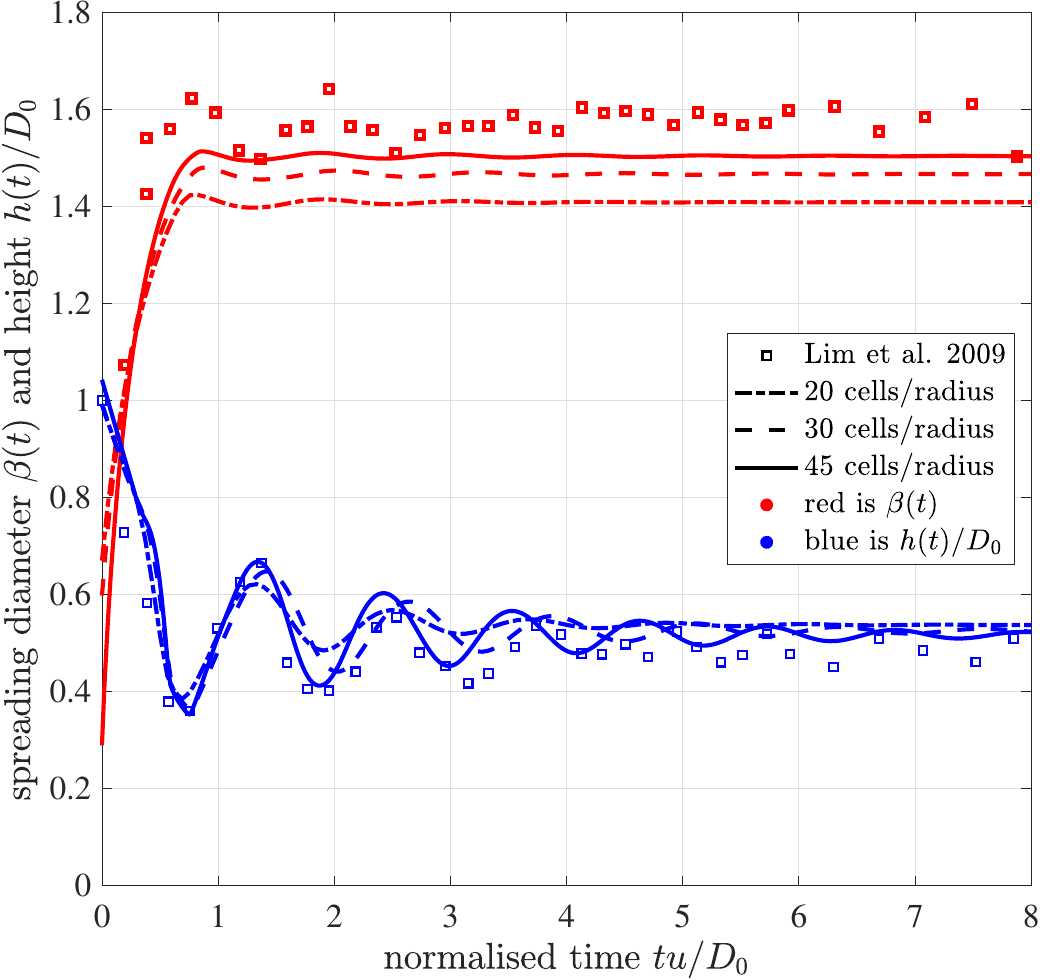}
	\caption{Comparison between the experimental results by \citet{Lim2008} and corresponding simulations of a 48.1\,\si{\micro\meter} droplet impacting a flat surface at 1.9\,\si{\meter\per\second} ($Re=107$; $We=2.4$) using different lattice resolutions. The data show the spreading diameter (in red) and height (in blue) of the droplet, all scaled by the initial droplet diameter.}
	\label{LimEtAlComparison}
\end{figure}

For the impact of a single droplet falling onto a smooth solid surface we compare our simulations with experimental data from \citet{Lim2008}, who examined droplets with in-flight diameter $D_0=48.1$\,\si{\micro\meter} hitting a smooth surface at speed $u=1.9$\,\si{\meter\per\second} corresponding to a Reynolds number $Re = uD_0/\nu=107$, Weber number $We = \rho u^2D_0/\gamma= 2.4$ and Ohnesorge number $Oh = \sqrt{We}/Re=0.015$, where $\gamma$ is the surface tension. The advancing and receding contact angles were reported to be $\theta_A=60^\circ$ and $\theta_R=40^\circ$ respectively. The normalised spreading diameter of the droplet $D_s/D_0$ and height $H_s/D_0$ were tracked over time. \cites{Lim2008} data are compared with equivalent simulations in figure \ref{LimEtAlComparison}. To give an indication of the sensitivity of the simulations to the resolution of the lattice, several different lattices were tested, with the resolution expressed in terms of the number of cells per initial droplet radius. Note that in the lattice Boltzmann method, testing the sensitivity to lattice resolution is not as straightforward as for direct discretisations of the Navier-Stokes equations, since the lattice discretises both coordinate space and the molecular velocity space. Hence modifications of the lattice node spacing require adjustments of other parameters to ensure that the same physical system is being simulated. 

In general terms, good agreement is achieved between the simulations and both sets of experimental data as the lattice resolution increases; oscillations in height agree well particularly for the first few periods, while the spreading rate and final diameter are close, with some small variation in experimental data due to experimental noise. There is still some sensitivity in the time scales of the simulations using different lattices, which becomes more evident at later times, but the same equilibrium state is reached in each case. In the non-axisymmetric simulations presented in \S\ref{s:mainresults}, the finest resolution was used and, for a random sample of the conditions considered, simulations were repeated with the other lattice resolutions. The same equilibrium shapes were obtained with each lattice.

\subsection{Comparison with analytical models}\label{ss:AnalyticalValid}
In addition to the models for the maximum spreading diameter (see table \ref{t:betamax}), theoretical models have been developed to predict the equilibrium diameter $\beta_{eq}$ for a given static contact angle $\theta$ (in the absence of hysteresis). Using conservation of volume, \citet{VanDam2004a} identified an expression for $D_{eq}$ in terms of the in-flight diameter of the droplet $D_0$ and $\theta$:
\begin{equation}
\beta_{eq}=\frac{D_{eq}}{D_0} = \left( \frac{8}{\tan\frac{\theta}{2}(3+\tan^2\frac{\theta}{2})} \right)^{1/3}.
\label{vanDamSpreadDiam}
\end{equation}
\begin{figure}
	\centering
    \begin{subfigure}[b]{0.49\textwidth}
		\centering
		\includegraphics[width=\textwidth, trim={0cm 0cm 0cm 0cm},clip]{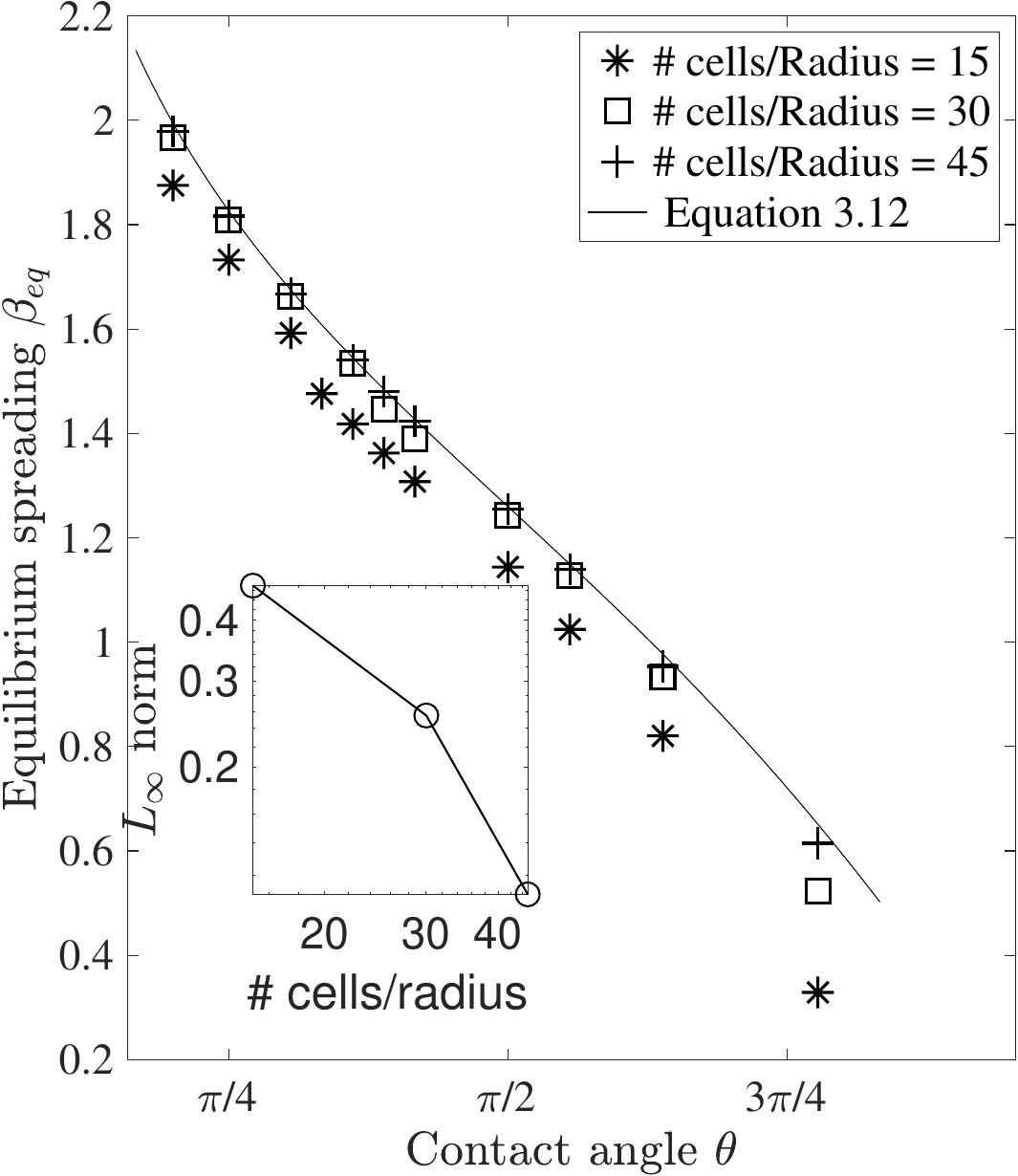}
		\caption{{}}
		\label{subfig:MeshIndependenceDeq}
	\end{subfigure}
  \begin{subfigure}[b]{0.5\textwidth}
		\centering
		\includegraphics[width=\textwidth, trim={0cm 0cm 0cm 0cm},clip]{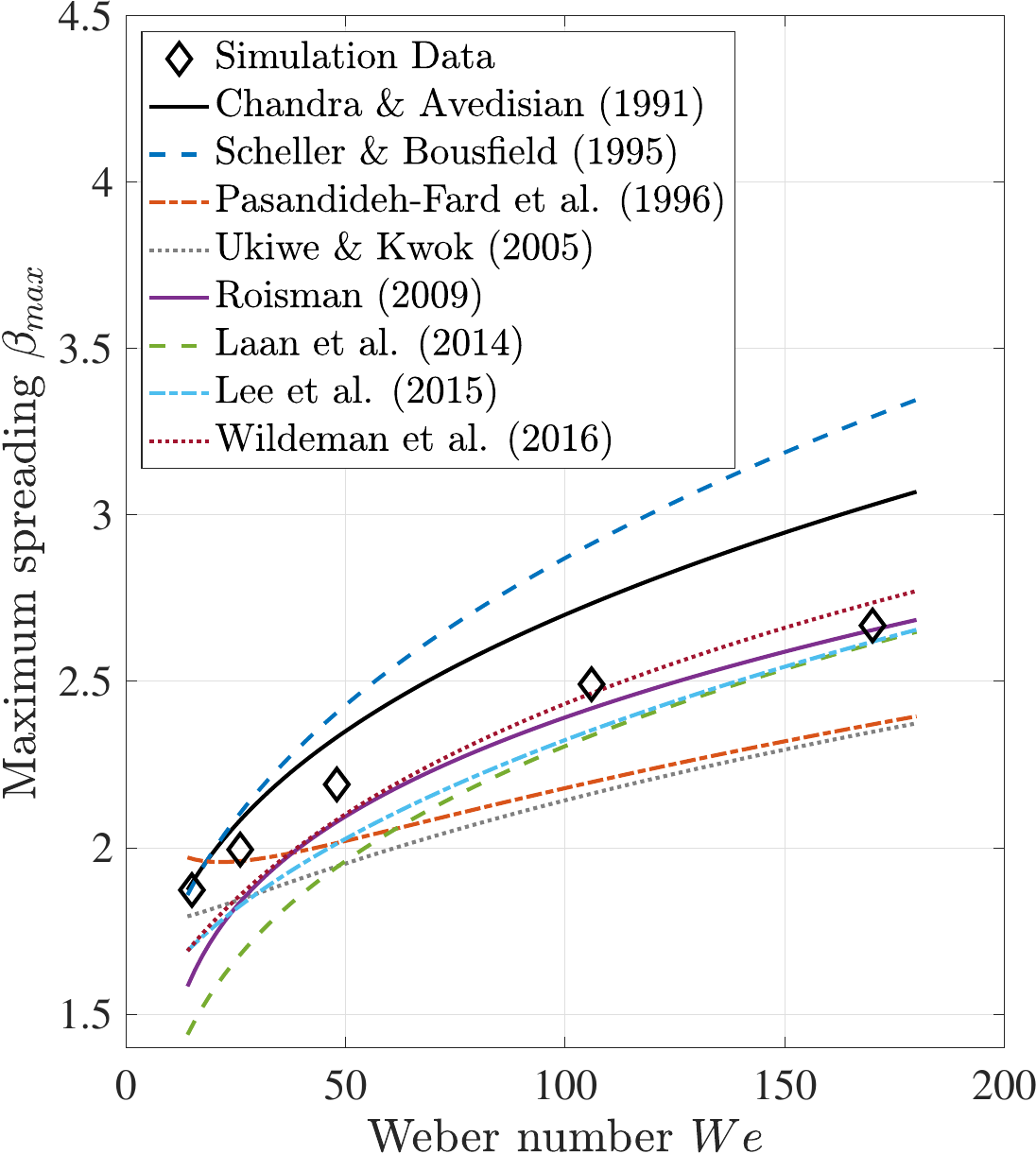}
		\caption{{}}
		\label{subfig:ValidationUsingDmax}
	\end{subfigure}
	\caption{Comparison of simulation predictions with analytical models for droplet spreading diameters on a flat surface. In (a), the equilibrium diameter is determined (without CAH) using different lattice resolutions and compared with equation \eqref{vanDamSpreadDiam} for different contact angles. The inset plot shows the $L_\infty$-norm of the error. In (b), the maximum spreading diameter from simulations using the finest lattice is shown for different Weber numbers, with $\theta_A=75^\circ$ and $\theta_R=1^\circ$.  This is compared with the predictions using the models in table \ref{t:betamax}.}
	\label{fig:analyticalcomparison}
\end{figure}

As further validation for the simulation methodology, and to provide a baseline simulation for more complex topographies, a typical inkjet droplet impacting a smooth surface was simulated. The droplet was 48.8\,\si{\micro\meter} in diameter falling at 3.74\,\si{\meter\per\second}, with density 1000\,\si{\kilogram\per\cubic\meter}, surface tension $26$\,\si{\newton\per\meter} and dynamic viscosity $9\times10^{-4}$\,\si{\pascal\second} (i.e.\ $Re=204$ and $We=26$). Several simulations were run without contact angle hysteresis for various $\theta$ and using different lattice resolutions. The resulting equilibrium diameters are compared with values predicted by equation \eqref{vanDamSpreadDiam} in figure \ref{fig:analyticalcomparison}(a), which shows very good convergence of the numerical simulations to the analytical result.

Figure \ref{fig:analyticalcomparison}(b) shows the maximum spreading diameter of a droplet (scaled by its initial diameter) obtained from numerical simulations with CAH included ($\theta_A=75^\circ$; $\theta_R=1^\circ$) for different values of the Weber number. For comparison, the models from table \ref{t:betamax} are plotted, showing that the simulations are consistent with the collective predicted behaviour. The more recent models \citep{Roisman2009,Laan2014a, Lee2015,Wildeman2016a} produce similar values to each other, and --- despite being developed for larger droplets --- agree quite well with the numerical simulations of micro-droplets, particularly for the larger values of Weber number. For the lowest Weber numbers considered, the model by \citet{Chandra1991} fits our data most closely, but agreement with this model is poor for larger $We$. Overall, based on our micro-droplet simulations, the model of \citet{Roisman2009} is arguably the best both in terms of its agreement but also because it is an explicit model that is easier to use. 

\begin{figure}
	\centering
	\includegraphics[width=0.9\textwidth, trim={0cm 0cm 0 0cm},clip]{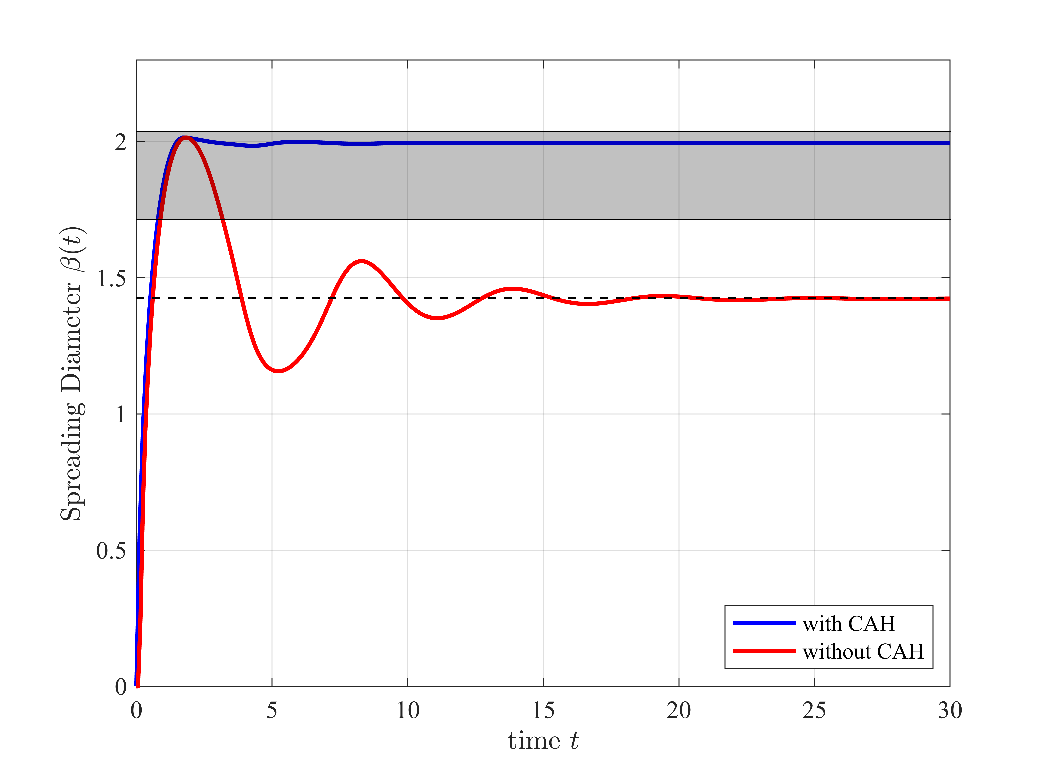}
	\caption{The spreading diameter as a function of time for a single droplet on a smooth surface. These are results from simulations with and without CAH compared with predictions of the maximum spread diameter from the models in table \ref{t:betamax} (grey shaded area) and the equilibrium diameter given by equation \eqref{vanDamSpreadDiam} (grey dashed line).}
	\label{f:DiameterWithTimeSmooth}
\end{figure}

Further comparison with the models in table \ref{t:betamax} is given in figure \ref{f:DiameterWithTimeSmooth} for the specific case $We=26$. The figure highlights the effect of CAH on the spreading. With hysteresis accounted for (again with $\theta_A=75^\circ$ and $\theta_R=1^\circ$), the contact line expands as the droplet spreads and becomes pinned essentially at the maximum spread diameter since the receding contact angle is so low. Such pinning is important in printing applications and is seen in practice with the colloidal inks used in the printed electronics industry \citep{DUINEVELD2003}. The grey shaded band again corresponds to the predictions of maximum spreading diameter from the models in table \ref{t:betamax}. In contrast, when CAH is not included in the simulation, the droplet recoils and contracts after reaching its maximum extent, then overshoots and oscillates in diameter as it settles to an equilibrium diameter consistent with equation \eqref{vanDamSpreadDiam}.

\section{Numerical simulations of single micro-droplet impact on a scratch \label{s:mainresults}}
The droplet deposition scenario considered is as described in \S\ref{s:geom} and shown in figure \ref{FigGeometry}. To provide a representative specific impact condition, simulation results are presented here for $Re=204$ and $We=26$, which are relevant to inkjet printing applications as indicated in figure \ref{fig:Re-We-map}(a). The advancing and receding contact angles are set to $\theta_A=75^\circ$ and $\theta_R=1^\circ$, again to be relevant to inkjet printing \citep{Davis1980,DUINEVELD2003}. The effect of variations in these four key parameters are considered in \S\ref{ss:EffectOfParameters}.

\subsection{Printed droplet morphologies\label{ss:morphs}}
\begin{figure}
	\centering
	\captionsetup{width=1\linewidth}
	\includegraphics[width=0.9\textwidth, trim={0cm 0cm 0 0cm},clip]{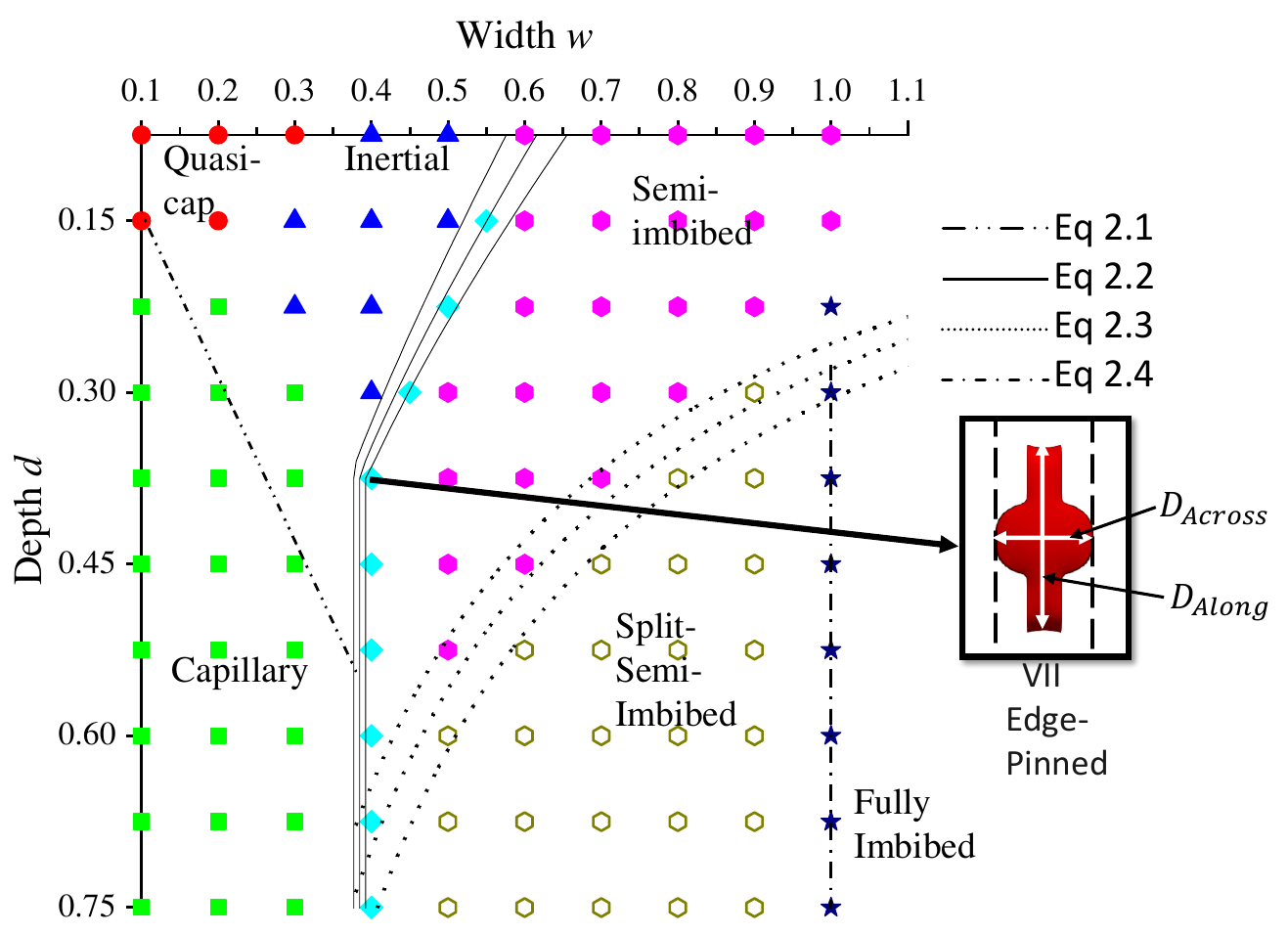}
	\caption{Numerically determined regime map for morphologies formed by printing a single droplet centred on the idealised scratch shown in figure \ref{FigGeometry}. Simulation parameters: $Re=204$, $We=26$, $\theta_A=75^\circ$ and $\theta_R=1^\circ$. The overlaid lines and curves give the theoretically estimated regime boundaries developed in \S\ref{s:geom}. Equations \eqref{eq:CriticalOverspilling} and \eqref{eq:ImbibitionLine} employ predictions of maximum spreading diameters from table  \ref{t:betamax}; the three curves for each equation represent the mean and standard deviation due to the models.}
	\label{fig:RegimeMapSingleSmooth}
\end{figure}
Figure \ref{fig:RegimeMapSingleSmooth} presents a map of the scratch width--depth parameter space showing the outcomes of droplet impact simulations that were run until an equilibrium morphology was formed. Different symbols indicate the morphology obtained at each width--depth combination, and corresponding examples of the equilibrium shapes for the same $Re$, $We$, $\theta_A$ and $\theta_R$ are given in figure \ref{fig:EightMorphos}. In addition, figure \ref{fig:RegimeMapSingleSmooth} shows an example of an `edge-pinned' final droplet, where the droplet has spread over the tops of the side ridges, reaching the outer edge with insufficient momentum to spill onto the original substrate surface. The contact line remains pinned on the outer edge as the very low receding contact angle prevents recession and the Gibbs pinning criterion is satisfied.  

The theoretically estimated regime boundaries developed in \S\ref{s:geom} are superimposed in figure \ref{fig:RegimeMapSingleSmooth} and show generally very good agreement with the outcomes of the numerical simulations. Recall that the three curves corresponding to equations \eqref{eq:CriticalOverspilling} and \eqref{eq:ImbibitionLine} give the mean and standard deviations of the predictions based on the five explicit maximum spreading models in table \ref{t:betamax}. There is slight deviation between equation \eqref{eq:CriticalYang} and the boundary of the capillary regime identified from simulations because equation \eqref{eq:CriticalYang} assumes an infinite source of liquid and does not capture the force due to the Laplace pressure resulting from curvature of the droplet, which becomes significant when the source droplet is of commensurate width to the groove. However, the criterion \eqref{eq:CriticalOverspilling} for pinning of the contact line on the edges of the side ridges captures very well the numerical predictions. Note that similar pinning has been observed even on rounded edges \citep{kant2017controlling}. 

The splitting boundary of equation \eqref{eq:ImbibitionLine} also fits well with the simulation results but has a slight deviation at larger depths. This is likely due to the assumption that the entire depth of the scratch is filled when deriving equation \eqref{eq:ImbibitionLine}, i.e.\ the simple geometrical argument does not account for the non-trivial shape of the liquid free-surface within the scratch. Similarly, the conditions for full imbibition of the droplet into the scratch are quite well identified by the criteria in \eqref{eq:FullyImbibed}, with some discrepancy in the critical depth due to the crude approximation of the free surface shape within the scratch. The delineation of the `quasi-spherical cap' region of figure \ref{fig:RegimeMapSingleSmooth} is somewhat subjective. Here, we define this to be where the final droplet shape deviates from a spherical cap by less than 10\% in the lateral dimensions.

\begin{figure}
	\includegraphics[width=1\textwidth, trim={0cm 0cm 0cm 0cm},clip]{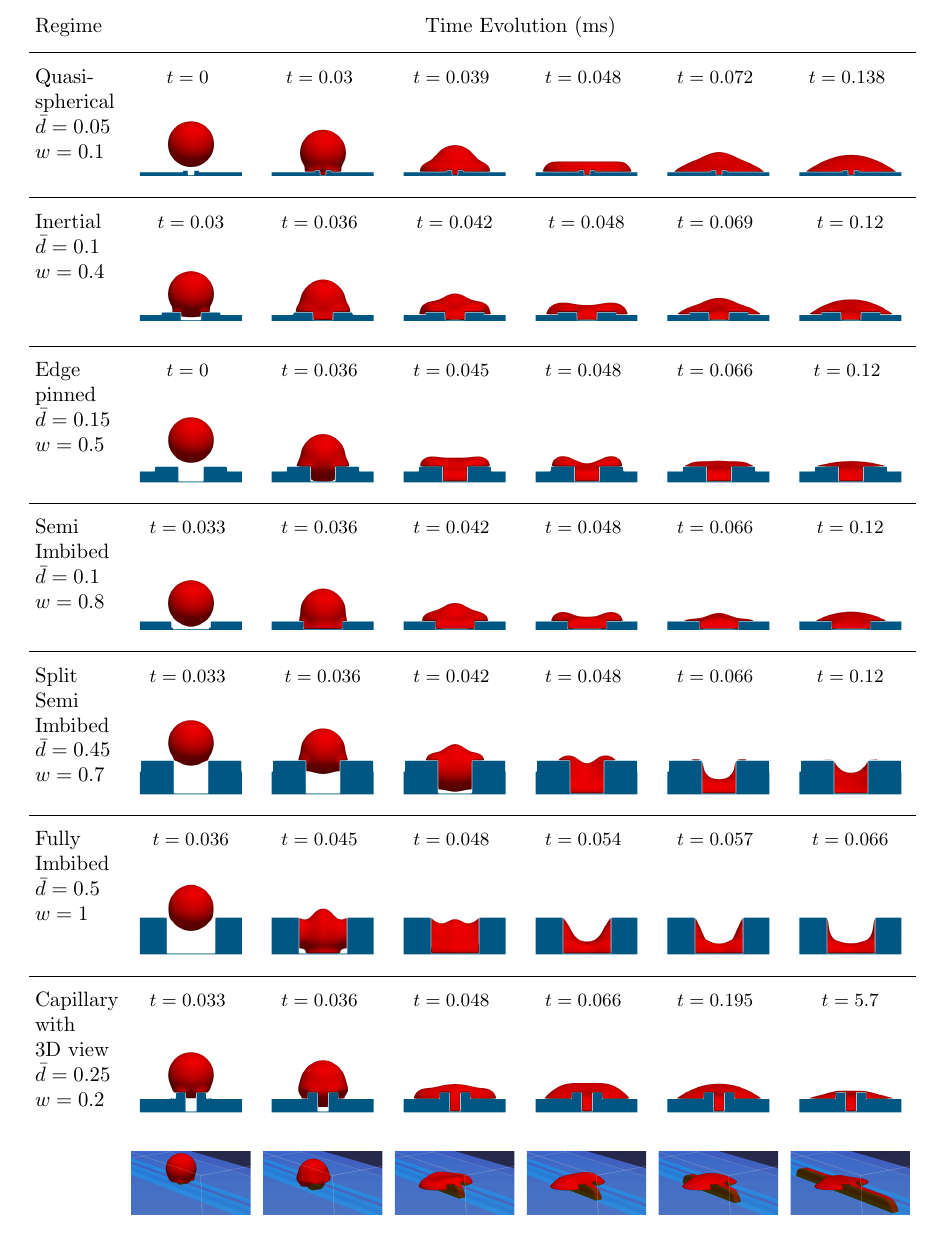} 
	 \caption{\label{t:TimeEvolutionOfRegimes} Front-view snapshots of the different regimes evolving in time measured in milliseconds. The images are to scale, the zoom scale is constant and the droplet diameter (hence also volume) is kept constant. The last row contains 3-D views to demonstrate the capillary flow along the scratch.}
\end{figure}

\subsection{Droplet spreading dynamics} \label{ss:Dynamics}
\begin{figure}
	\centering
    \begin{subfigure}[b]{0.3\textwidth}
		\centering
		\includegraphics[width=\textwidth, trim={0cm 0cm 0cm 0cm},clip]{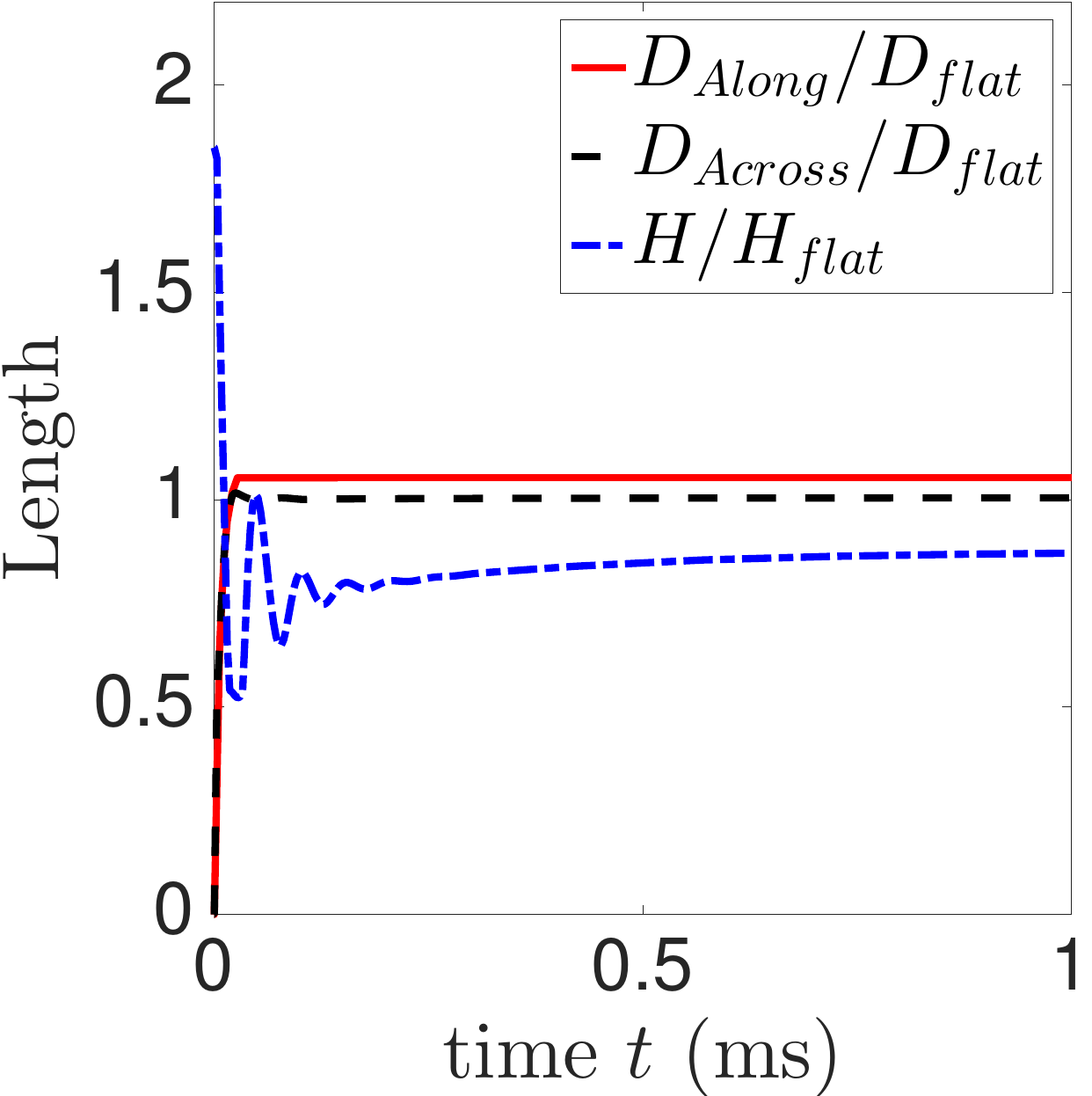}
		\caption{{}}
		\label{subfig:Edge-Pinned Dynamics}
	\end{subfigure}
  \begin{subfigure}[b]{0.3\textwidth}
		\centering
		\includegraphics[width=\textwidth, trim={0cm 0cm 0cm 0cm},clip]{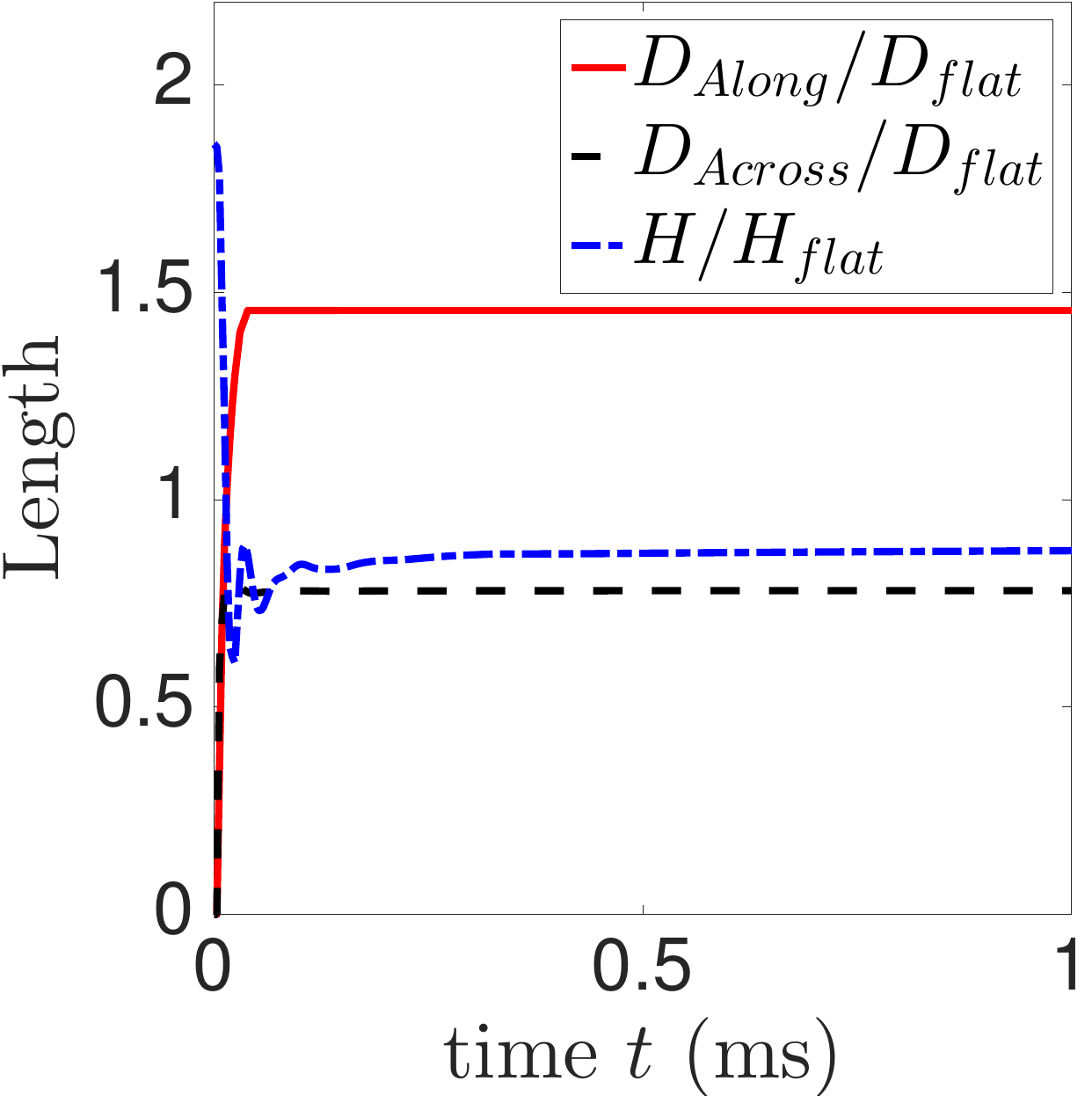}
		\caption{{}}
		\label{subfig:Edge-Pinned Dynamics}
	\end{subfigure}
	\begin{subfigure}[b]{0.3\textwidth}
		\centering
		\includegraphics[width=\textwidth, trim={0cm 0cm 0cm 0cm},clip]{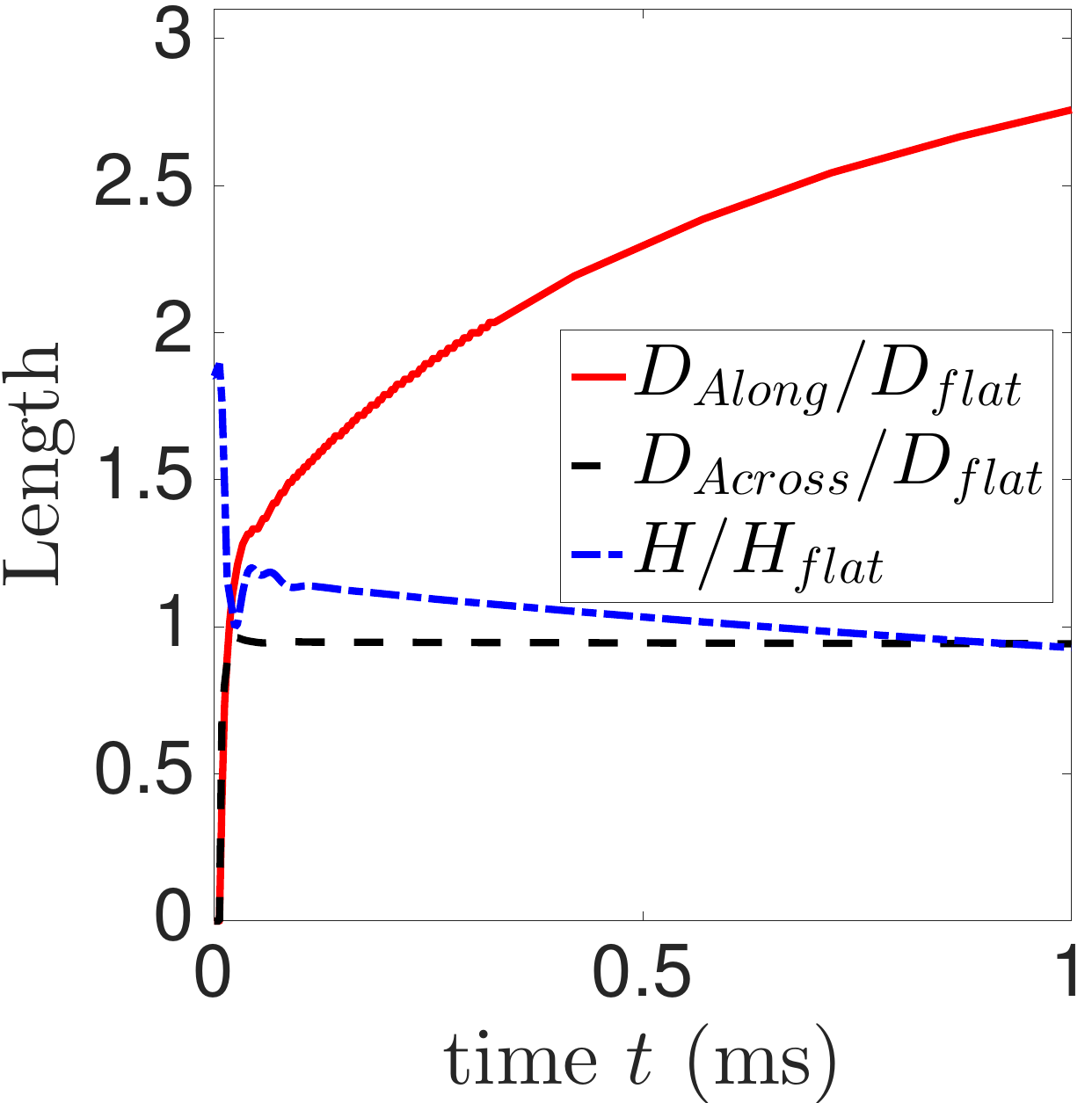}
		\caption{{}}
		\label{subfig:CapillaryDynamics}
	\end{subfigure}	
	\caption{Evolution with time of the morphology height and dimensions along and across a scratch for the (a) quasi-spherical cap, (b) edge-pinned and (c) capillary flow cases shown in figure \ref{t:TimeEvolutionOfRegimes}.}
	\label{fig:DimensionsVsSctrachDynamics}
\end{figure}

Although seven different equilibrium morphologies have been identified in figures \ref{fig:EightMorphos} and \ref{fig:RegimeMapSingleSmooth}, essentially only two types of flow dominate the spreading dynamics, namely inertia-driven spreading and capillary flow. Front views of the droplet spreading process at different times are shown in figure \ref{t:TimeEvolutionOfRegimes} for a selection of scratch widths and depths that lead to each of the seven equilibrium morphologies, while figure \ref{fig:DimensionsVsSctrachDynamics} illustrates how the horizontal and vertical dimensions of the droplet change in time for different morphologies. There, $D_{Along}$ refers to the length of the liquid in the direction of the scratch, $D_{Across}$ is the diameter perpendicular to the scratch, and $H$ is the droplet height measured at the centre of the scratch from the bottom of the scratch to the free surface. These quantities are normalised by the equilibrium spreading diameter of an equivalent droplet on a flat surface, $D_{flat}$ and height $H_{flat}$ of a corresponding droplet after impact on a smooth flat surface, i.e.\ with no scratch, from section \ref{ss:AnalyticalValid}. Note that in most cases considered here $D_{flat}\approx \beta_{max}=D_{max}/D_0$ (the expected maximum spreading diameter, $D_{max}$, of the droplet on a flat substrate with the same advancing and receding contact angles normalised by the in-flight diameter $D_0$) since the very low receding contact angle prevents contraction of the contact line, but for other values of $\theta_R$ these values are generally different.

In the quasi-spherical cap regime, the scratch is filled and covered very quickly, the droplet spreads to reach a maximum diameter and then oscillates, as seen in the $H$ curve in figure \ref{fig:DimensionsVsSctrachDynamics}(a), before relaxing more slowly to its equilibrium shape. The resulting horizontal dimensions $D_{Along}$ and $D_{Across}$ are similar to the corresponding spherical cap formed on a smooth flat surface. However, the slight extension of the droplet along the scratch results in a lower equilibrium height.

    In the inertial regime, the droplet touches the side ridges, then penetrates to the bottom surface of the scratch. The liquid spreads on the side ridges and spills over onto the original substrate surface, as seen in the second row of figure \ref{t:TimeEvolutionOfRegimes}. Compared with the quasi-spherical cap case, there is greater spreading along the scratch and slightly shorter spreading in the direction perpendicular to the scratch. The wider scratch results in a greater volume of liquid occupying the scratch, and consequently a slightly reduced final droplet height. In the edge-pinned regime, i.e.\ for wider scratches, the droplet contact line reaches the outer edges of the side ridges, but, as noted in \S\ref{ss:morphs}, there is insufficient momentum to carry the free surface past the edges, and the contact line remains pinned as seen in figure \ref{t:TimeEvolutionOfRegimes}. Figure \ref{fig:DimensionsVsSctrachDynamics}(b), which corresponds to $w=0.5$, shows the increased inertia-driven spreading along the scratch and the subsequent halting of the contact line producing a constant $D_{Along}$ as a result of the very low receding contact angle. The extent of the droplet in the direction across the scratch is reduced by the enhanced flow along the direction of the scratch and the pinning on the ridge edges.

    Under conditions leading to the semi-imbibed morphology, the droplet impacts the bottom of the scratch and spreads along and across the scratch and then impacts onto the side ridges, see figure \ref{t:TimeEvolutionOfRegimes}. The droplet spreads on the top of the side ridges but the contact line does not reach the outer edges; it becomes pinned somewhere on the top of the ridges because of the very low receding contact angle. In the split semi-imbibed regime, the droplet impacts the side ridges first, and then penetrates into the scratch, reaching the bottom and spreading along it. As the droplet spreads into the scratch, it splits along the inner edges of the side ridges, leaving separated droplets sitting on the top of the side ridges. These droplets are almost flat because of the very low receding contact angle. In the fully imbibed regime, the droplet never impacts on the top surface of the side ridges but falls into the scratch and spreads until an equilibrium is reached.

    Finally, as shown in the 3-D view included in the bottom row of figure \ref{t:TimeEvolutionOfRegimes}, for scratches producing capillary flow, the droplet initially spills over the side ridges onto the original substrate as in the quasi-spherical cap and inertial regimes. However, the liquid then spreads along the scratch by capillary action until there is no more liquid to supply further flow. This can be clearly seen in the evolution of both $D_{Along}$ and the droplet height in figure \ref{fig:DimensionsVsSctrachDynamics}(c). Note the different scale on the vertical axis in this plot compared with the others in figure \ref{fig:DimensionsVsSctrachDynamics}.

    The rate of capillary propagation has been studied extensively starting with work by \citet{washburn1921dynamics}, \citet{bell1906flow} and \citet{lucas1918ueber} on cylindrical capillaries. The main finding is that the propagation length $x$ increases as the square root of time, that is $x\propto \sqrt{t}$. More recently this analysis has been extended to other geometries including capillary flow in open rectangular micro-channels, studied theoretically and experimentally by \citet{yang2011dynamics}. They followed a similar approach to that used to derive the Washburn model except that they applied the procedure to an open rectangular micro-channel. The reservoir droplet in their experiments was large enough relative to the micro-channel to ignore its Laplace pressure in the model. Although the capillary flow seen here initially follows the propagation rate predicted by their model, the limited volume of fluid in the droplet supplying the flow soon results in a more rapid decrease in the speed of propagation and the corresponding flattening of the $D_{Along}$ curve in figure \ref{fig:DimensionsVsSctrachDynamics}(c).

\FloatBarrier

\subsection{Groove without ridges}\label{ss:SecNoRidges}
\begin{figure}
	\centering
	\includegraphics[width=0.9\textwidth, trim={0cm 0cm 0 0cm},clip]{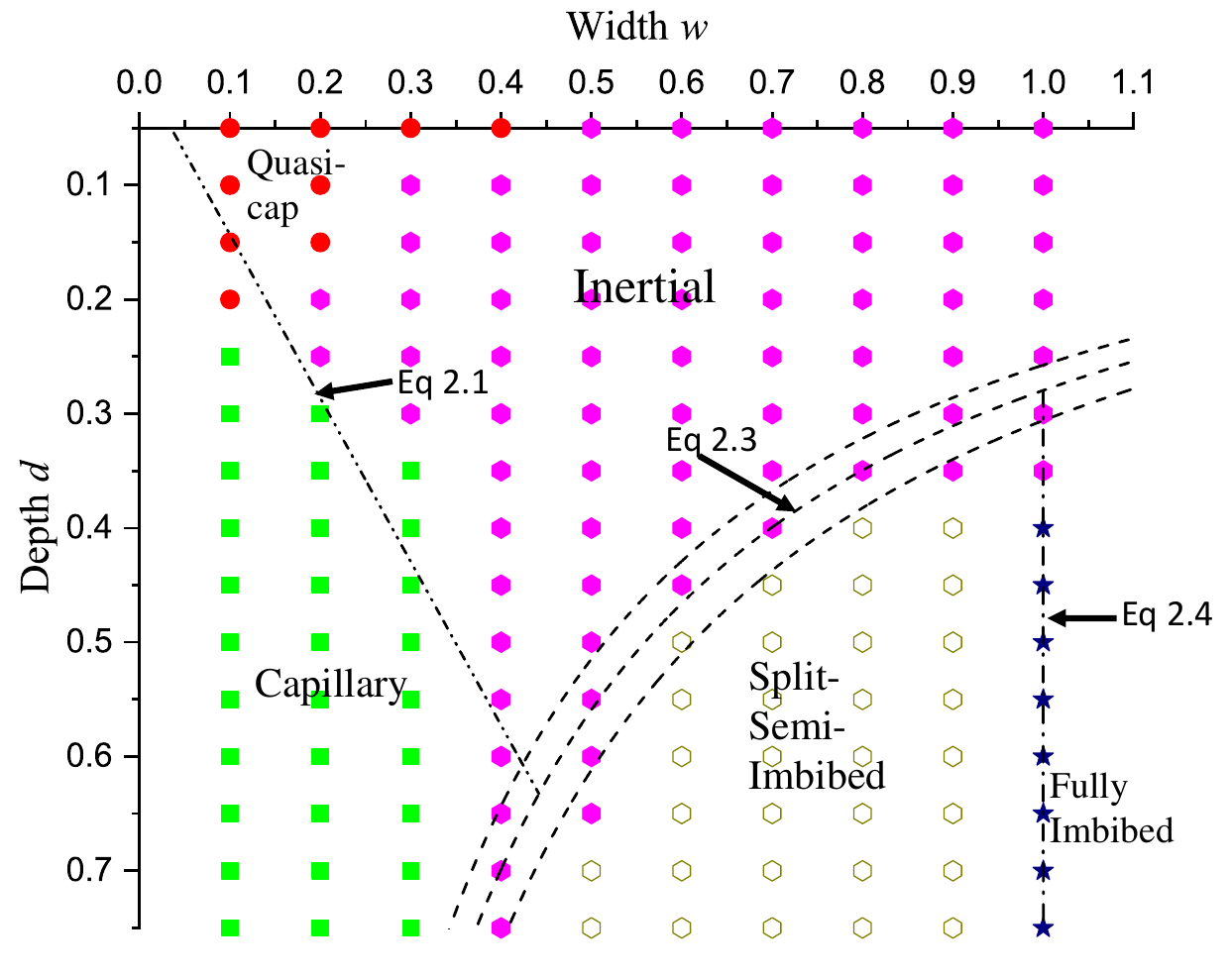} 
	\caption{Regime map of equilibrium morphologies formed by printing a single droplet centred on a groove of width $w$ and depth $d$ with no side ridges. Simulation parameters: $Re=204$, $We=26$, $\theta_A=75^\circ$ and $\theta_R=1^\circ$. The overlaid lines and curves are theoretical estimates of the regime boundaries based on conditions \eqref{eq:CriticalYang}, \eqref{eq:ImbibitionLine} and \eqref{eq:FullyImbibed}. Equation \eqref{eq:ImbibitionLine} employs predictions of maximum spreading diameters from table  \ref{t:betamax}, the three curves representing the mean and standard deviation of the predictions.}
	\label{fig:RegimeMapSingleSmoothNoRidge}
\end{figure}

Removing the side ridges from the topography in figure \ref{FigGeometry} reduces the number of possible equilibrium morphologies from seven to five, and alters the combinations of groove width and depth at which the morphologies arise. The corresponding regime map, constructed from simulations for groove widths ranging from 0.1--1 and depths of 0.05--0.75 with increments of 0.1 and 0.05 respectively, is shown in figure \ref{fig:RegimeMapSingleSmoothNoRidge}. The morphologies caused by the presence of the ridges, namely edge pinned and inertial (where the droplet spills over the ridges), are not seen with this topography. These both merge into the semi-imbibed morphology, where the equilibrium shape of the droplet occupies both the groove and the nearby region of the substrate surface, and is the dominant morphology for sufficiently shallow grooves. For narrow and shallow grooves, the morphology can still be classed as a quasi-spherical cap, and for narrow and deep grooves, the capillary morphology is still seen. The fully imbibed, and split semi-imbibed morphologies are again seen for sufficiently wide and deep grooves.

The theoretical estimates \eqref{eq:CriticalYang}, \eqref{eq:ImbibitionLine} and \eqref{eq:FullyImbibed} for the boundaries of the regions in the regime map are readily adapted by using the appropriate depth $\bar{d}$, and these are included in figure \ref{fig:RegimeMapSingleSmoothNoRidge}. Again, good agreement is seen between these estimates and the results of the simulations.

\section{Effect of flow and substrate parameters}\label{ss:EffectOfParameters}
The regime maps in figures \ref{fig:RegimeMapSingleSmooth} and \ref{fig:RegimeMapSingleSmoothNoRidge} were constructed using a single set of material and flow parameters $\theta_A$, $\theta_R$, $Re$, and $We$ to illustrate the effects of the scratch geometry and the relationships among the morphologies. For this case, the boundaries of the regions within the regime maps are found to be represented very well by the theoretical estimates \eqref{eq:CriticalYang}--\eqref{eq:FullyImbibed} that are more broadly applicable. Apart from the criterion for capillary flow, these expressions are based on the maximum spreading diameter, for which many models exist that account for the effects of Reynolds number, Weber number and advancing contact angle (see table \ref{t:betamax}). Hence, it is possible to predict new regime boundary estimates for other values of these parameters. As noted in \S\ref{ss:AnalyticalValid}, the \citet{Roisman2009} model appears to be most suitable for our micro-droplets, but all models show the same trends with $We$ in figure \ref{subfig:ValidationUsingDmax}.

\subsection{Effect of Reynolds number and Weber number}
\begin{figure}
	\includegraphics[width=1\textwidth, trim={1cm 0cm 1cm 0.5cm},clip]{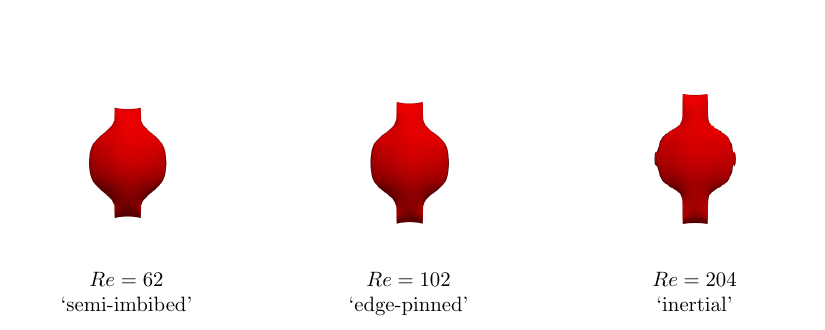} 
	 \caption{Effect of $Re$ on the morphology resulting from a droplet impact on a scratch of dimensions $d=0.15$ and $w=0.5$ at $We=26$, with $\theta_A=75^\circ$ and $\theta_R=1^\circ$. }
    \label{f:Re-effect}
\end{figure}

Increasing $Re$ and/or $We$ promotes greater spreading of the droplet on impact and the droplet is therefore able to spill over wider side ridges. Hence $w_{pin}$ --- the critical scratch width for edge pinning in \eqref{eq:CriticalOverspilling} --- increases and the `inertial' region of figure \ref{fig:RegimeMapSingleSmooth} will expand to the right. This effect is illustrated in the specific case shown in figure \ref{f:Re-effect}, for a scratch geometry given by $d=0.15$ and $w=0.5$. A droplet impact at $Re=62$ results in a `semi-imbibed' morphology, whereas an impact on the same scratch at $Re=102$ produces the `edge-pinned' morphology, and at $Re=204$ the `inertial' morphology arises, consistent with the increase in the value of $w_{pin}$.

The greater inertia of the droplet will also result in greater penetration into and along the scratch, resulting in a smaller liquid height within the scratch and consequently a greater tendency for the droplet to split along the inner edges of the scratch and leave separate small droplets on the outer surface. From the form of equation \eqref{eq:ImbibitionLine} it is clear that the boundary of the `split semi-imbibed' region in both figures \ref{fig:RegimeMapSingleSmooth} and \ref{fig:RegimeMapSingleSmoothNoRidge} will expand upwards and to the left --- i.e.\ towards narrower, shallower scratches. The critical condition \eqref{eq:CriticalYang} for the onset of capillary flow is independent of $Re$ and $We$ since capillary flow is not inertia driven and continues long after the initial spreading of the droplet. It is of course greatly influenced by $\theta_A$, and this is discussed below.

\subsection{Effect of the advancing contact angle, $\theta_A$}\label{subsubsection:EffectThetA}
\begin{figure}
	\includegraphics[width=1\textwidth, trim={0cm 0.3cm 0cm 0cm},clip]{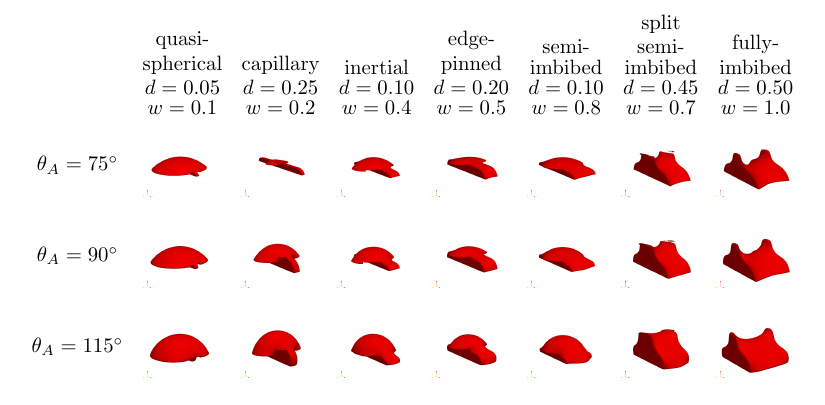} 
	\caption{Effect of $\theta_A$ on final droplet shape for $Re=204$, $We=26$ and $\theta_R=1^\circ$. The images show just the liquid volume(s) within and around each scratch.}
\label{f:effect-thetaA}
\end{figure}

The effect of $\theta_A$ on the initial spreading of a droplet on a flat surface is captured in some of the models of maximum spreading diameter in table \ref{t:betamax}. It is well known that contact angles below 90$^\circ$ promote spreading and those above hinder it. Hence increasing $\theta_A$ results in reduced spreading both along and perpendicular to the scratch, and a consequently increased droplet height (unless of course the droplet is fully imbibed). However, $\theta_A$ also influences the shape of the liquid free surface within the scratch, as can be seen in figure \ref{f:effect-thetaA}, which shows just the liquid volume with the confining solid made invisible.

The contact line on the bottom of the scratch, which is generally concave at $\theta_A=75^\circ$, becomes convex when $\theta_A=115^\circ$ as the sidewalls of the scratch then act to hinder rather than assist spreading. The non-trivial shape of the liquid free surface in the `semi-imbibed' and `split semi-imbibed' cases illustrate why there is a small discrepancy between the predicted regime boundary given by equation \eqref{eq:ImbibitionLine} and that observed from the simulations.

In the `quasi-spherical cap' regime, the liquid volume that runs along the scratch decreases with increasing $\theta_A$ until it no longer extends beyond the diameter perpendicular to the scratch, and for $\theta_A=115^\circ$ (figure \ref{f:effect-thetaA}) the liquid inside the scratch does not extend as far as that above the scratch. Moreover, capillary flow will no longer occur for $\theta_A\geqslant 90^\circ$, and as a consequence the region in figures \ref{fig:RegimeMapSingleSmooth} and \ref{fig:RegimeMapSingleSmoothNoRidge} where capillary flow exists moves to smaller values of $w/d$ as $\theta_A$ increases, until the capillary flow region vanishes.

For a scratch that produces the `inertial' morphology when $\theta_A=75^\circ$, the reduced spreading that occurs when $\theta_A$ is increased can mean that the droplet no longer has sufficient momentum to spill over the side ridges, and the droplet becomes `edge-pinned' or `semi-imbibed'.  A similar effect is seen for wider scratches, and we conclude that the boundaries within the regime map in figure \ref{fig:RegimeMapSingleSmooth} shift towards lower widths as $\theta_A$ increases, with the capillary flow region eventually disappearing.

\subsection{Effect of the receding contact angle, $\theta_R$}\label{subsubsection:EffectThetR}
\begin{figure}
	\includegraphics[width=0.95\textwidth, trim={0cm 0.5cm 0cm 0cm},clip]{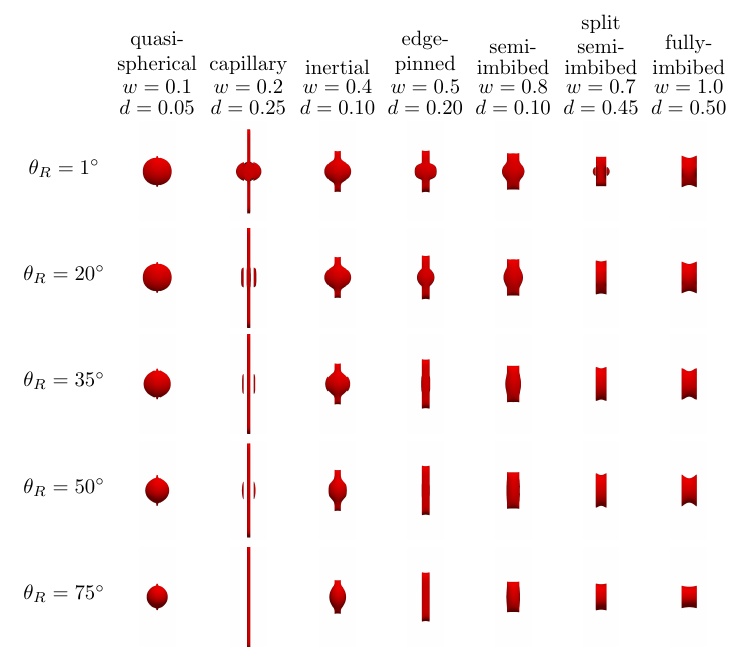} 
	\caption{Effect of $\theta_R$ on final droplet shape. The images show top views for $\theta_A=75^\circ$, $Re=204$ and $We=26$.}
\label{f:effect-thetaR}
\end{figure}

If, as the droplet shape changes, the contact angle falls below the receding contact angle, the contact line will recede. Changes in $\theta_R$ are therefore be expected to have a significant influence on the droplet morphologies discussed above. Figure \ref{f:effect-thetaR} highlights this via top-view plots of the final droplet shape for five different values of $\theta_R$ from the 1$^\circ$ value used to create figures \ref{fig:RegimeMapSingleSmooth} and \ref{fig:RegimeMapSingleSmoothNoRidge} up to $\theta_R=75^\circ$, which corresponds to the case where there is no CAH.
For scratch dimensions leading to the quasi-spherical cap morphology, the droplet remains quasi-spherical, but its footprint becomes smaller as $\theta_R$ increases because the contact line recedes until an equilibrium is reached where the contact angle is everywhere equal to or greater than $\theta_R$. The conditions for capillary flow to occur are independent of $\theta_R$, but as $\theta_R$ increases, the contraction of the droplet above the scratch means that more liquid is available to feed the capillary flow and the final extent of the capillary flow is therefore increased. However, for intermediate $\theta_R$, as the droplet recedes from the top of the side ridges and original surface, small amounts of liquid are left behind in the corners between the outer wall of the side ridges and the original surface of the substrate. When there is no contact angle hysteresis ($\theta_R=\theta_A=75^\circ$), the entire droplet is pulled into the scratch by capillary action.

The `inertial' morphology in figure \ref{fig:RegimeMapSingleSmooth} arises when the droplet spills over the side ridges of the scratch onto the original surface, but no capillary action occurs. When $\theta_R$ is larger, the contact line recedes after reaching its maximum spread, and climbs back onto the side ridges to reach a different equilibrium. Hence the `inertial' morphology becomes `edge pinned' and `semi-imbibed' for $\theta_R$ values of $50^\circ$ and $75^\circ$, respectively. The liquid inside the scratch also recedes with higher $\theta_R$. The `semi-imbibed' morphology becomes `fully imbibed' due to the recession of the footprint outside the groove into it. The `edge-pinned' morphology turns into a `semi-imbibed' and then `fully-imbibed' as $\theta_R$ approaches $\theta_A$.  The fully imbibed morphology only changes in its extent along the scratch.  Increasing the receding contact angle generally results in an equilibrium morphology that is less spread in the direction perpendicular to the scratch, and also shorter in the direction along the scratch. The exceptions are the capillary and edge-pinned regimes, where the lack of contact-line pinning on the upper substrate releases more liquid to penetrate along the scratch. All the regime map boundaries shift to lower widths, except the capillary regime because it is governed by $\theta_A$.

\section{Implications for printing applications}\label{s:Implications}

In printing an electrical circuit, a key requirement is continuity of the printed track. Achieving a stable, continuous printed line requires consistent droplet spreading behaviour and a careful balance of droplet generation frequency and printing speed to achieve the correct droplet spacing \citep{Stringer2010}. Variations in the droplet spreading caused by a scratch could therefore destabilise the line or cause a break in continuity. Hence quantitative measures of the spreading behaviour on a scratch are potentially useful. Here, for simplicity, the two simplest configurations will be considered, namely printing perpendicular to a scratch and printing along a scratch.

\subsection{Line printing across a scratch}

\begin{figure}
	\centering
    \begin{subfigure}[b]{0.495\textwidth}
		\centering
		\includegraphics[width=\textwidth, trim={0cm 0cm 0cm 0cm},clip]{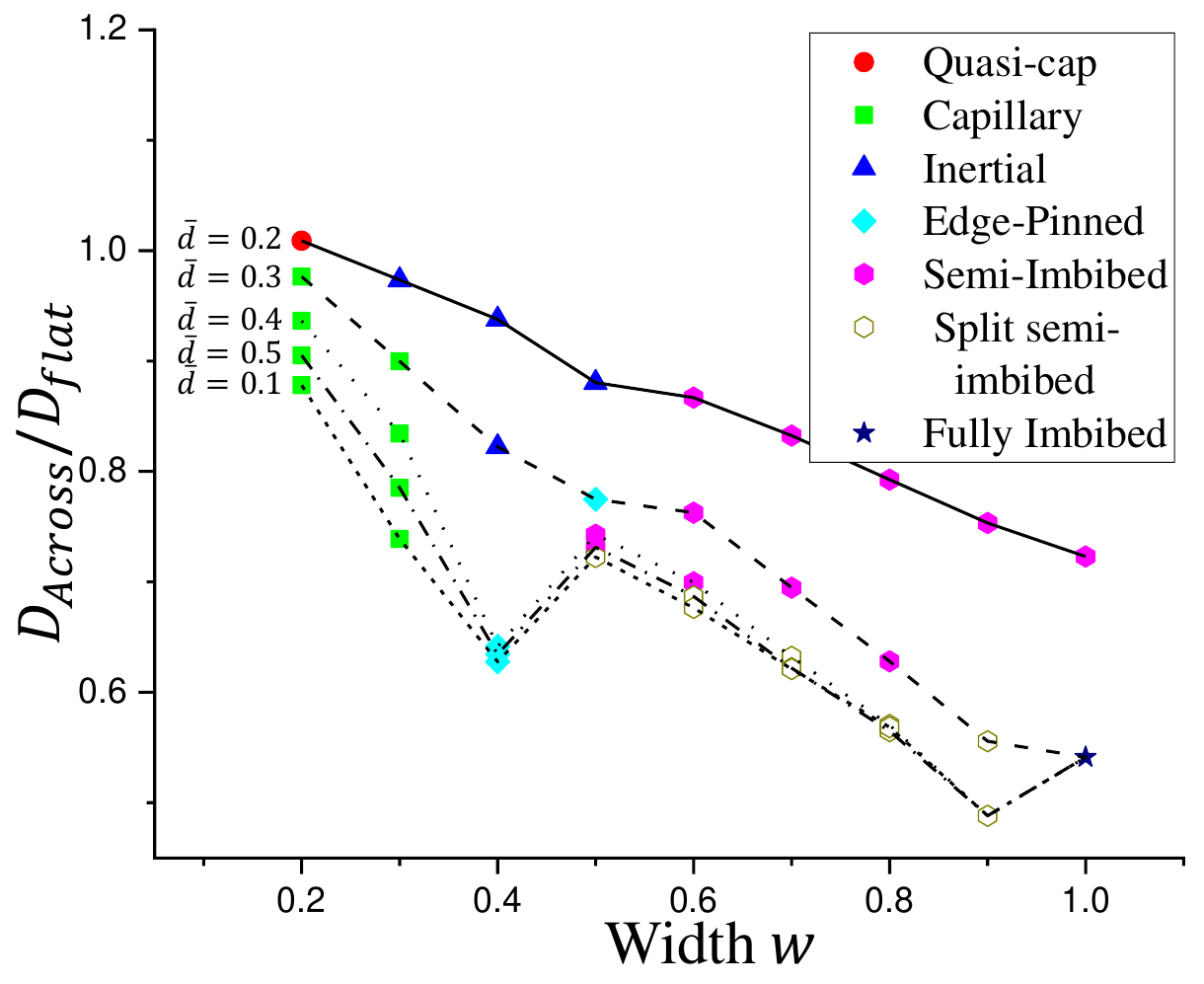}
		\caption{{}}
		\label{subfig:Edge-Pinned Dynamics}
	\end{subfigure}
  \begin{subfigure}[b]{0.495\textwidth}
		\centering
		\includegraphics[width=\textwidth, trim={0cm 0cm 0cm 0cm},clip]{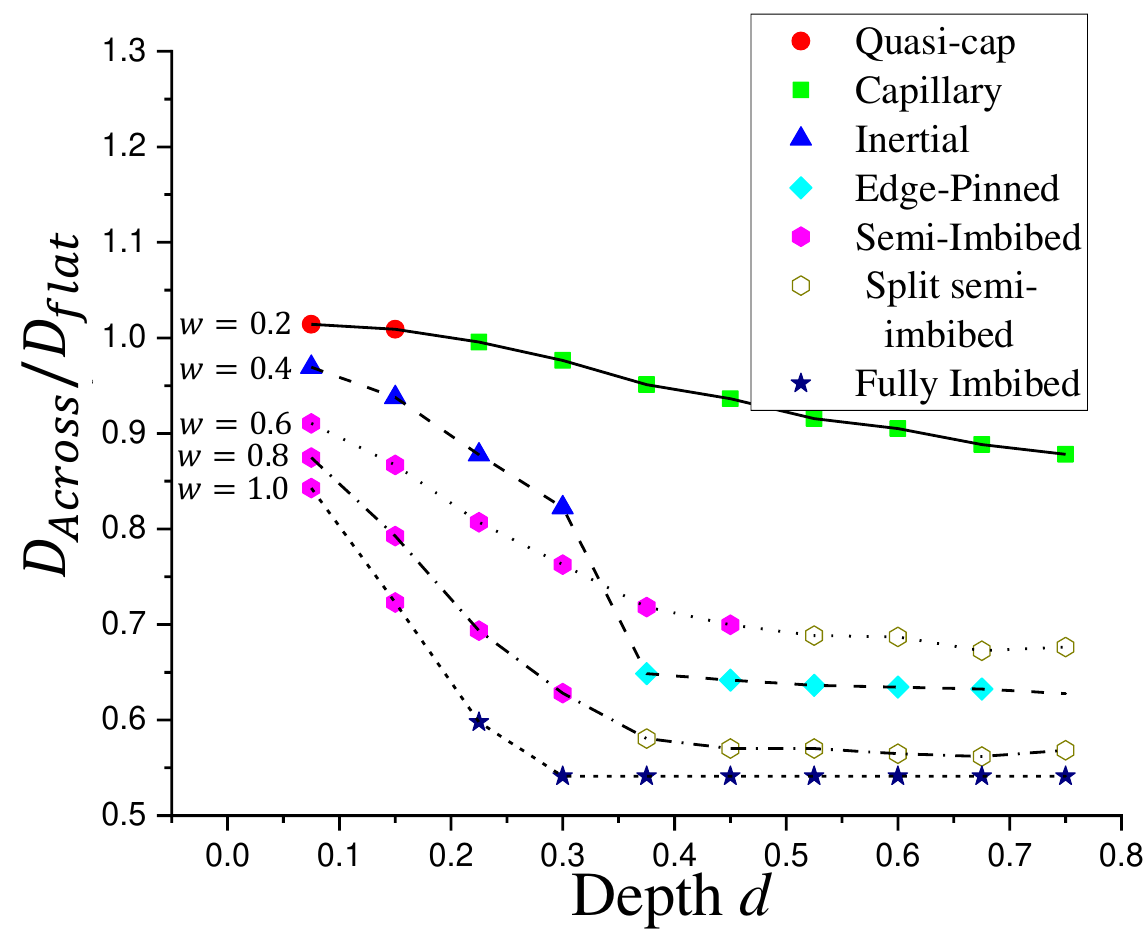}
		\caption{{}}
		\label{subfig:Edge-Pinned Dynamics}
	\end{subfigure}
	\caption{Value of $D_{Across}$, the length of the final single-droplet morphology in the direction perpendicular to a scratch following deposition of a droplet on the scratch shown in figure \ref{FigGeometry} at $Re=204$, $We=26$, with $\theta_A=75^\circ$ and $\theta_R=1^\circ$. The plots show variation with scratch width and depth, and the coloured symbols indicate the type of morphology following the same labelling as in figure \ref{fig:RegimeMapSingleSmooth}.}
	\label{fig:Dacross}
\end{figure}

With the prospect of a break in line continuity, a key quantity to consider is the extent of single-droplet spreading in the direction perpendicular to the scratch. Figure \ref{fig:Dacross} shows the behaviour of $D_{Across}$, i.e.\ the length of the resulting single-droplet morphology in the direction perpendicular to the scratch (see figure \ref{fig:RegimeMapSingleSmooth}), as a function of scratch width and depth. The plots show $D_{Across}$ normalised by $D_{flat}$, the equilibrium spreading diameter of an equivalent droplet on a flat surface. Recall that the scratch width and depth are scaled by the impacting droplet's in-flight diameter. While narrow, shallow scratches cause only a small change in the extent of spreading, it is clear that as the droplet and scratch become similar in size, a significant shortfall in the spread length occurs. Depending on the degree of overlap between consecutive droplets, which determines the printed line width \citep{Stringer2010}, such a reduction in spreading could prevent coalescence of a droplet with the rest of the line and hence a break in continuity. Note that the convergence of the lines in figure \ref{fig:Dacross}(a) and the sudden drop in the $w=0.4$ line in figure \ref{fig:Dacross}(b) are a result of pinning of the contact line on the outer edge of the side ridges. The upturn between $w=0.9$ and $w=1.0$ in figure \ref{fig:Dacross}(a) is because for the `fully imbibed' morphology, $D_{Across}=w$.

\begin{figure}
	\centering
	\begin{subfigure}[b]{0.23\textwidth}
		\centering
		\includegraphics[width=\textwidth, trim={2cm 4.5cm 2.5cm 9cm},clip]{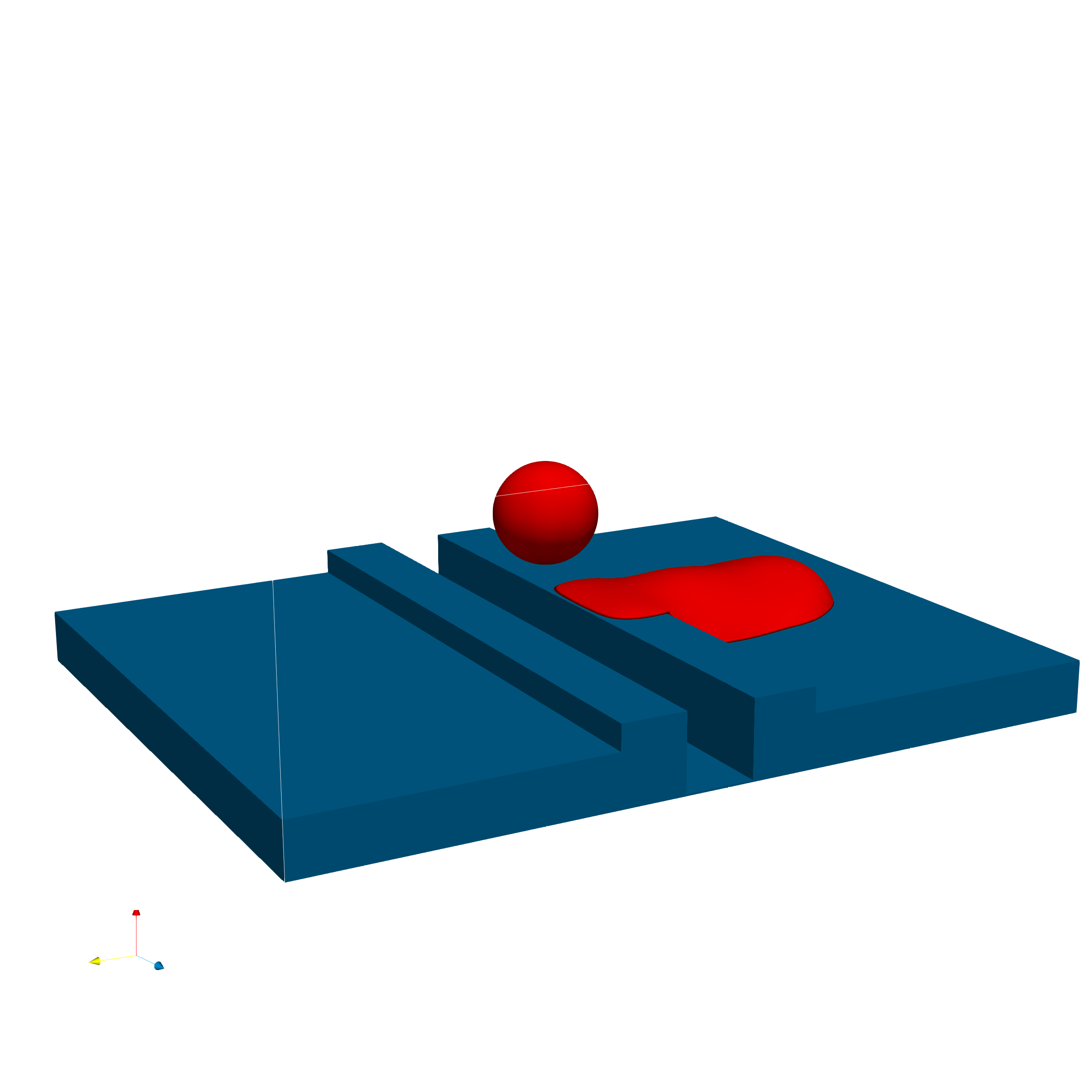}
		\caption{}
		\label{subfig:Across2impact2T1}
	\end{subfigure}
	\begin{subfigure}[b]{0.23\textwidth}
		\centering
		\includegraphics[width=\textwidth, trim={2cm 4.5cm 2.5cm 9cm},clip]{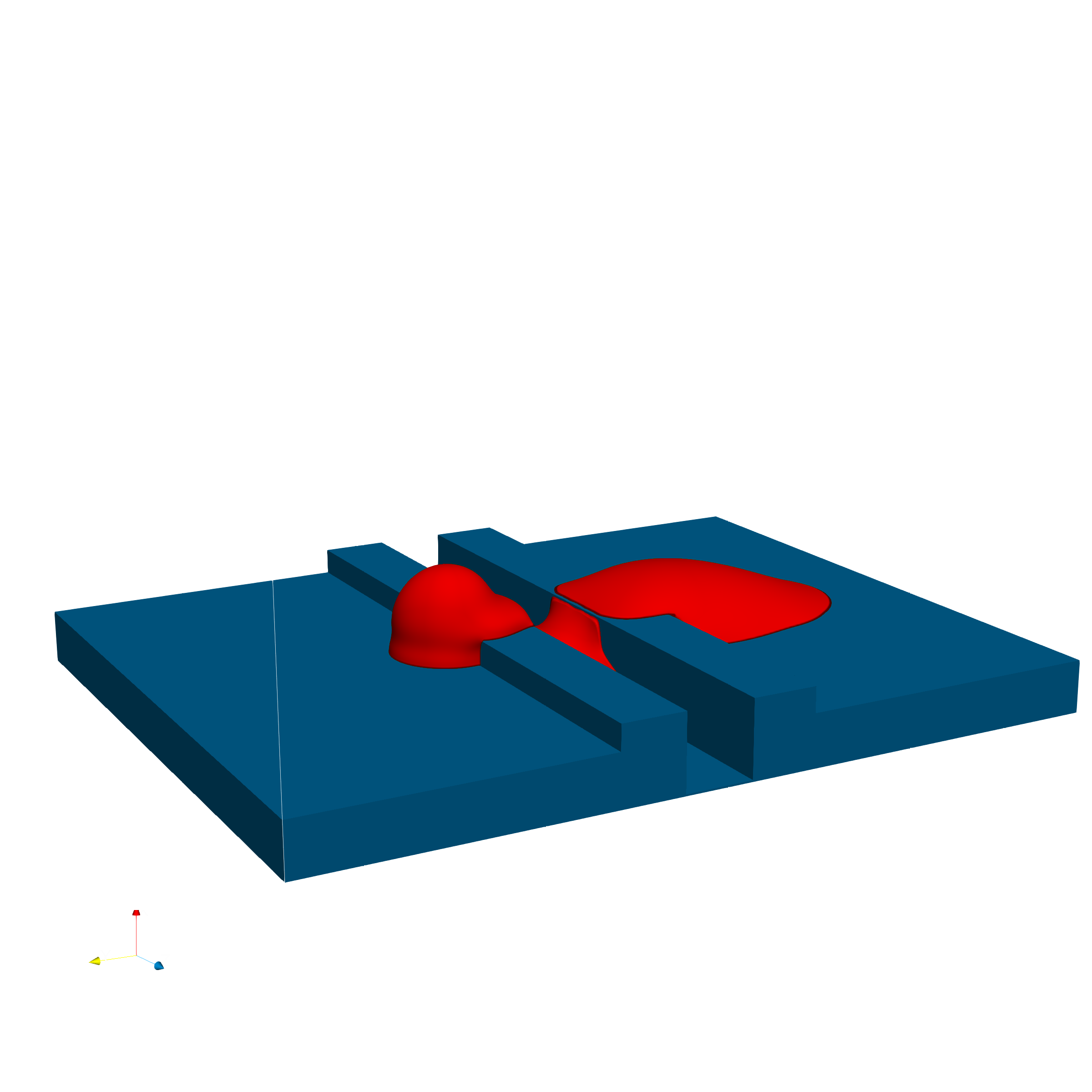}
		\caption{}
		\label{subfig:Across2impact2T2}
	\end{subfigure}
	\begin{subfigure}[b]{0.23\textwidth}
		\centering
		\includegraphics[width=\textwidth, trim={2cm 4.5cm 2.5cm 9cm},clip]{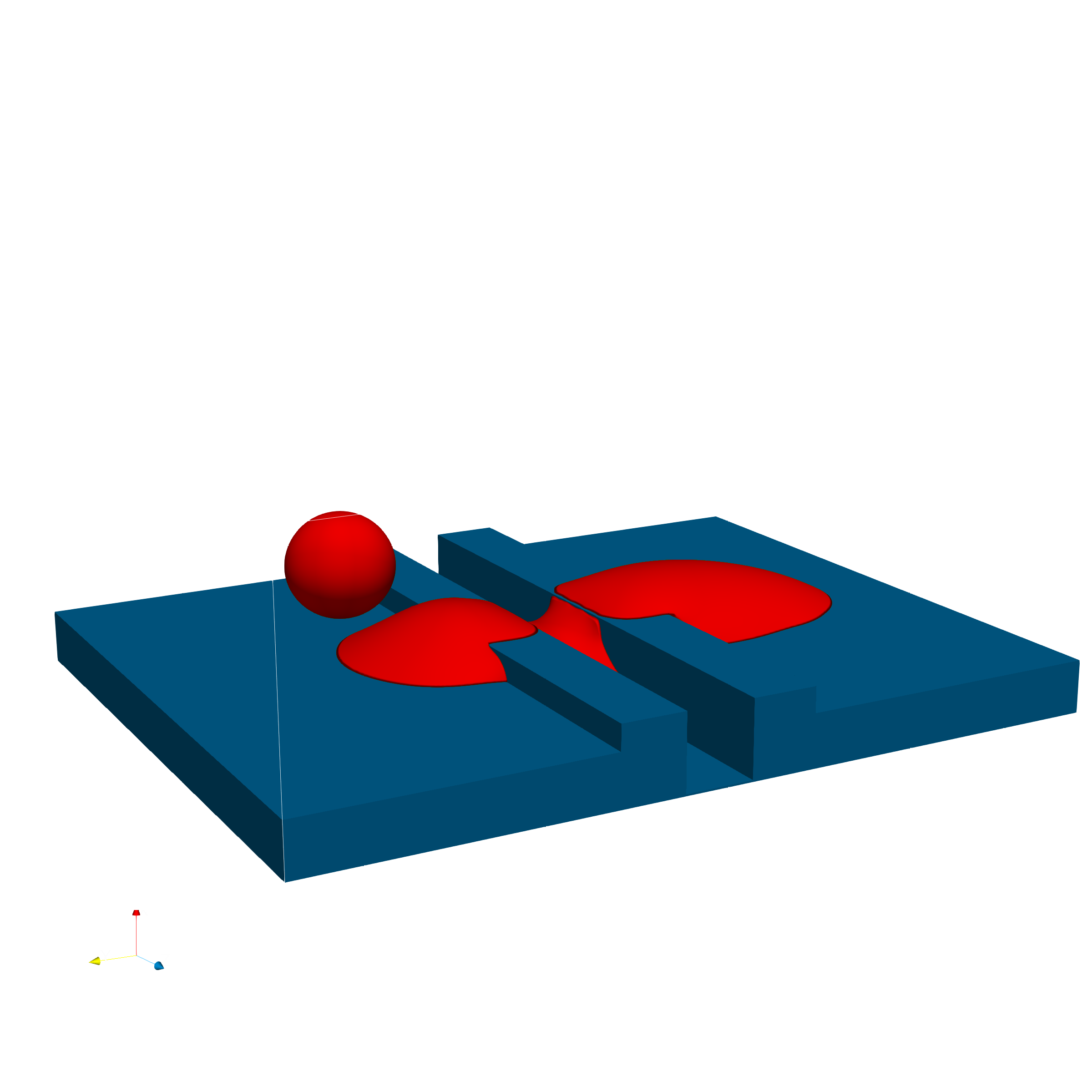}
		\caption{}
		\label{subfig:Across2impact2T3}
	\end{subfigure}
	\begin{subfigure}[b]{0.23\textwidth}
		\centering
		\includegraphics[width=\textwidth, trim={2cm 4.5cm 2.5cm 9cm},clip]{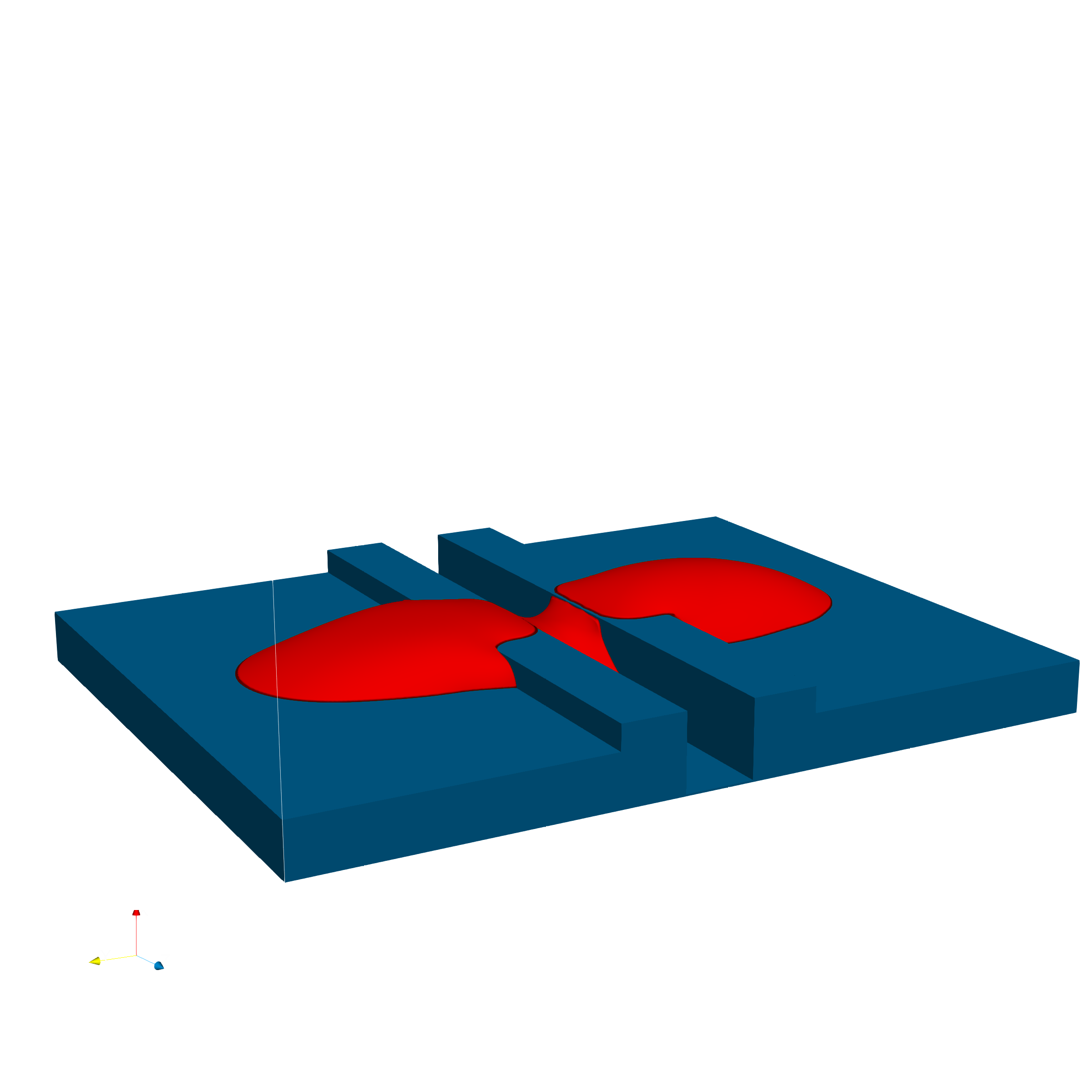}
		\caption{}
		\label{subfig:Across2impact2T4}
	\end{subfigure} \\
	\begin{subfigure}[b]{0.23\textwidth}
		\centering
		\includegraphics[width=\textwidth, trim={2cm 4.5cm 2.5cm 9cm},clip]{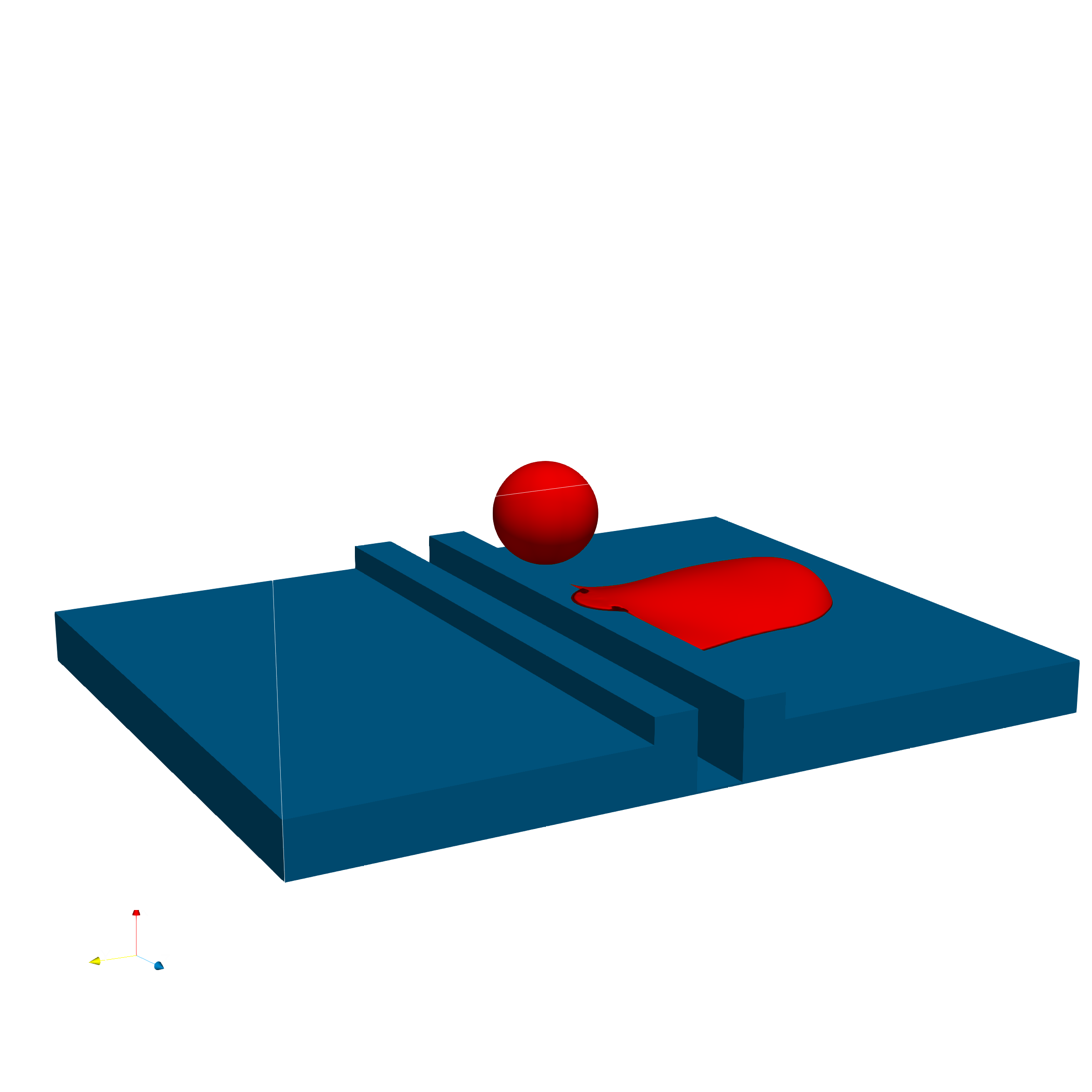}
		\caption{}
		\label{sf:PrintingAcrossFiveDrops1a}
	\end{subfigure}
	\begin{subfigure}[b]{0.23\textwidth}
		\centering
		\includegraphics[width=\textwidth, trim={2cm 4.5cm 2.5cm 9cm},clip]{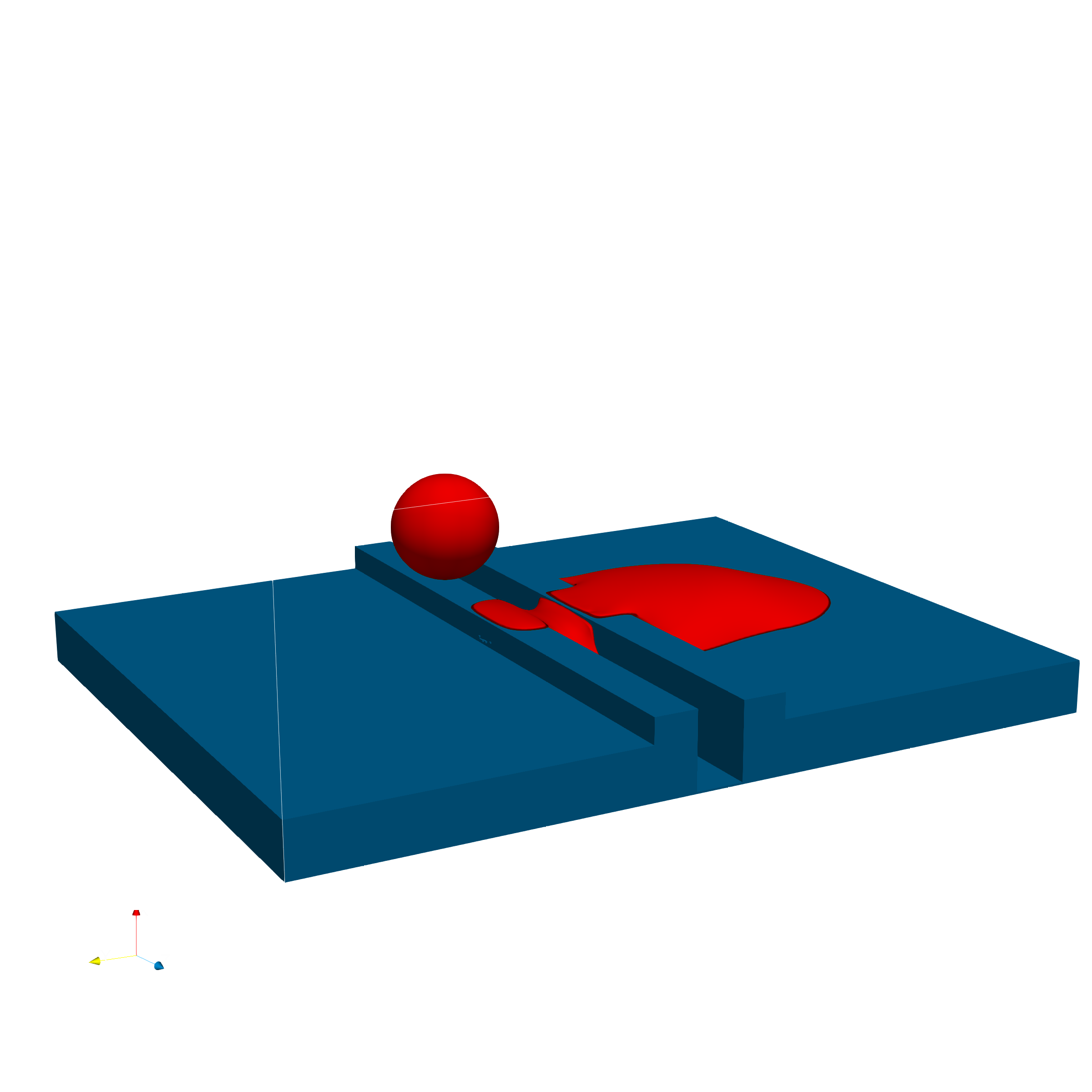}
		\caption{}
		\label{sf:PrintingAcrossFiveDrops1b}
	\end{subfigure}
	\begin{subfigure}[b]{0.23\textwidth}
		\centering
		\includegraphics[width=\textwidth, trim={2cm 4.5cm 2.5cm 9cm},clip]{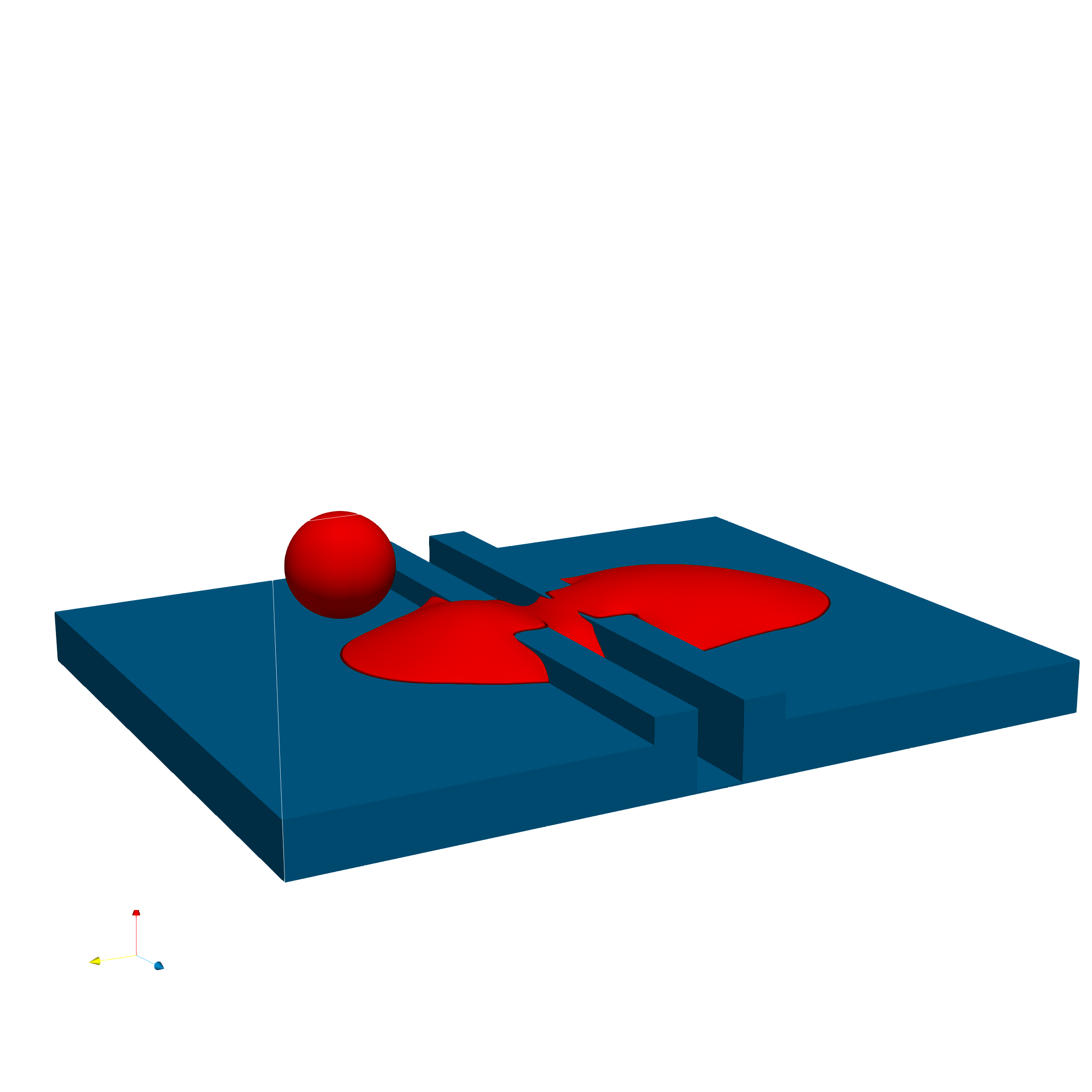}
		\caption{}
		\label{sf:PrintingAcrossFiveDrops1c}
	\end{subfigure}
	\begin{subfigure}[b]{0.23\textwidth}
		\centering
		\includegraphics[width=\textwidth, trim={2cm 4.5cm 2.5cm 9cm},clip]{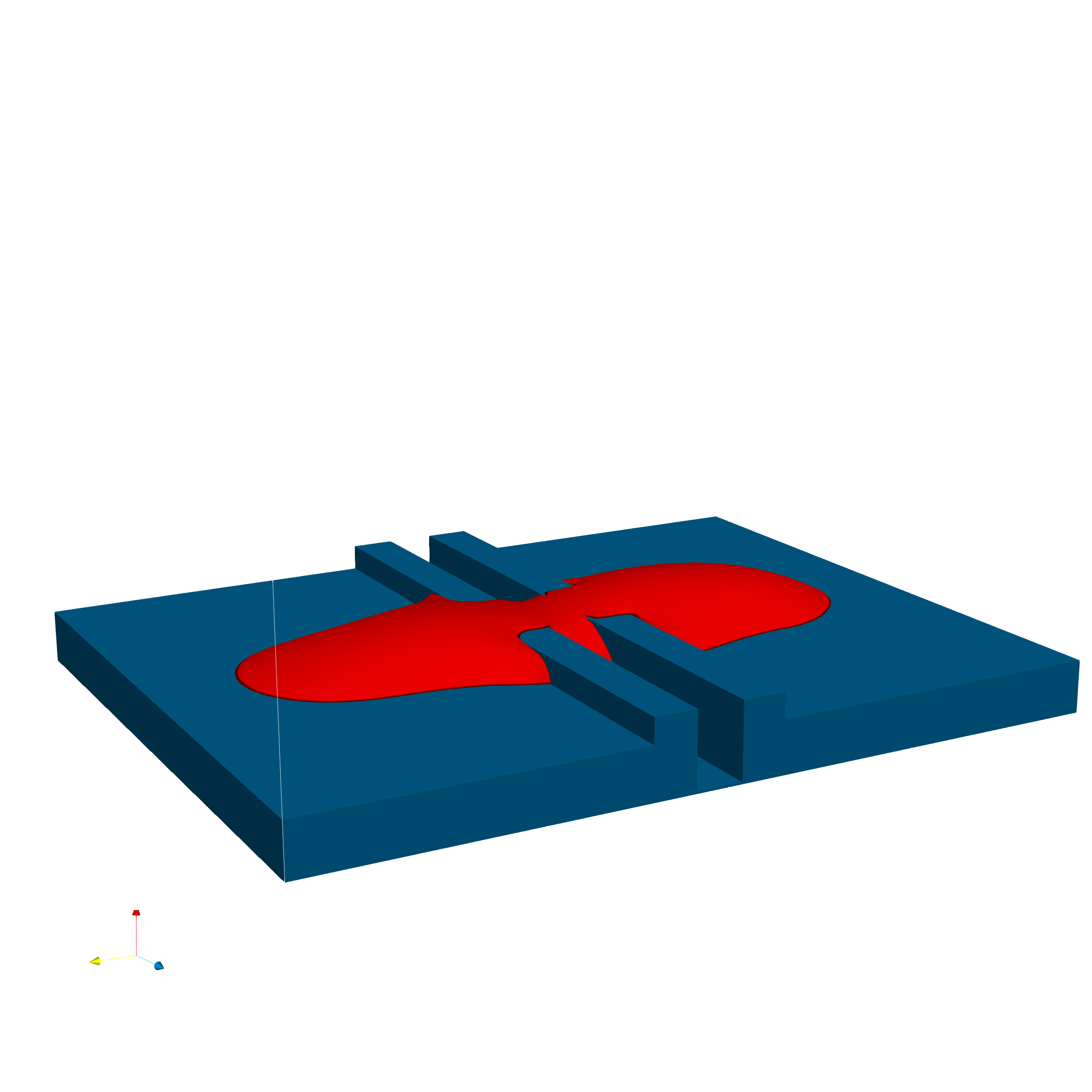}
		\caption{}
		\label{sf:PrintingAcrossFiveDrops1d}	\end{subfigure}
	\caption{Simulations of printing consecutive droplets across scratches of depth $\bar{d}=0.4$ ($d=0.6$) and widths $w=0.6$ (a-d) and $w=0.45$ (e-h). }
	\label{f:PrintingAcrossFiveDrops2}
\end{figure}

Beside the prevention of coalescence due to the reduction in spreading, another mechanism by which line continuity could be broken is through the splitting of the droplet along the inner edge of the scratch. This is illustrated in figure \ref{f:PrintingAcrossFiveDrops2}, which shows simulations of printing a series of five droplets across two scratches with the same depth but different widths. Both scratches appear in the `split semi-imbibed' region of figure \ref{fig:RegimeMapSingleSmooth}, and the third droplet, landing on the scratch centre, splits along the inner edges of the scratch, as expected --- see figure \ref{f:PrintingAcrossFiveDrops2}(b) and (f). For the wider of the two scratches, this splitting is irrecoverable and the printing continues with a separate line on the other side of the scratch. However, for the slightly narrower scratch, when the fourth droplet is printed, it pushes back on the pre-existing liquid on the substrate, making the liquid coalesce and become continuous again, see figure \ref{f:PrintingAcrossFiveDrops2}(g). This illustrates that the dynamics of the consecutively printed and coalescing droplets can be subtly different from the single-droplet dynamics. Note that while the narrower scratch does not cause a break in continuity, the thinner parts of the printed line could cause problems such as higher resistance or local heat generation.

\begin{figure}
	\centering
    \begin{subfigure}[b]{0.495\textwidth}
		\centering
		\includegraphics[width=\textwidth, trim={0cm 0cm 0cm 0cm},clip]{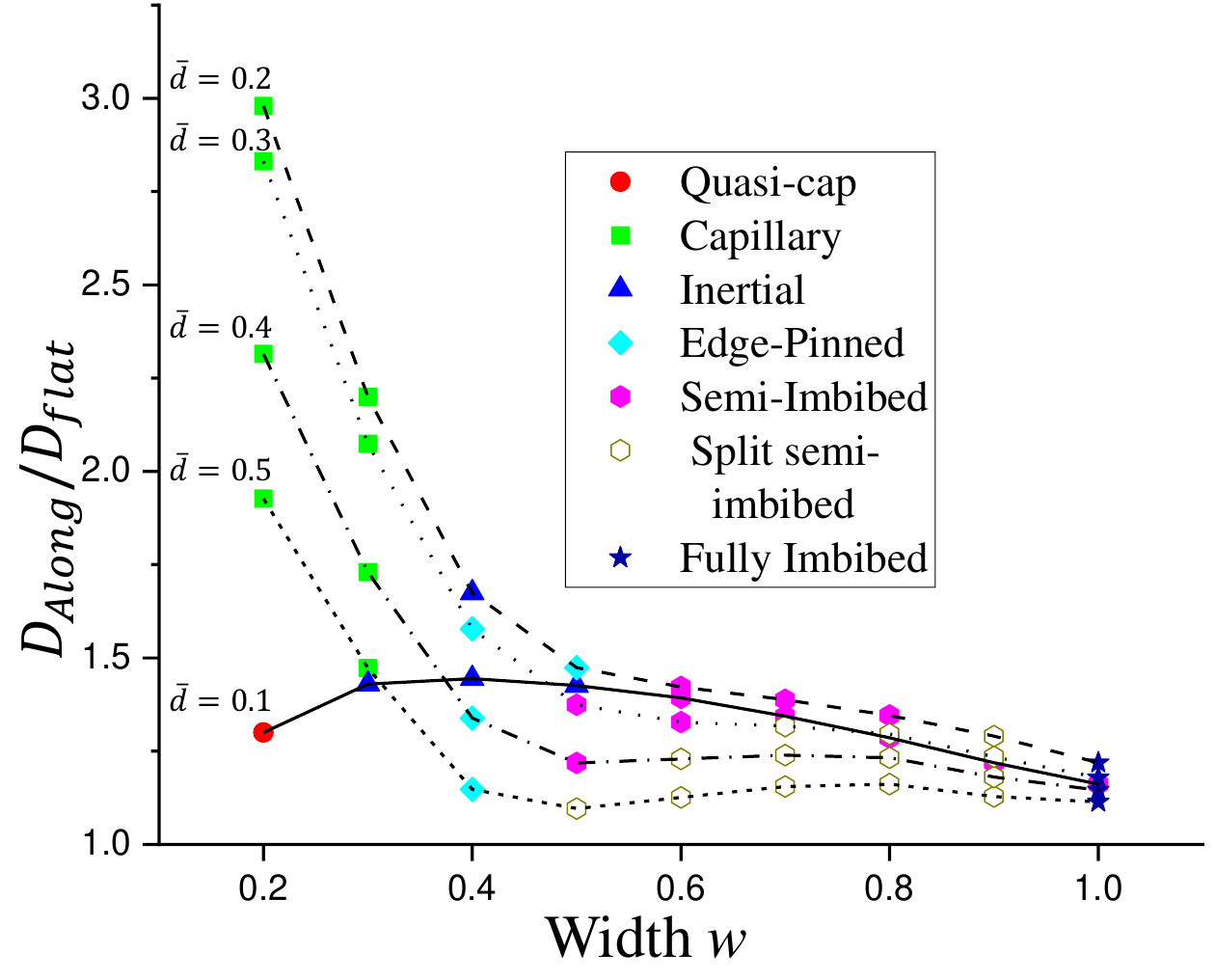}
		\caption{{}}
		\label{subfig:Edge-Pinned Dynamics}
	\end{subfigure}
  \begin{subfigure}[b]{0.495\textwidth}
		\centering
		\includegraphics[width=\textwidth, trim={0cm 0cm 0cm 0cm},clip]{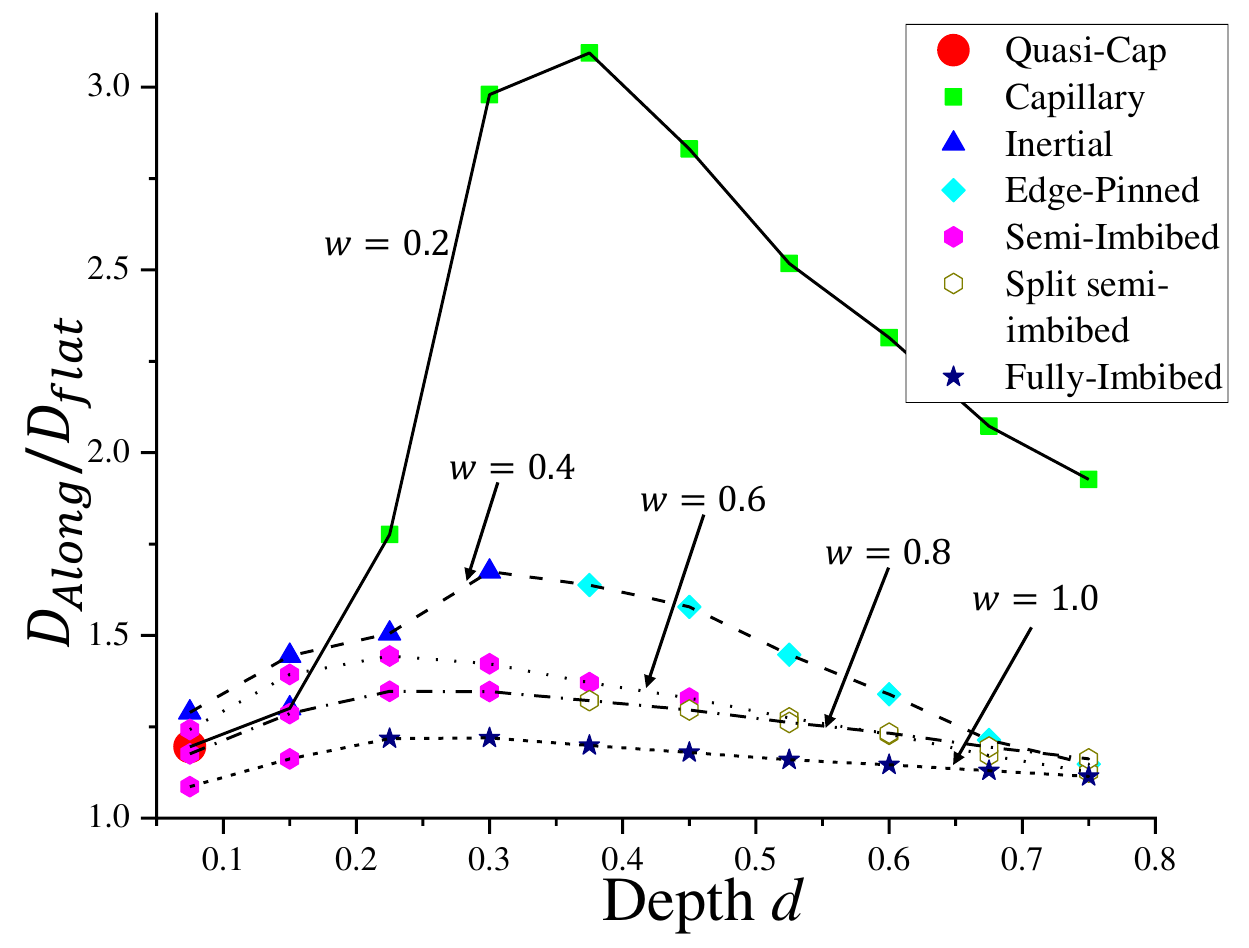}
		\caption{{}}
		\label{subfig:Edge-Pinned Dynamics}
	\end{subfigure}
	\caption{Value of $D_{Along}$, the length of the final single-droplet morphology in the direction along a scratch following deposition of a droplet on the scratch shown in figure \ref{FigGeometry} at $Re=204$, $We=26$, with $\theta_A=75^\circ$ and $\theta_R=1^\circ$. The plots show variation with scratch width and depth, and the coloured symbols indicate the type of morphology following the same labelling as in figure \ref{fig:RegimeMapSingleSmooth}.}
	\label{f:Dalong}
\end{figure}

The extent of liquid spreading along a scratch due to a single droplet impact, $D_{Along}$ (see figure \ref{fig:RegimeMapSingleSmooth}), is also an important consideration in line printing, and this is captured in figure \ref{f:Dalong}. For a line crossing the scratch, spreading along the scratch would create a variation in the thickness of the line that could potentially lead to instability and formation of bulges along the line. As can be seen in figure \ref{f:Dalong}, all dimensions of scratch lead to enhanced spreading along the scratch direction, with the most significant extent being of course that corresponding to the deep, narrow scratches where capillary flow occurs. The long filaments of the capillary morphology could be problematic if, for instance, two parallel lines are being printed in close proximity for a printed circuit: they could cause the two lines to connect unintentionally, resulting in a short circuit. However, because of the slower time scale of the capillary flow, as seen in figure \ref{fig:DimensionsVsSctrachDynamics}(c), this issue could perhaps be avoided using a fast enough curing mechanism or by creating a larger advancing contact angle. On the other hand, this morphology can be exploited to connect two lines by designing such a feature and using a slowly evaporating ink.

\subsection{Printing along a scratch}
\begin{figure}
	\centering
	\begin{subfigure}[b]{0.23\textwidth}
		\centering
		\includegraphics[width=\textwidth, trim={2.2cm 4cm 2.2cm 9cm},clip]{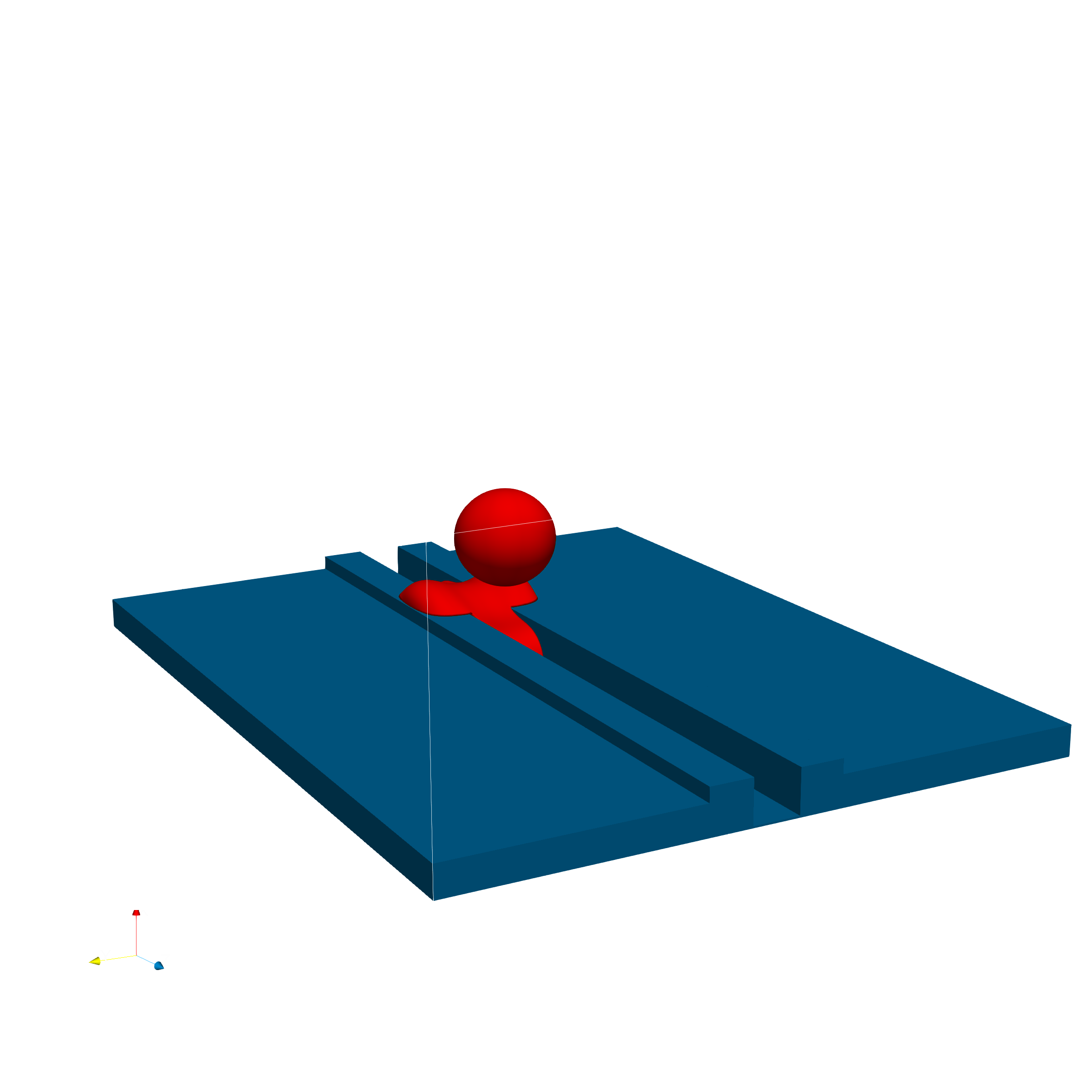}
		\caption{}
		\label{subfig:Along1impact2T1}
	\end{subfigure}
	\begin{subfigure}[b]{0.23\textwidth}
		\centering
		\includegraphics[width=\textwidth, trim={2.2cm 4cm 2.2cm 9cm},clip]{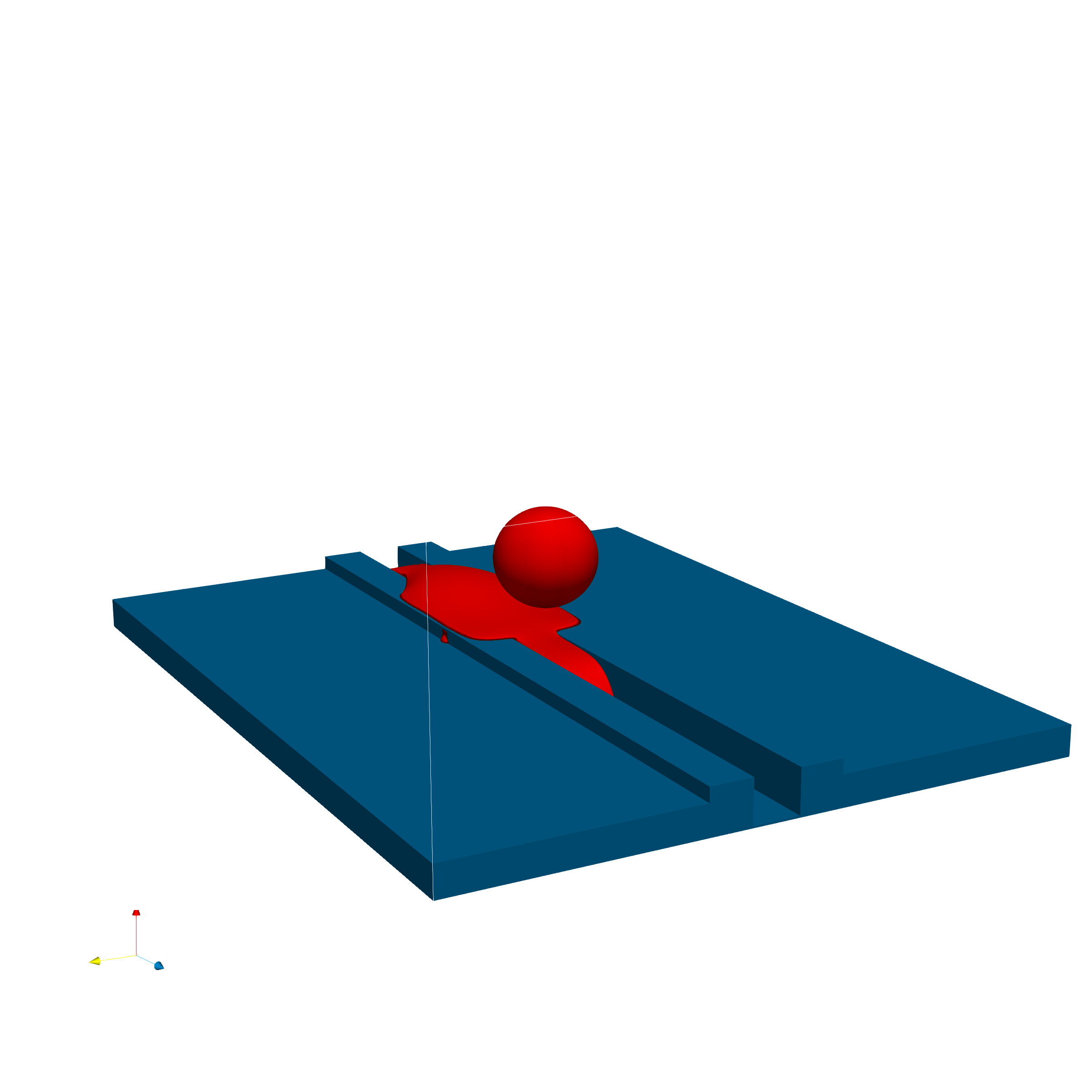}
		\caption{}
		\label{subfig:Along1impact2T2}
	\end{subfigure}
	\begin{subfigure}[b]{0.23\textwidth}
		\centering
		\includegraphics[width=\textwidth, trim={2.2cm 4cm 2.2cm 9cm},clip]{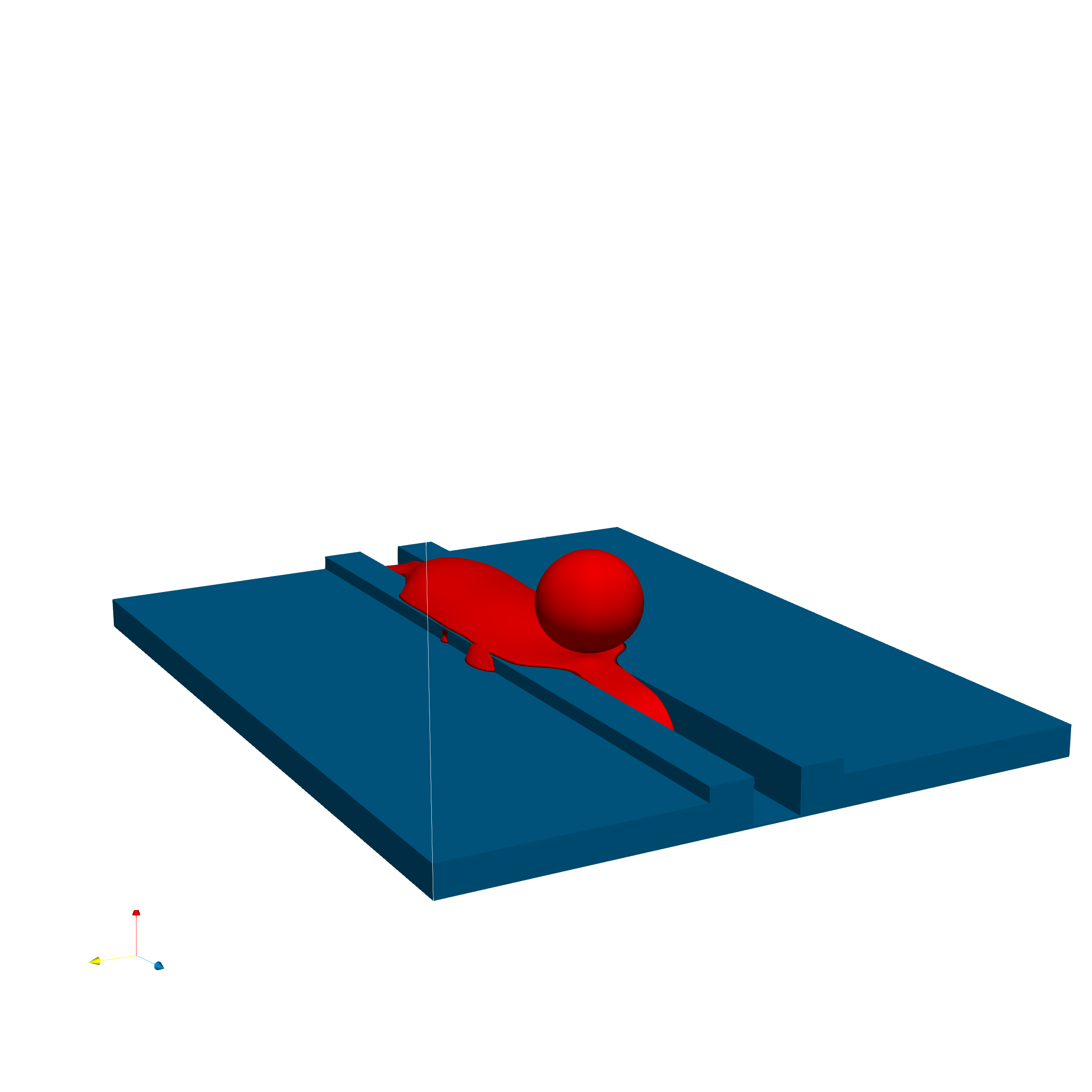}
		\caption{}
		\label{subfig:Along1impact2T3}
	\end{subfigure}
	\begin{subfigure}[b]{0.23\textwidth}
		\centering
		\includegraphics[width=\textwidth, trim={2.2cm 4cm 2.2cm 9cm},clip]{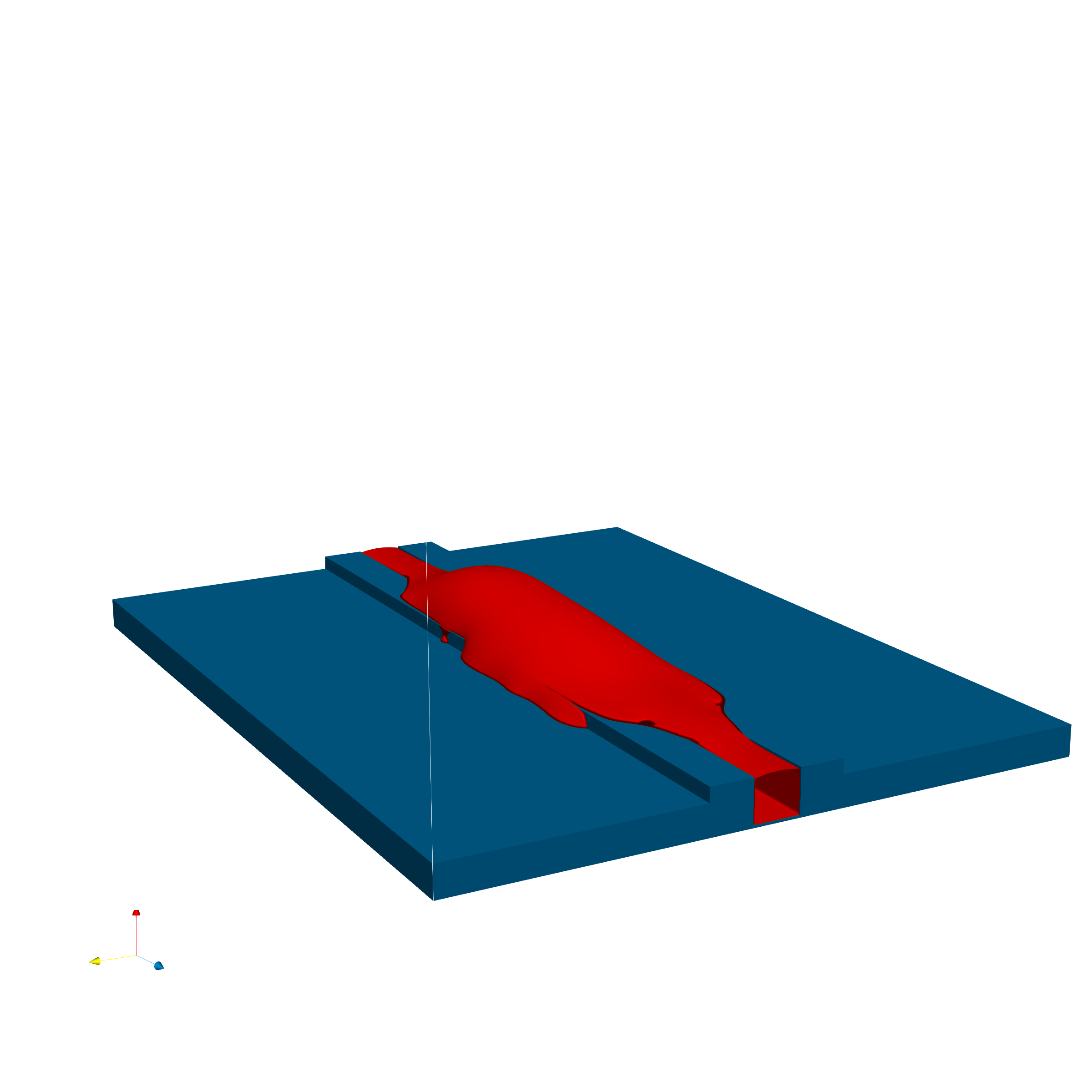}
    \caption{}
    \label{subfig:Along1impact2T4}
	\end{subfigure}
    \\
	\begin{subfigure}[b]{0.23\textwidth}
		\centering
		\includegraphics[width=\textwidth, trim={2.2cm 4cm 2.2cm 9cm},clip]{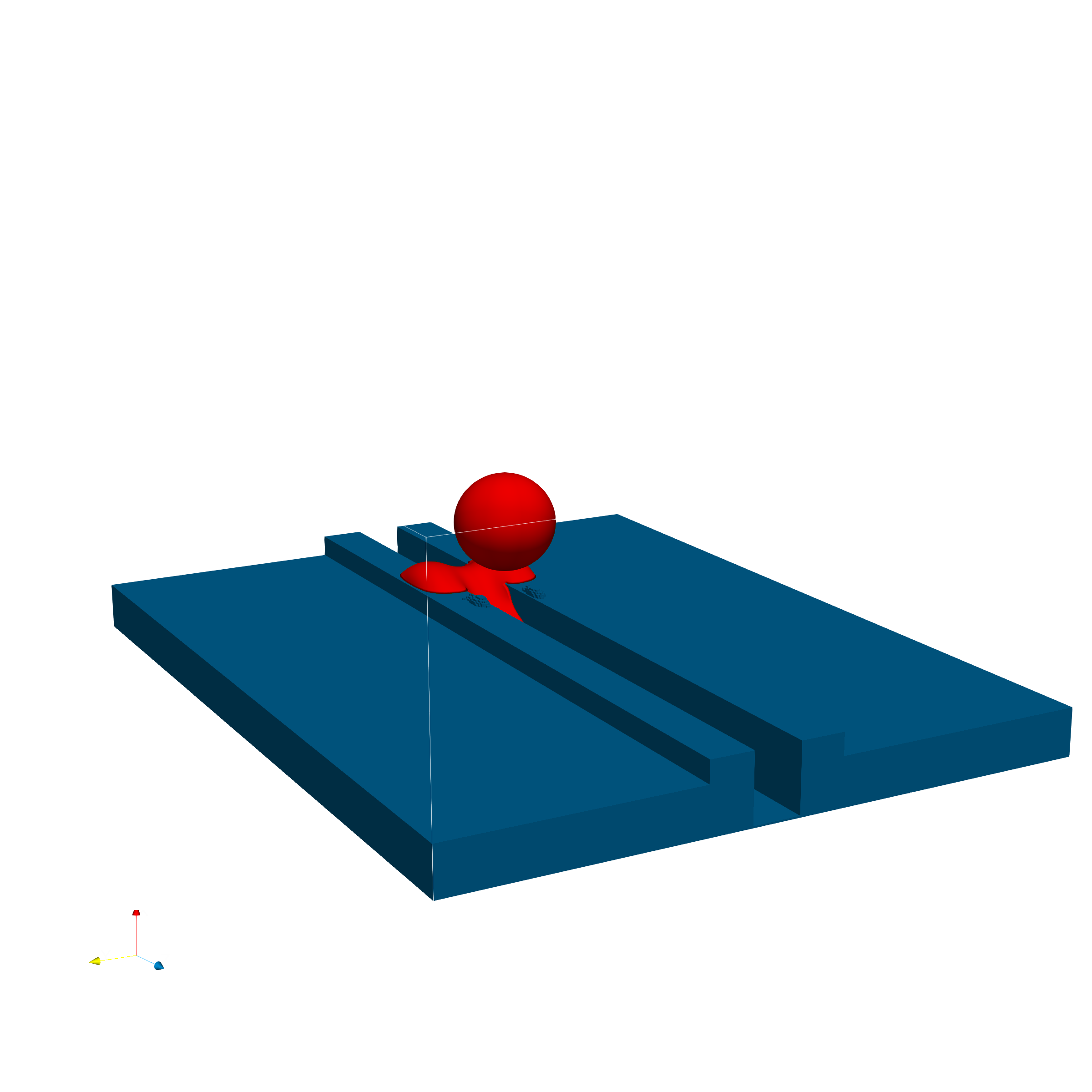}
		\caption{}
		\label{subfig:Along2impact2T1}
	\end{subfigure}
	\begin{subfigure}[b]{0.23\textwidth}
		\centering
		\includegraphics[width=\textwidth, trim={2.2cm 4cm 2.2cm 9cm},clip]{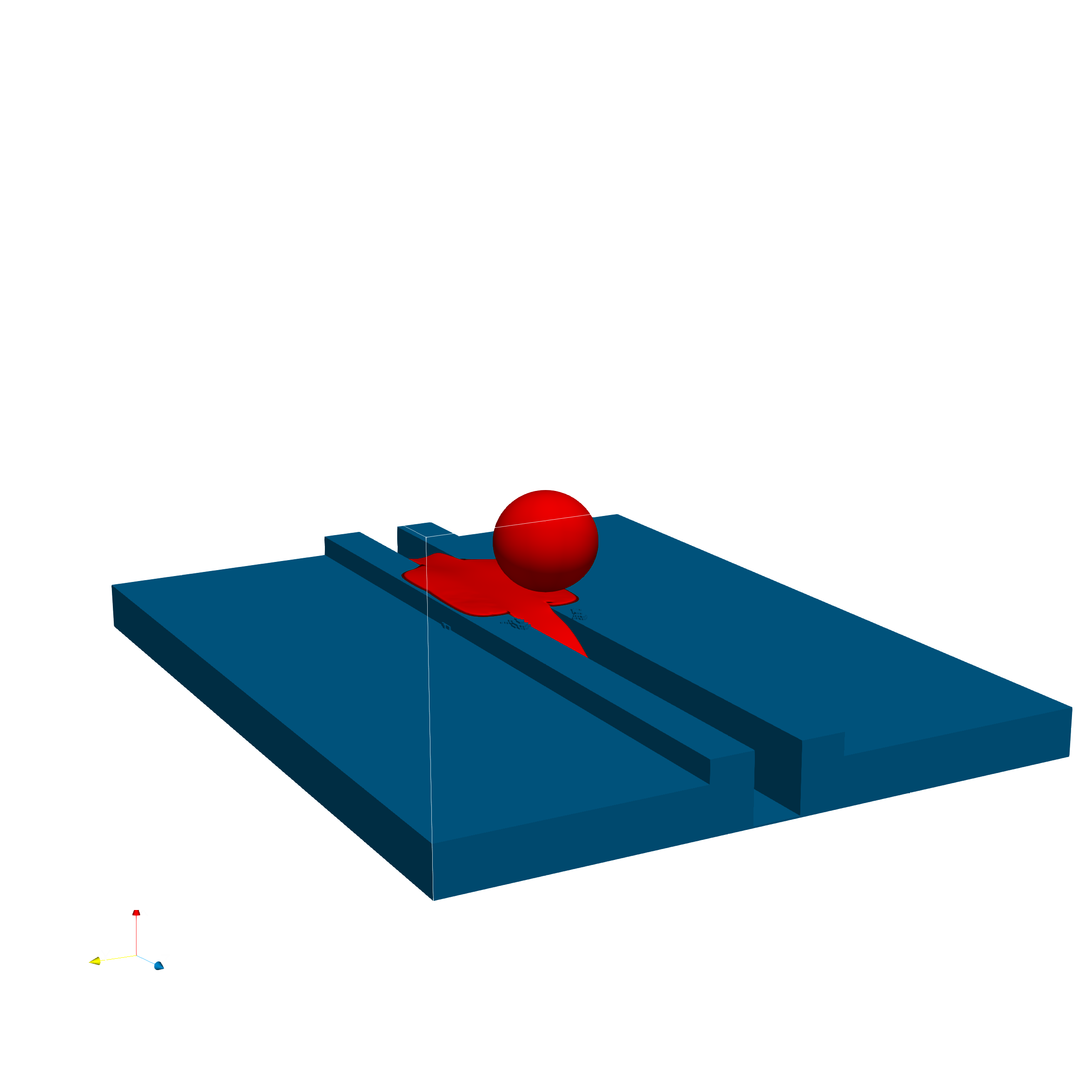}
		\caption{}
		\label{subfig:Along2impact2T2}
	\end{subfigure}
	\begin{subfigure}[b]{0.23\textwidth}
		\centering
		\includegraphics[width=\textwidth, trim={2.2cm 4cm 2.2cm 9cm},clip]{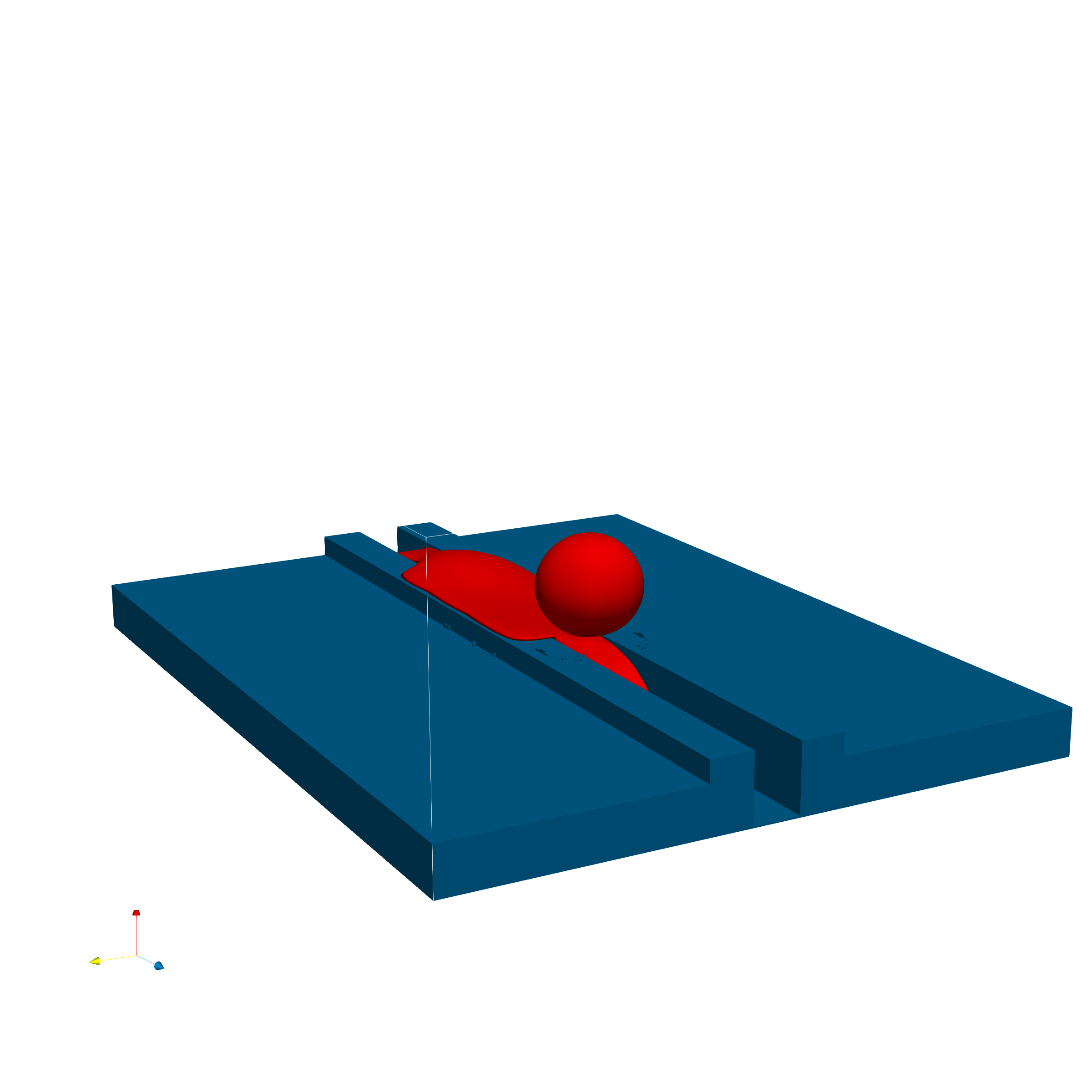}
		\caption{}
		\label{subfig:Along2impact2T3}
	\end{subfigure}
	\begin{subfigure}[b]{0.23\textwidth}
		\centering
		\includegraphics[width=\textwidth, trim={2.2cm 4cm 2.2cm 9cm},clip]{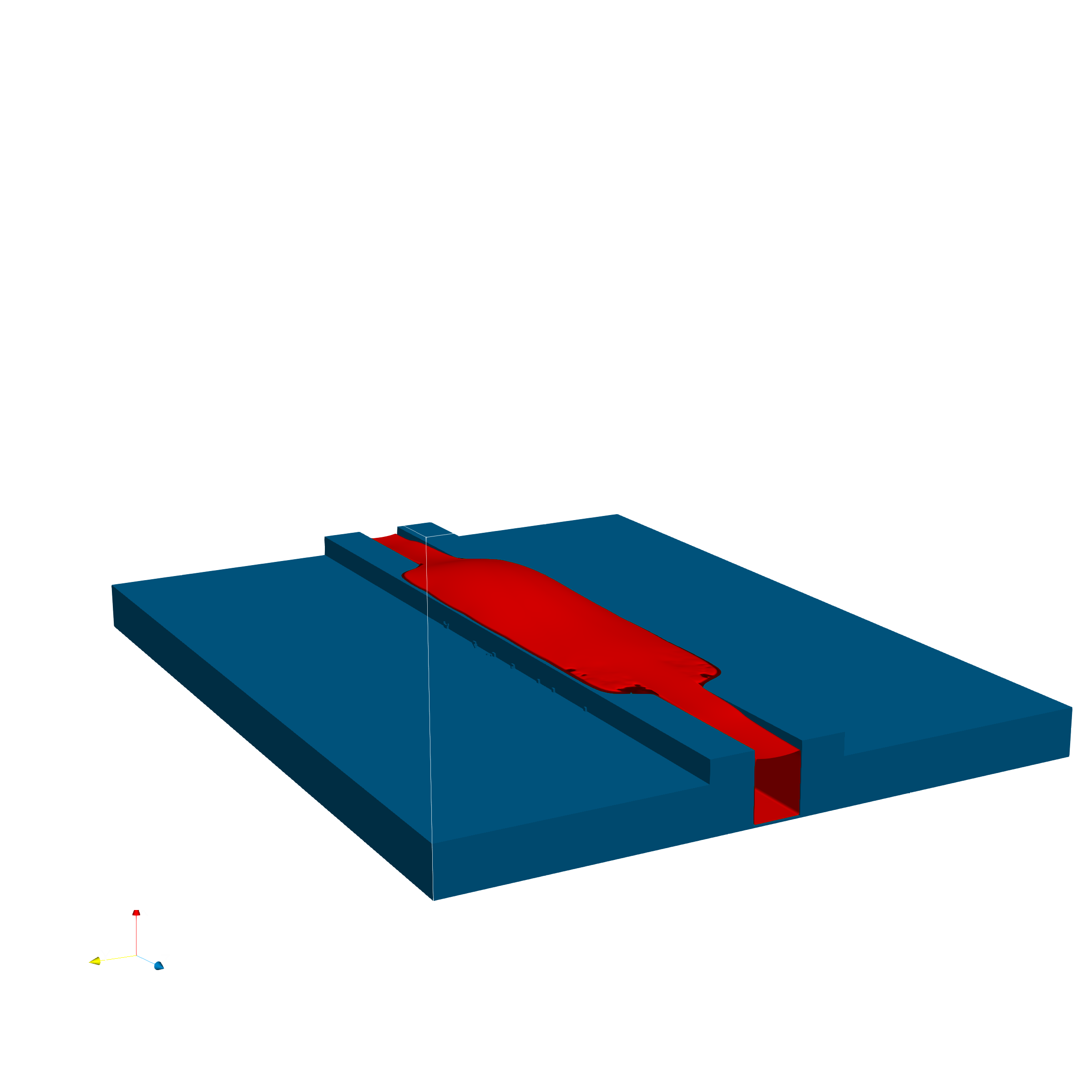}
		\caption{}
		\label{subfig:Along2impact2T4}
	\end{subfigure}
	\caption{Simulations of printing five consecutive droplets along scratches with width $w=0.4$ and depths $\bar{d}=0.3$ (a-d) and $\bar{d}=0.45$ (e-h). }
	\label{f:PrintingAlongFiveDrops2}
\end{figure}

The edge-pinned morphology is an example of how structured substrates can be exploited to print lines with sharp edges. To demonstrate this effect, five droplets are printed into a groove with depths of $\bar{d}=0.3$ and $\bar{d}=0.45$ and width $w=0.4$, both corresponding to the edge-pinned morphology. The results are visualised in figure  \ref{f:PrintingAlongFiveDrops2}. In the first simulation, ($\bar{d}=0.3$ and $w=0.4$), although the first droplet forms a sharp edge as expected, subsequent droplets spill over as seen in figure \ref{f:PrintingAlongFiveDrops2}(c) and (d). This occurs because the precursor droplet inside the groove is in the spreading path of the subsequent droplet; this causes it to spill over. Increasing the depth to $0.45$, however, allows more volume for droplet spreading inside the groove, so overspill does not occur and a sharp line is formed along the outer edges of the side ridges, as seen in figure \ref{f:PrintingAlongFiveDrops2}(h). These two simulations demonstrate how challenging printing a sharp line can be.
 
The simulations of printing along and across a scratch show that topographical features of commensurate size to droplets have a significant effect not only on a single droplet but also a series of droplets.

\section{Conclusions}\label{conclusionsSection}
The deposition of micro-droplets onto a scratched substrate has been investigated using an idealised scratch comprising of a groove of rectangular cross-section, with rectangular side ridges representing material displaced from the groove. Seven distinct equilibrium morphologies arise as a result of inertial spreading, contact-line pinning on various features of the topography, imbibition of the droplet into the scratch and capillary flow along it. These morphologies occur for distinct ranges of scratch depth and width, relative to the droplet size, which define regions of a regime map.

Adapting existing models for the maximum spreading diameter of a droplet on a flat surface to account for liquid entering the scratch, theoretical estimates of the boundaries within the regime map have been obtained that show good agreement with numerical predictions using a 3-D multiphase lattice Boltzmann model implemented on a GPU architecture and validated against relevant previously published experiments. 

Despite being developed for much larger droplets, some of the theoretical models for droplet spreading diameter showed good agreement with the micro-droplet simulation predictions, suggesting that they can be successfully applied to micro-droplet impact. In particular, the model of \citet{Roisman2009} is arguably the best both in terms of its predictions and its explicit nature, making it easier to use than more complicated models. 

From a practical perspective, the interaction of droplets with a scratch of commensurate size can be detrimental in different ways. When inkjet printing a track across a scratch, the shortening of the spreading in the direction perpendicular to the scratch means droplets may not join as intended, resulting in line breaks. Alternatively, for sufficiently deep and narrow scratches, capillary flow along the scratches could lead to unintentional connections between parallel tracks; in the case of printed electronics this can result in malfunctioning circuits. However, this also suggests opportunity to exploit intentional features (such as those described by  \citet{nie2008patterning}, or \citet{seemann2005wetting}) to control spreading and maintain a uniform track on the substrate. As printing resolutions improve, and droplet sizes decrease, the results show that consideration of the substrate’s topographical features becomes increasingly important in achieving desired printing outcomes.
\FloatBarrier
\section*{Acknowledgements}
The authors are grateful to the Engineering and Physical Sciences Research Council (EPSRC) Centre for Doctoral Training in Fluid Dynamics at the University of Leeds (Grant No. EP/L01615X/1) and Dupont Teijin Films (DTF) for financial support. We thank Andrew Bates and Kieran Looney from DTF and Karrar Al-Dirawi from the University of Leeds for useful discussions and insight. We also thank the High Performance Computing facilities at the University of Leeds, where the simulations in this work were performed on machines ARC3 and ARC4.

\section*{Declaration of Interest}
The authors report no conflict of interest.

\appendix
\section{Computational details}\label{appA}
\subsection{Discrete velocities}
The discrete velocities used are the D3Q19 ones written as,
\footnotesize
\setlength{\arraycolsep}{2.5pt}
\medmuskip = 1mu 

\[\boldsymbol{e}_\alpha= \left[\begin{array}{lcccccccccccccccccr}
0& \hspace{0.3cm}1&               -1&  \hspace{0.3cm} 0&  \hspace{0.3cm}0& \hspace{0.3cm}0&  \hspace{0.3cm}0&                1&               -1&  \hspace{0.3cm}1&                -1&  \hspace{0.3cm}1&               -1&  \hspace{0.3cm}1&               -1& \hspace{0.3cm} 0&  \hspace{0.3cm} 0& \hspace{0.3cm}0& \hspace{0.3cm}0 \\
0& \hspace{0.3cm}0&  \hspace{0.3cm}0&  \hspace{0.3cm} 1&               -1& \hspace{0.3cm}0&  \hspace{0.3cm}0&                1&               -1&               -1&   \hspace{0.3cm}1&  \hspace{0.3cm}0&  \hspace{0.3cm}0&  \hspace{0.3cm}0&  \hspace{0.3cm}0& \hspace{0.3cm} 1&                -1& \hspace{0.3cm}1&              -1 \\
0& \hspace{0.3cm}0&  \hspace{0.3cm}0&  \hspace{0.3cm} 0&  \hspace{0.3cm}0& \hspace{0.3cm}1&               -1&  \hspace{0.3cm}0&  \hspace{0.3cm}0&  \hspace{0.3cm}0&   \hspace{0.3cm}0&  \hspace{0.3cm}1&               -1&               -1&  \hspace{0.3cm}1& \hspace{0.3cm} 1&                -1&              -1& \hspace{0.3cm}1
\end{array}\right]\]

\normalsize
  The source term used to incorporate the interaction force can be written as,

\begin{equation}
\setlength{\arraycolsep}{0pt}
\renewcommand{\arraystretch}{1.3}
\boldsymbol{S} = \left[
\begin{array}{c}
0 \\
F_x \\
F_y\\
F_z\\
2\boldsymbol{F}\cdot\boldsymbol{u}+\frac{6\sigma |\boldsymbol{F} |^2}{\psi^2(s_\zeta^{-1}-0.5)}\\
2(2F_xu_x - F_yu_y - F_zu_z) \\
2(F_yu_y-F_zu_z)\\
F_xu_y+F_yu_x \\
F_xu_z+F_zu_x \\
F_yu_z+F_zu_y \\
c_s^2F_y \\
c_s^2F_x \\
c_s^2F_z \\
c_s^2F_x \\
c_s^2F_z \\
c_s^2F_y \\
2c_s^2(u_xF_x+u_yF_y)\\
2c_s^2(u_xF_x+u_zF_z)\\
2c_s^2(u_yF_y+u_zF_z)\\
\end{array}  \right] .
\label{defQc}
\end{equation}
where $\boldsymbol{F}$ is the total force with components $F_x$, $F_y$ and $F_z$.

\subsection{Geometric Wetting Boundary Condition} \label{AppendixGeomCondition}
\begin{figure}
	\centering
	\includegraphics[width=0.8\textwidth, trim={0cm 0cm 0 0cm},clip]{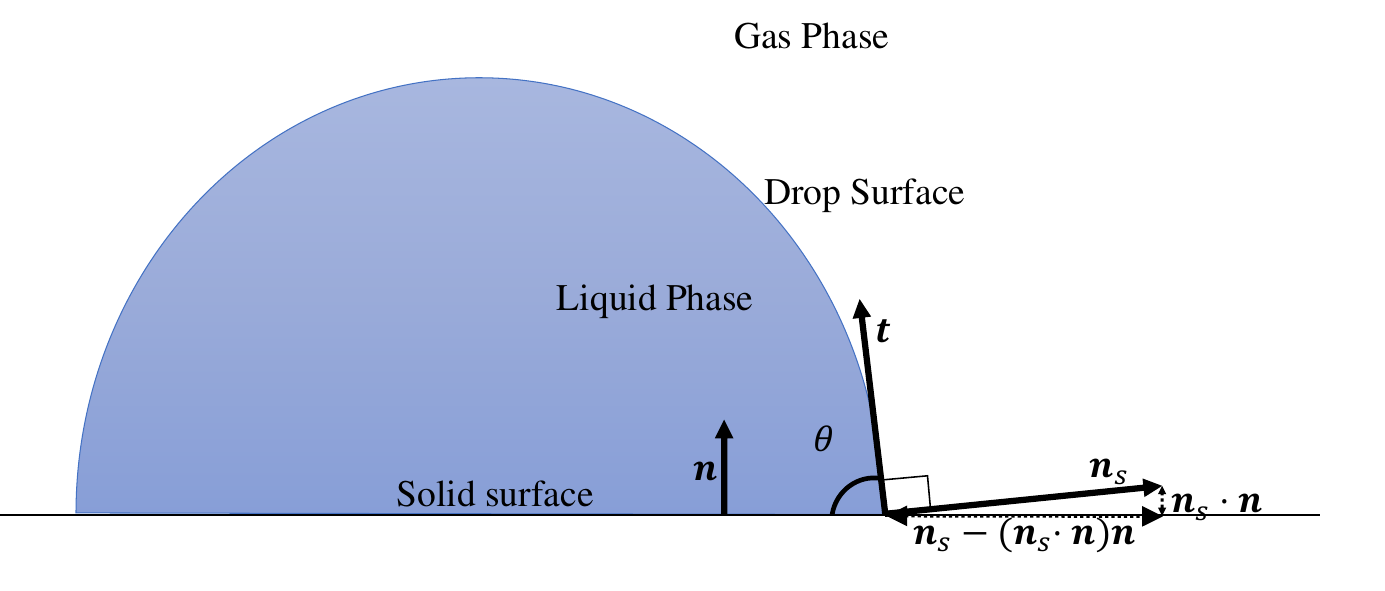}
	\caption{Schematic of the geometry at the three-phase contact line where $\boldsymbol{n}$ is the unit normal to the solid surface, and $\boldsymbol{n}_s$ and $\boldsymbol{t}$ are the unit normal and unit tangent to droplet surface. The unit normal to the fluid surface can be calculated using the density field which in turn can be used in geometric arguments to calculate the density on the solid surface to satisfy the contact angle $\theta$. }
	\label{GeomBoundCondition}
\end{figure}

Consider a droplet spreading on a solid surface with unit normal $\boldsymbol{n}$. The unit normal and unit tangent to the droplet surface are $\boldsymbol{n}_s$ and $\boldsymbol{t}$ respectively. Since the droplet is made of the liquid (heavy) phase submerged in the gas (light) phase, the density gradient at the droplet surface will point in the direction of  $-\boldsymbol{n}_s$. Therefore,

\begin{equation}
\boldsymbol{n}_s = - \frac{\boldsymbol{\nabla}\rho}{ | \boldsymbol{\nabla} \rho | }.
\label{NormalToDropInTermsOfTheta}
\end{equation}
Looking at figure \ref{GeomBoundCondition}, an expression for $\theta$ is derived,

\begin{equation}
\tan\left( \frac{\upi}{2} - \theta \right) = \frac{ \boldsymbol{n_s \cdot n} }{ |\boldsymbol{n}_s - (\boldsymbol{n}_s \boldsymbol{\cdot} \boldsymbol{n})\boldsymbol{n}   | }.
\label{FirsEqForTheta}
\end{equation}
Substituting equation \eqref{NormalToDropInTermsOfTheta} into \eqref{FirsEqForTheta} and simplifying,
\begin{equation}
\tan\left( \frac{\upi}{2} - \theta \right) = \frac{-\boldsymbol{\nabla}\rho \boldsymbol{\cdot} \boldsymbol{n}}{|\boldsymbol{\nabla}\rho- (\boldsymbol{\nabla}\rho \boldsymbol{\cdot} \boldsymbol{n})\boldsymbol{n}| }.
\label{ContGeomCond}
\end{equation}

\begin{figure}
	\centering
	\includegraphics[width=0.9\textwidth, trim={0cm 0cm 0 0cm},clip]{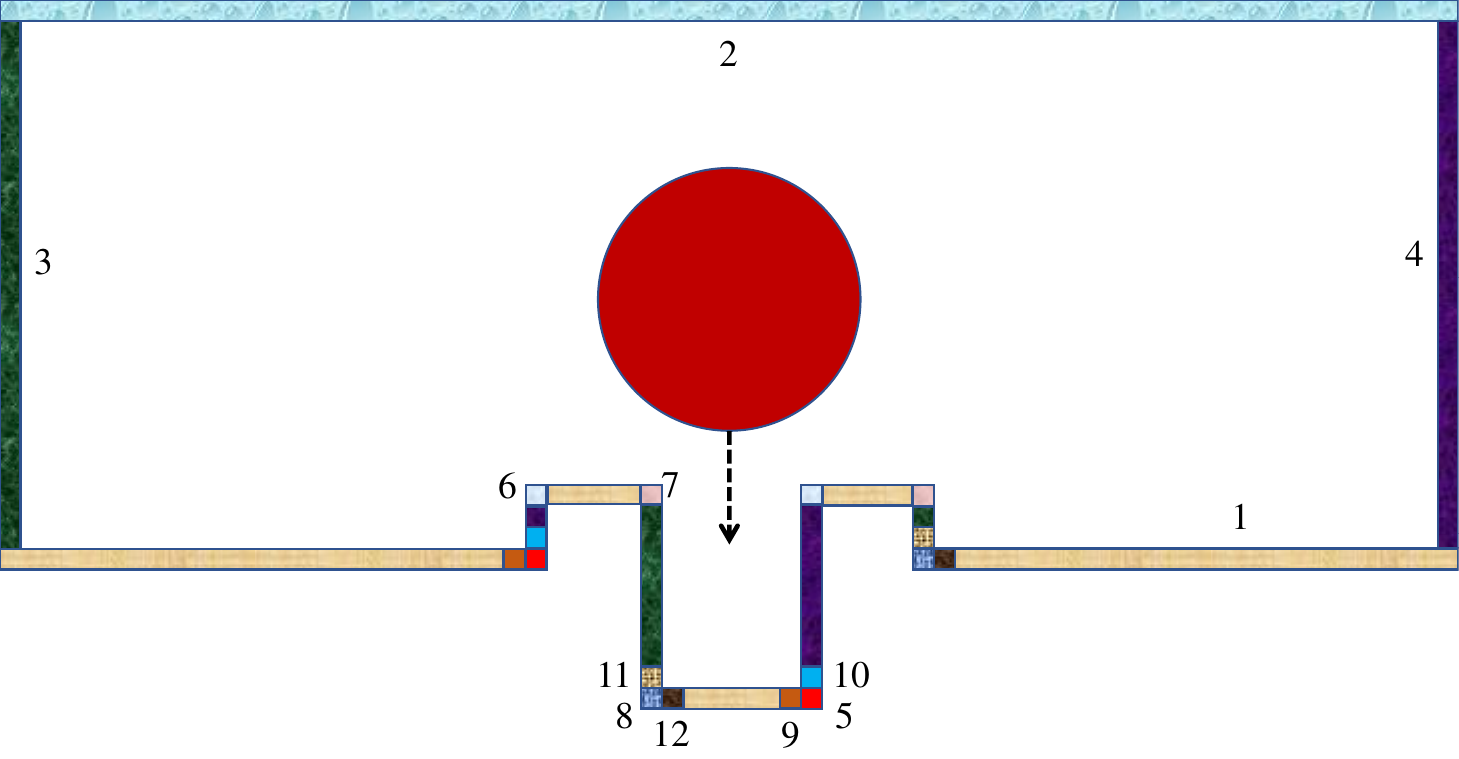}
	\caption{The fluid domain is surrounded by adjacent ghost lattice sites to apply the boundary conditions. }
	\label{BoundariyTypes}
\end{figure}

Equation \eqref{ContGeomCond} was discretised differently for the various parts of the geometry depending on the local normal $\boldsymbol{n}$. The geometry is illustrated through a cross-section seen in figure \ref{BoundariyTypes}, with the various geometry types numbered. Similar boundary conditions are labelled with similar patterns. The fluid domains are surrounded by ghost lattice sites whose density is calculated to satisfy a pre-determined contact angle. The density at ghost lattice sites of type 1 can be calculated using equation \eqref{ContGeomCond}, giving
\begin{equation}
\rho_{ijk} = \rho_{ij+2k} + \tan\left(\frac{\pi}{2}-\theta\right)\zeta,
\end{equation}
where,
\begin{equation}
\zeta = \sqrt{(\rho_{i+1jk}-\rho_{i-1jk})^2 + (\rho_{ijk+1}-\rho_{ijk-1})^2}.
\end{equation}
A similar form can be used for lattice sites of type 2,3 or 4. For corner 7,
\begin{equation}
\rho_{ijk} = \rho_{i+2j+2k} + \tan\left(\frac{\pi}{2}-\theta\right)\zeta,
\label{cornersBoundCond}
\end{equation}
where,
\begin{equation}
\zeta = \sqrt{(\rho_{i+1j+3k}-\rho_{i+3j+1k})^2+2(\rho_{i+2j+2k+1}-\rho_{i+2j+2k-1})^2}.
\label{zetaBoundCond}
\end{equation}
A similar form is used for all other corners. Equations \eqref{cornersBoundCond} and \eqref{zetaBoundCond} cannot be used for solid lattice sites directly adjacent to corner lattice sites because a solid lattice site might be used to update another solid lattice site. Instead, a second-order accurate forward difference scheme is used for lattice sites of type 11,
\begin{equation}
\rho_{ijk} = \rho_{ij-2k} + \tan\left(\frac{\pi}{2}-\theta\right)\zeta,
\label{cornersBoundCond2}
\end{equation}
where,

\begin{equation}
\zeta = \sqrt{(-\rho_{i+3j-1k}+4\rho_{i+2j-1k})^2-3(\rho_{i+1j-1k+1}-\rho_{ij-1k-1})^2}.
\label{zetaBoundCond2}
\end{equation}
A similar form was used for all solid lattice sites neighbouring a corner.

\subsection{Constants in the Carnahan-Starling equation of state}\label{appendix:constantEOS}

\begin{figure}
  \centering
      \includegraphics[width=0.7\textwidth]{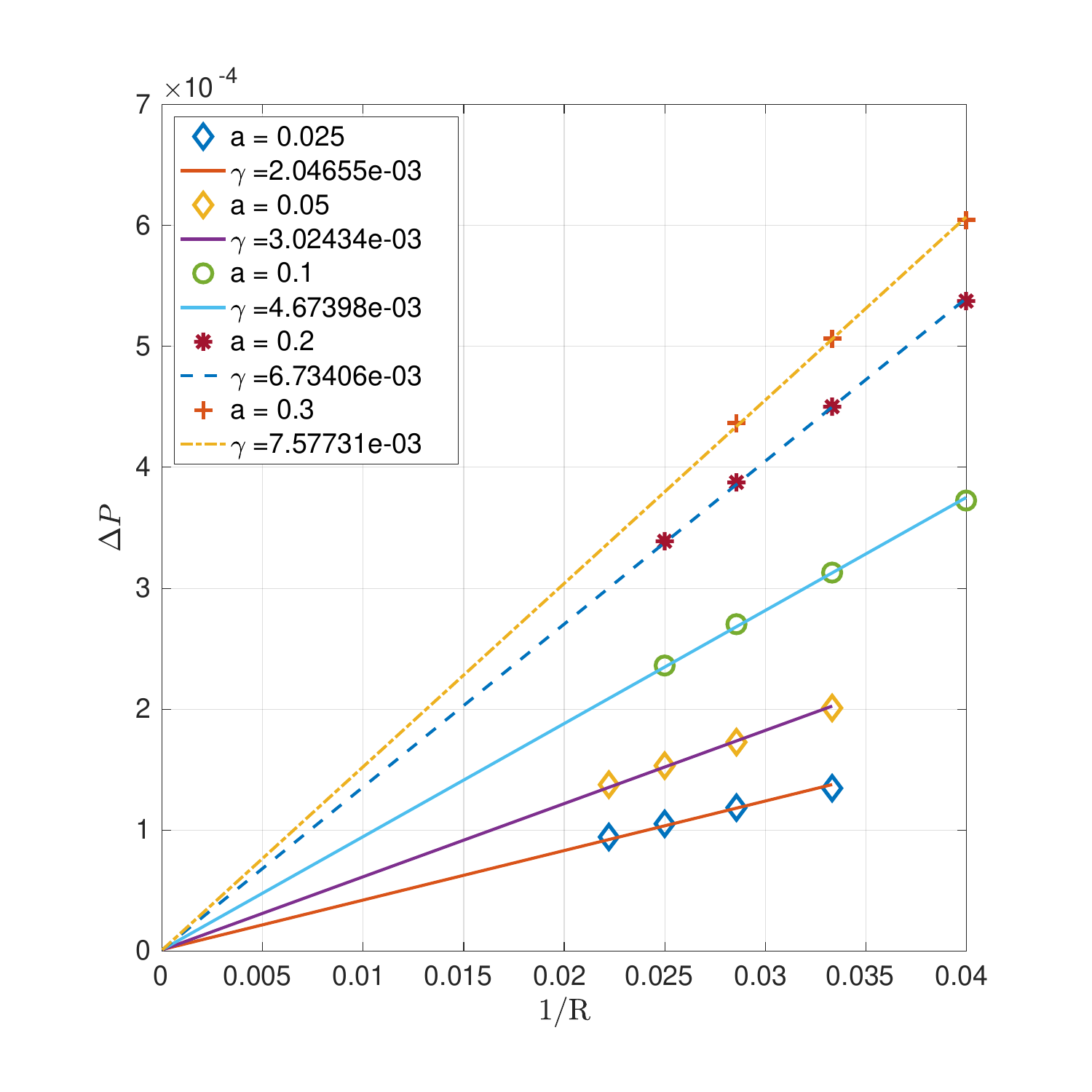}
  \caption{Young--Laplace test of the pseudo-potential model with the modifications from \citet{Li2013b,Li2018}; $a$ is the parameter from the Carnahan-Starling equation of state \eqref{CSEOS}. Increasing $a$ increases surface tension $\gamma$ (all quantities given in lattice units).}
  \label{f:SurfaceTensionTestLi}
\end{figure}

Here, we clarify our choice of values for the constants $a$ and $b$ in the equation of state \eqref{CSEOS}. Surface tension is an emergent property in the pseudo-potential model, rather than an input parameter, so to calculate surface tension, the Young--Laplace law is used. A series of droplets with varying radii $R$ are simulated and the pressure difference ($\Delta p$) between the inside of the droplets and their surrounding is calculated. Plotting $\Delta p$ as a function of $1/R$, a linear relationship is seen, with the gradient being twice the surface tension $\gamma$ in three dimensions. Doing this for various values of the $a$ constant in equation \eqref{CSEOS}, we can see in figure \ref{f:SurfaceTensionTestLi} that surface tension increases with $a$, because $a$ represents the strength of interaction between molecules. It was also observed that the simulations are more stable for lower $a$ values. We therefore choose a value of $a=0.05$ that is low enough to be stable and high enough to capture sufficiently large surface tension to simulate micro-droplets in the inkjet parameter space. For more information on the constant $a$, refer to \citet{Li2018}. The value of $b$ is set to 4 in this work. A different value can be chosen and it should not affect the scheme. The particular choice of $b=4$ simplifies equation \eqref{CSEOS} and reduces the number of arithmetic operations; it is the conventionally used value in multiphase pseudo-potential LBM simulations that employ the Carnahan-Starling equation of state.

\FloatBarrier
\bibliographystyle{jfm}
\bibliography{main.bib}
\end{document}